\documentclass[aps, prc, twocolumn, floatfix, superscriptaddress, nofootinbib, preprintnumbers, longbibliography, amsmath, amsthm, amssymb]{revtex4-2}
\usepackage{graphicx}
\usepackage{amsfonts}
\usepackage{color}
\usepackage[colorlinks=true,linkcolor=blue,citecolor=blue]{hyperref}
\usepackage{dcolumn}% Align table columns on decimal point
\usepackage{bm}% bold math
\usepackage{braket}
\usepackage{empheq}
\usepackage[version=4]{mhchem}
\usepackage[normalem]{ulem}  % \sout{old text} for strikeout

\begin{document}
\allowdisplaybreaks[1]
\title{Nuclear Responses to Two-Body External Fields Studied with the Second Random-Phase-Approximation}
\author{Futoshi Minato}
\email[E-mail: ]{minato.futoshi.009@m.kyushu-u.ac.jp}
\affiliation{Department of Physics, Kyushu University, Fukuoka 819-0395, Japan}
\affiliation{RIKEN Nishina Center for Accelerator-Based Science, Wako, Saitama 351-0198, Japan}
%\affiliation{Nuclear Data Center, Japan Atomic Energy Agency, Tokai, Ibaraki 319-1195, Japan}
%
\date{\today}
\begin{abstract}
This study investigates nuclear responses to two-body external fields, interpreted as double-phonon excitations, within the subtracted second random-phase approximation (SSRPA) for $^{16}$O and $^{40}$Ca. 
To clarify the underlying characteristics of these modes, Hartree–Fock~(HF) and SSRPA with the diagonal approximation are first examined. 
The resulting strength distributions are nearly identical, indicating that residual interactions in the 1p-1h sector contribute only weakly.
This behavior contrasts with that of one-body excitations, where coupling between 1p-1h and 2p-2h configurations is essential for generating collectivity. 
In the full SSRPA calculation, which incorporates the residual interaction among 2p-2h configurations, the strength distributions are substantially modified. 
The double IS $0^{+}$ and $2^{+}$ modes show pronounced redistribution, with peaks shifted to lower energies and additional strength emerging at higher energies, whereas the double IV $1^{-}$ mode shifts predominantly to higher energies due to a largely repulsive interaction, resulting in a resonance around twice the peak energy of the single giant dipole resonance.
Analysis of single transition amplitudes reveals that low-lying resonances are formed coherently through constructive neutron-neutron, proton-proton, and neutron-proton configurations, while high-lying resonances are dominated by neutron-proton configurations, reflecting their higher state density. 
These results demonstrate that double-phonon excitations cannot be described by simple folding of one-body responses; a fully microscopic treatment of 2p-2h mixing, as provided by SSRPA, is essential.
\end{abstract}
\maketitle
\section{Introduction}
\label{intro}
Collective motions are general properties of many-body systems and can be found in electronic systems, biological systems, human society, and so on.
The nucleus is also one of the many-body systems in nature.
Its characteristic feature is that it consists of two types of barions, protons and neutrons, and its dynamics are governed by quantum mechanics with spin and isospin degrees of freedom.
In addition, the nucleus is considered to be an isolated system, allowing one to study quantum many-body problems without external interference.
As a result, nuclei have exhibited many intriguing collective motions and have provided important insights into the behavior of quantum many-body systems.
\par
A typical collective motion of the nucleus is surface vibration and rotation.
In particular, surface vibrations have attracted considerable interests from researcher because of their similarity to collective motions in other many-body systems.
In the first order, nuclear surface vibrations are generally interpreted as collective one-particle-one-hole~(1p--1h) excitations.
One of the most succeeded models in this context is the random-phase approximation~(RPA)~\cite{BohmPines, BohrMottelson, RingandSchuck}.
In particular, self--consistent RPA approaches built on top of the energy density functional~(EDF), which have the advantage of avoiding additional phenomenological parameters besides those determined from the nuclear bulk properties, have achieved remarkable successes over the last two decades.
RPA has also been extended to the description of nuclear reactions in both continuum form~\cite{Bertsch1975, Ichimura2013, Matsuo2015, Saito2023, Sasaki2025} and discrete form~\cite{Dupuis2017, Dupuis2024}.
It has provided valuable insights into the mechanisms of nuclear excitation modes and has played a key role in advancing nuclear structure studies.  
\par
A model that goes beyond the RPA is the second random-phase approximation~(SRPA)~\cite{Providencia1965}, which extends the description of nuclear excitations by including up to two-particle-two-hole~(2p-2h) configurations.
A major advantage of SRPA is that it can account for the spreading width (damping effect), thereby enabling a more direct comparison with experimental data.
This approach has been studied and developed by several groups~\cite{Wambach1988, Drozdz1990, Ait-Tahar1993}.
With the increase in computational resources, this approach has been extended to a self-consistent framework~\cite{Gambacurta2011, Papa2009, Papa2010}.
However, when applied with effective interactions, SRPA encounters the problem of double--counting of higher-order correlations.
To address this issue, the subtracted SRPA (SSRPA) was proposed, introducing a subtraction method~\cite{Tselyaev2007, Gambacurta2015}.
This approach has proven successful in describing experimental strength distributions and has been widely applied to studies of the spreading widths of giant resonances~\cite{Gambacurta2015, Yang2021} and the quenching of Gamow-Teller transitions~\cite{Gambacurta2018, Gambacurta2020, Yang2022, Gambacurta2022, Yang2023}.
Here, I should also mention that another class of approaches beyond RPA is provided by particle--vibrational coupling~\cite{Colo2010, Niu2012}, relativistic quasiparticle time--blocking approximation~\cite{Litvinova2008}, two-phonon models~\cite{Catara1992, Severyukhin2004, Litvinova2013, Gambacurta2016, Arsenyev2023}, and the small amplitude limit of the time-dependent density--matrix theory~\cite{Tohyama2001}.
\par
Models beyond RPA are frequently applied to the study of double–phonon excitations, that is, excitations in which one phonon is built on top of another~\cite{Emling1994, Chomaz1995}.
Such excitations are considered to be induced either by the sequential action of one-body external fields or by direct action of two-body external fields, and have been investigated both experimentally and theoretically in proton-capture reactions~\cite{Dowell1983}, heavy-ion reactions~\cite{Bertulani1988, Bertulani1999, Esbensen2005}, double-charge-exchange reactions~\cite{Auerbach1990, Muto1992, Sagawa2016, Sakaue2024}.
Theoretical approaches based on multiphonon models~\cite{Catara1992, Lanza1997, Severyukhin2004, Litvinova2013, Gambacurta2016, Arsenyev2023, Andreozzi2007, Ponomarev1992, Ponomarev1994}, a semiclassical formalism~\cite{Bertulani1994}, and double time Green's function method~\cite{Dang1999} have also been developed independently.
Some of them are also applied to calculate heavy-ion reactions~\cite{Bertulani1994, Ponomarev1994} (For further details, see also Refs.~\cite{Chomaz1995, Dang1999}).
The SSRPA framework enables the study of nuclear excitations induced by both the sequential operation of one-body external fields and the direct action of two-body external fields~\cite{Tohyama2001, Papa2010}.
The former corresponds to a specific channel of sequential excitation (e.g. double quadrupole excitations coupled to a certain angular momentum), while the latter directly probes the doorway state of nuclear resonances.
\par
Although nuclear excitations due to one-body external fields have been extensively studied, those induced by two-body external fields remain poorly understood.
A comprehensive microscopic understanding of nuclear excitations driven by two-body external fields is therefore essential.
In this regard, Nishizaki and Wambach investigated double-dipole excitations in $^{40}$Ca and $^{208}$Pb within the SRPA framework~\cite{Nishizaki1995, Nishizaki1998}; however, they adopted a Woods-Saxon potential and a relatively limited model space.
The present work shows that the residual interaction between 2p-2h configurations plays an essential role in properly describing double-phonon excitations.
In particular, the present analysis demonstrates that low-lying resonances are coherently formed through constructive neutron--neutron, proton--proton, and neutron--proton configurations, while high-lying resonances are dominated by neutron--proton configurations, reflecting their higher state density.
The present analysis also shows that a sufficiently large model space is essential for a proper description of double-phonon excitations, although double IV dipole excitations are somewhat less sensitive to the model-space size.
\par
Experiments on double-phonon excitations have been carried out mainly in medium- and heavy-mass nuclei; in particular, tin, xenon, gold, lead, and uranium isotopes have been extensively studied (see, e.g., Refs.~\cite{Yeh1996, Bargioni1995, Caballero2005, Pysmenetska2006, Isaak2023} and the review in Ref.~\cite{Emling1994}).
In principle, these nuclei can be treated within the SSRPA framework. 
In practice, however, the required model spaces are prohibitively large with currently available computational resources.
The present work therefore focuses on the $0^{+}$ components of double multipole excitations in $^{16}$O, which are numerically feasible.
Results for $^{40}$Ca are also presented, within a feasible model-space size, to examine whether the qualitative features found in $^{16}$O are generic rather than specific to that nucleus.
Although the present study is limited to two nuclei and a single channel $0^{+}$, the findings are expected to provide valuable insight.
\par
Two-body external fields are also closely related to electromagnetic and neutrino reactions on nuclei, muon capture, and $\beta$-decays through two--body currents~\cite{Lifshitz1980, Minato2023, Menendez2011, Bacca2014, Brase2026}.
Although I do not explicitly address two-body current effects in this paper, I expect that the basic findings obtained here will contribute to a deeper understanding of such effects. 
The present framework can also be straightforwardly extended to describe charge-exchange responses.
\par
The structure of this paper is as follows. 
In Sect.~\ref{sect:theo}, the theoretical framework used in this study is described. 
Sect.~\ref{sect:result} presents and discusses the result, and Sect.~\ref{sect:summary} provides a summary of the findings and conclusions.
\section{Theoretical Framework of SSRPA}
\label{sect:theo}
\subsection{SSRPA formalism}
Although the basic framework of SRPA has been introduced in many literature~\cite{Providencia1965, Yannouleas1987, Ait-Tahar1993, Wambach1988, Drozdz1990, Gambacurta2015, Papa2010, Yang2021}, a brief review is given here because some of quantities are essential for discussing our results.
Excited states with spin $J$ and its projection on $z$-axis $M$ are approximated by $|\nu; JM\rangle=Q_{\nu;JM}^{\dagger}|\mathrm{SRPA}\rangle$ with $Q_{\nu; JM}^{\dagger}$ being the SRPA phonon creation operator defined as
\begin{align}
\nonumber
Q^{\dagger}_{\nu;JM}
&=\sum_{ph}\left(
X_{\nu;ph}^{(J)} O_{ph}^{(JM) \dagger}-(-1)^{J+M}Y_{\nu;ph}^{(J)} O_{ph}^{J-M}
\right)\\
\nonumber
&+\sum_{\substack{p_{1}<p_{2}\\ h_{1}<h_{2}}}
\left(\mathcal{X}_{\nu;p_{1}p_{2}h_{1}h_{2}}^{(J_{p}J_{h})J} O_{p_{1}p_{2}h_{1}h_{2}}^{(J_{p}J_{h})JM \dagger}
\right.\\
&\left.
\qquad
-(-1)^{J+M}\mathcal{Y}_{\nu;p_{1}p_{2}h_{1}h_{2}}^{(J_{p}J_{h})J} O_{p_{1}p_{2}h_{1}h_{2}}^{(J_{p}J_{h})J-M}
\right).
\label{eq:Phonon}
\end{align}
and $|\mathrm{SRPA}\rangle$ is the SRPA correlated ground state.
Particle and hole states of single-particle states are denoted as $p$ and $h$, respectively.
The single-particle energies and wave functions are calculated within the Skyrme energy-density functionals (EDFs) under the spherical symmetry using the SGII force~\cite{Giai1981}.
The corresponding equation for the Skyrme EDFs is 
\begin{equation}
\begin{split}
&\frac{\hbar^{2}}{2m_{q}^{*}}\left(-u''_{\alpha}(r)
+\frac{l(l+1)}{r^{2}}u_{\alpha}(r)\right) 
-\left(\frac{\hbar^{2}}{2m_{q}^{*}}\right)' u'_{\alpha}(r)\\
&+\left(
U_{q}+\frac{1}{r}\left(\frac{\hbar^{2}}{2m_{q}^{*}}\right)'+V_{coul}+\frac{1}{r}W_{q} \langle \vec{l}\cdot\vec{\sigma} \rangle
\right)
u_{\alpha}(r)\\
&=\epsilon_{\alpha} u_{\alpha}(r),
\end{split}
\label{eq:SHF}
\end{equation}
where $\alpha$ denote the set of quantum numbers $n, l, j$ with $n$ the principal quantum number, $l$ the orbital angular momentum, and $j$ the total angular momentum.
The index $q$ denotes a proton or a neutron, and $\langle \vec{l}\cdot\vec{\sigma} \rangle=j(j+1)-l(l+1)-3/4$.
The effective mass $m_{q}^{*}$, the Hartree-Fock potential $U_{q}=U_{q}(r)$, the Coulomb potential $V_{coul}=V_{coul}(r)$, and the spin-orbit potential $W_{q}=W_{q}(r)$ are given in Ref.~\cite{VautherinBrink}.
For the exchange term of the Coulomb potential, the Slater approximation~\cite{Slater1951} is adopted.
The continuum states are discretized by assuming a box size of $20$ fm with a step of $0.1$ fm. 
The operators $O_{ph}^{(JM)\dagger}$ and $O_{p_{1}p_{2}h_{1}h_{2}}^{(J_{p}J_{h})JM\dagger}$ create 1p-1h and 2p-2h states, respectively, and $J_{p}$ and $J_{h}$ are the spins formed by 2 particles and 2 holes, respectively.
The detailed form can be found in Ref.~\cite{Papa2010, Ait-Tahar1993}.
It is worth noting that one needs to avoid double counting in particle and hole configurations so that the sums in Eq.~\eqref{eq:Phonon} run over $p_{1}<p_{2}$ and $h_{1}<h_{2}$.
Neglecting the terms of $\mathcal{X}_{\nu;p_{1}p_{2}h_{1}h_{2}}^{(J_{p}J_{h})J}$ and $\mathcal{Y}_{\nu;p_{1}p_{2}h_{1}h_{2}}^{(J_{p}J_{h})J}$, the formalism corresponds to the standard 1p-1h RPA.
The coefficients of $X_{\nu;ph}^{(J)}, Y_{\nu;ph}^{(J)}, \mathcal{X}_{\nu;p_{1}p_{2}h_{1}h_{2}}^{(J_{p}J_{h})J}, \mathcal{Y}_{\nu;p_{1}p_{2}h_{1}h_{2}}^{(J_{p}J_{h})J}$, and SRPA phonon energy are determined by solving the SRPA equation:
\begin{equation}
\left(
\begin{array}{cc}
A & B\\
-B^{*} & -A^{*}\\
\end{array}
\right)
\left(
\begin{array}{c}
X^{(J)}_{\nu}\\
Y^{(J)}_{\nu}
\end{array}
\right)
=\hbar\omega
\left(
\begin{array}{c}
X^{(J)}_{\nu}\\
Y^{(J)}_{\nu}
\end{array}
\right)
\label{eq:SRPA}
\end{equation}
with
\begin{equation}
A=\left(
\begin{array}{cc}
A_{11'} & A_{12'}\\
A_{21'} & A_{22'}\\
\end{array}
\right),\quad
B=\left(
\begin{array}{cc}
B_{11'} & B_{12'}\\
B_{21'} & B_{22'}\\
\end{array}
\right)
\label{eq:block}
\end{equation}
and
\begin{equation}
X^{(J)}_{\nu}=\left(
\begin{array}{c}
X^{(J)}_{\nu}\\
\mathcal{X}^{(J_{p}J_{h})J}_{\nu}
\end{array}
\right),\quad
Y^{(J)}_{\nu}=\left(
\begin{array}{c}
Y^{(J)}_{\nu}\\
\mathcal{Y}^{(J_{p}J_{h})J}_{\nu}
\end{array}
\right).
\end{equation}
The index numbers $1$ and $2$ denote 1p-1h ($ph$) and 2p-2h ($p_{1}p_{2}h_{1}h_{2}$) configurations, respectively.
The block matrices of $A_{11'}$ and $B_{11'}$ represent matrix elements of the particle-hole residual interaction that couple different 1p-1h configurations, and have the same form as in the ordinary 1p-1h RPA~\cite{RingandSchuck}.
On the other hand, the block matrices $A_{12'}$, $A_{21'}$, $B_{12'}$, and $B_{21'}$ describe the coupling between 1p-1h and 2p-2h configurations.
The block matrices of $A_{22'}$ and $B_{22'}$ represent matrix elements that couple different 2p-2h configurations.
In this work, the matrix elements for $A_{11'}$ and $B_{11'}$ are estimated by the second derivative of energy density with respect to density, while those for other block matrices are calculated directly from the Skryme force, being $B_{12'}=B_{21'}=B_{22'}=0$~\cite{Gambacurta2011}.
\par
It is known that strength distributions calculated by SRPA are shifted toward lower energies compared to those obtained in 1p-1h RPA,
because some of matrix elements of the two-body force are already taken into account in the Skyrme-HF~\cite{Tselyaev2007}.
To overcome this problem, the subtraction method is applied to SRPA~\cite{Gambacurta2015}, and denoted as SSRPA.
The present paper does not go into detail about SSRPA in this paper; however the validity has been discussed in some papers~\cite{Gambacurta2015, Gambacurta2018, Yang2021}.
In the subtracted method, $A_{11'}$ block matrix of SSRPA is modified as
\begin{equation}
A_{11'}^{(\mathrm{S})}=A_{11'}+\sum_{22'}A_{12}A_{22'}^{-1}A_{2'1'},
\label{eq:subtract}
\end{equation}
while the other block matrices remain the same. 
\par
If the coupling among the 2p-2h configurations is neglected, the $A_{22'}$ block will become diagonal.
In this situation, it is given by
\begin{equation}
A_{22'}^{(\mathrm{D})}=
\delta_{p_{1}p_{1}'}\delta_{p_{2}p_{2}'}
\delta_{h_{1}h_{1}'}\delta_{h_{2}h_{2}'}
E_{p_{1}h_{1}p_{2}h_{2}},
\end{equation}
where the unperturbed 2p-2h energy is $E_{p_{1}h_{1}p_{2}h_{2}}=\varepsilon_{p_{1}}+\varepsilon_{p_{2}}
-\varepsilon_{h_{1}}-\varepsilon_{h_{2}}$. 
Here, $\varepsilon_{p}$ and $\varepsilon_{h}$ are the single-particle energies of particle and hole states, respectively, obtained from the Skyrme-HF calculation of Eq.~\eqref{eq:SHF}.
This diagonal approximation is effective in reducing the computational cost because it allows us to avoid calculating the inverse matrix in Eq.~\eqref{eq:subtract}.
To examine the impact of the diagonal approximation within the SSRPA framework, the present study considers two cases:
(1) SSRPA$_{\mathrm{F}}$, the full SSRPA calculation without any diagonal approximation;
(2) SSRPA$_{\mathrm{D}}$, in which the diagonal approximation is applied both to the subtraction method in Eq.~\eqref{eq:subtract} and to $A_{22'}$ in Eq.~\eqref{eq:block}.
In the next section, these results are also compared with those obtained from the unperturbed calculation, which neglects all residual interactions.
Namely, $A_{11'}=(\varepsilon_{p}-\varepsilon_{h})\delta_{pp'}\delta_{hh'}$, $A_{12'}=A_{21'}=0$, $B_{11'}=0$, and $A_{22'}=A_{22'}^{(\mathrm{D})}$.
\par
The adopted model space is defined as follows.
Single-particle levels with energies up to $\varepsilon_{p} < 100$~MeV are included.
Unperturbed 1p-1h excitation energies are taken into account up to $100$~MeV, and 2p-2h configurations are included up to $100$, $80$, $60$, and $60$~MeV for the $0^{+}$, $1^{-}$, $2^{+}$, and $3^{-}$ states of $^{16}$O, respectively.
Within this model space, the resulting strength distributions are sufficiently converged.
In particular, a large model space is essential to achieve adequate convergence for the strength distributions of two-body excitations.
This point is discussed in detail for the $0^{+}$ case of $^{16}$O in Sec.~\ref{sec:III-B-4}.
For $^{40}$Ca, the same model space for the 1p-1h configurations as that used for $^{16}$O is adopted, while unperturbed 2p-2h configurations are included up to $70$~MeV.
\subsection{Strength distribution}
Provided that $J$ is the multipolarity of the external field and $M$ its projection onto the $z$-axis, the strength distribution of a one-body external field without a spin operator is given by
\begin{equation}
\left|\langle \nu,J || F_{J} || \mathrm{SRPA} \rangle\right|^{2}
\approx
\left|\sum_{ph} C_{\nu;ph}^{(J)}\right|^{2},
\label{eq:1tran}
\end{equation}
where $F_{JM}$ may represent the isoscalar (IS) monopole ($r^{2}Y_{00}$), isovector (IV) dipole  ($rY_{1M}\tau_{z}$), IS quadrupole ($r^{2}Y_{2M}$), or IS octupole ($r^{3}Y_{3M}$) operator.
The strength distribution associated with a two-body external field, 
\begin{equation}
\mathcal{F}_{JM;L_{1}L_{2}}\equiv\left[F_{L_{1}}G_{L_{2}}\right]^{JM},
\label{eq:2bodyext}
\end{equation}
is given by
\begin{equation}
\left|\langle \nu,J||\mathcal{F}_{J;L_{1}L_{2}}||\mathrm{SRPA}\rangle\right|^{2}
\approx
\left|\sum_{\substack{p_{1}<p_{2}\\h_{1}<h_{2}}}\sum_{J_{p}J_{h}}
C_{\nu;p_{1}p_{2}h_{1}h_{2}}^{(J_{p}J_{h};L_{1}L_{2})J}\right|^{2},
\label{eq:2tran}
\end{equation}
respectively, where $G$ is another one-body external operator.
The single 1$p$-1$h$ transition amplitude $C_{\nu;ph}^{(J)}$ and the single 2$p$-2$h$ transition amplitude $C_{\nu;p_{1}p_{2}h_{1}h_{2}}^{(J_{p}J_{h};L_{1}L_{2})J}$ represent the contribution to resonance strength from 1$p$-1$h$ and 2$p$-2$h$ configurations, respectively, and defined as
\begin{equation}
C_{\nu;ph}^{(J)}=\Big(X_{\nu;ph}^{(J)}+(-)^{J}Y_{\nu;ph}^{(J)}\Big)F_{ph}^{(J)}
\label{eq:t1p1hamp}
\end{equation}
\begin{equation}
\begin{split}
&C_{\nu;p_{1}p_{2}h_{1}h_{2}}^{(J_{p}J_{h};L_{1}L_{2})J}\\
&=\Big(\mathcal{X}_{\nu;p_{1}p_{2}h_{1}h_{2}}^{(J_{p}J_{h})J}
+(-)^{L_{1}+L_{2}}\mathcal{Y}_{\nu;p_{1}p_{2}h_{1}h_{2}}^{(J_{p}J_{h})J}\Big)\mathcal{F}_{p_{1}p_{2}h_{1}h_{2}}^{(J_{p}J_{h};L_{1}L_{2})J}.
\end{split}
\label{eq:t2p2hamp}
\end{equation}
In the following discussion, the superscripts and subscripts of $C$ are omitted for brevity.
The reduced transition strength $F_{ph}^{(J)}$ is
\begin{equation}
F_{ph}^{(J)}=\langle p||F_{J}||h\rangle
\equiv
\langle j_{p} l_{p} || F_{J} || j_{h} l_{h} \rangle,
\end{equation}
and
\begin{equation}
\begin{split}
&\mathcal{F}_{p_{1}p_{2}h_{1}h_{2}}^{(J_{p}J_{h}J)}=
2\sqrt{\frac{(2J_{p}+1)(2J_{h}+1)}{(1+\delta_{p_{1}p_{2}}))(1+\delta_{h_{1}h_{2}})}}
\left[\left(
\begin{array}{ccc}
j_{p_{1}} & j_{p_{2}} & J_{p}\\
j_{h_{1}} & j_{h_{2}} & J_{h}\\
L_{1} & L_{2} & J\\
\end{array}
\right)\right.\\
&\left.\times F^{(L_{1})}_{p_{1}h_{1}}G^{(L_{2})}_{p_{2}h_{2}}
-(-)^{j_{h_{1}}+j_{h_{2}}-J_{h}}
\left(
\begin{array}{ccc}
j_{p_{1}} & j_{p_{2}} & J_{p}\\
j_{h_{2}} & j_{h_{1}} & J_{h}\\
L_{1} & L_{2} & J\\
\end{array}
\right)
F_{p_{1}h_{2}}^{(L_{1})} G^{(L_{2})}_{p_{2}h_{1}} \right].
\end{split}
\label{eq:amp2}
\end{equation}
Although Eq.~\eqref{eq:2bodyext} is written in a separable form, two-body operators with non-separable forms can also be calculated within the same framework.
\par
Analyzing the single transition amplitudes defined in Eqs.~\eqref{eq:t1p1hamp} and ~\eqref{eq:t2p2hamp} provides insight into the collectivity of a given resonance state $\nu$.
A notable feature of two-body external fields is that their collectivity differs from that of one-body external fields, which excite only proton 1p-1h, neutron 1p-1h, or a mixture of the two. 
Since a 2p-2h configuration consists of two 1p-1h excitations, and each 1p-1h excitation may involve either proton or neutron transitions, the resulting 2p-2h configuration can be built from neutron-neutron ($nn$), proton-proton ($pp$), or neutron-proton ($np$) components.
Here, $nn$ and $pp$ denote configurations in which both 1p-1h excitations involve neutrons or protons, respectively, while np denotes configurations in which one excitation involves neutrons and the other involves protons.
\par
In this work, I investigate the $J^{\pi}=0^{+}$ components of two-body external fields, i.e., double multipole excitations defined as
\begin{eqnarray}
\mathcal{F}_{0;00}=&\Big[[r^{2}Y_{0}] [r^{2}Y_{0}]\Big]^{0} \label{eq:DISM}\\
\mathcal{F}_{0;11}=&\Big[[c_{q}\, rY_{1}\tau_{z}] [c_{q}\, rY_{1}\tau_{z}]\Big]^{0} \label{eq:DIVD}\\
\mathcal{F}_{0;22}=&\Big[[r^{2}Y_{2}] [r^{2}Y_{2}]\Big]^{0} \label{eq:DISQ}\\
\mathcal{F}_{0;33}=&\Big[[r^{3}Y_{3}] [r^{3}Y_{3}]\Big]^{0}, \label{eq:DISO}
\end{eqnarray}
where the coefficients $c_{q}$ in Eq.~\eqref{eq:DIVD} depend on the nucleon type, with $c_{p}=N/A$ for protons and $c_{n}=Z/A$ for neutrons.
These modes may be interpreted either as sequential actions of the corresponding one-body external fields or as direct actions of the two-body external fields.
For convenience, the present paper refers to the external fields defined in Eqs.~\eqref{eq:DISM}, \eqref{eq:DIVD}, and \eqref{eq:DISQ} as the double IS monopole ($0^{+}$), double IV dipole ($1^{-}$), double IS quadrupole ($2^{+}$), and double IS octupole ($3^{-}$) modes, respectively.
\par
In principle, the center-of-mass (CoM) motion exactly appears at zero energy within a fully self-consistent RPA framework~\cite{Rowe1968, Nakada2009}.
However, the SRPA violates the stability condition, although it still satisfies Thouless' theorem for the energy-weighted sum-rule~\cite{Papa2014}. 
This violation can be partially remedied by the subtraction method~\cite{Tselyaev2007, Gambacurta2015}; however this does not guarantee that the CoM mode appears exactly at zero energy including numerical accuracy and model space dependence~\cite{Gambacurta2018}.
In fact, one observes a finite strength accumulated at low energies in both the single and double isoscalar dipole modes, indicating a presence of a contamination from the CoM mode in the intrinsic modes.
Nevertheless, it is expected that the impact of this contamination on intrinsic collective modes is rather limited, as reported in RPA~\cite{Mizuyama2012}, the finite amplitude method~\cite{Inakura2009}, and the equation of motion phonon method~\cite{Knapp2023}.
In this work, the strength of the lowest state that strongly couples to the spurious CoM motion the most is simply removed when calculating the strength distributions.
\par
The strength distribution folded by a Lorentzian function of width $\Gamma_{0}$ is defined as
\begin{equation}
    L_{J}(E)=\sum_{\nu} \left|\langle \nu,J||\mathcal{O}_{J}||\mathrm{SRPA}\rangle \right|^{2}
    \frac{1}{\pi}\frac{\Gamma_{0}/2}{(E-E_{\nu})^2+(\Gamma_{0}/2)^{2}},
    \label{eq:foldL}
\end{equation}
where $\mathcal{O}_{J}$ takes $F_{J}$ and $\mathcal{F}_{J;L_{1}L_{2}}$.
The centroid energy is defined as $m_{1}/m_{0}$, where $m_{k}$ is the $k$-th order energy-weighted sum-rule
\begin{equation}
m_{k}=\sum_{\nu} (\hbar\omega_{\nu})^{k}|\langle \nu,J||\mathcal{O}_{J}||\mathrm{SRPA}\rangle|^{2}.
\label{eq:kthsum}
\end{equation}
\section{Result}
This section presents the strength distributions for both one-body and two-body external fields for $^{16}$O and $^{40}$Ca. 
Since these two types of excitations are closely related, discussing them together provides a more coherent physical picture of the strength distributions induced by two-body external fields.
The discussion begins with the results for $^{16}$O and then examines whether the main qualitative features persist in $^{40}$Ca.

\label{sect:result}
\subsection{Strength distributions of one-body external fields for $^{16}$O}
First, I discuss the ordinary IS monopole ($0^{+}$), IV dipole ($1^{-}$), IS quadrupole ($2^{+}$), and IS $3^{-}$ excitations for $^{16}$O before turning to the double multipole excitations.
These modes have already been analyzed in Refs.~\cite{Papa2009, Papa2010, Gambacurta2010, Gambacurta2015, Yang2021}, but they are presented here as a reference for the subsequent discussion.
After their general properties are described, the results are compared with available experimental data and briefly discussed.
\par
\begin{figure}
\centering
\includegraphics[width=0.99\linewidth]{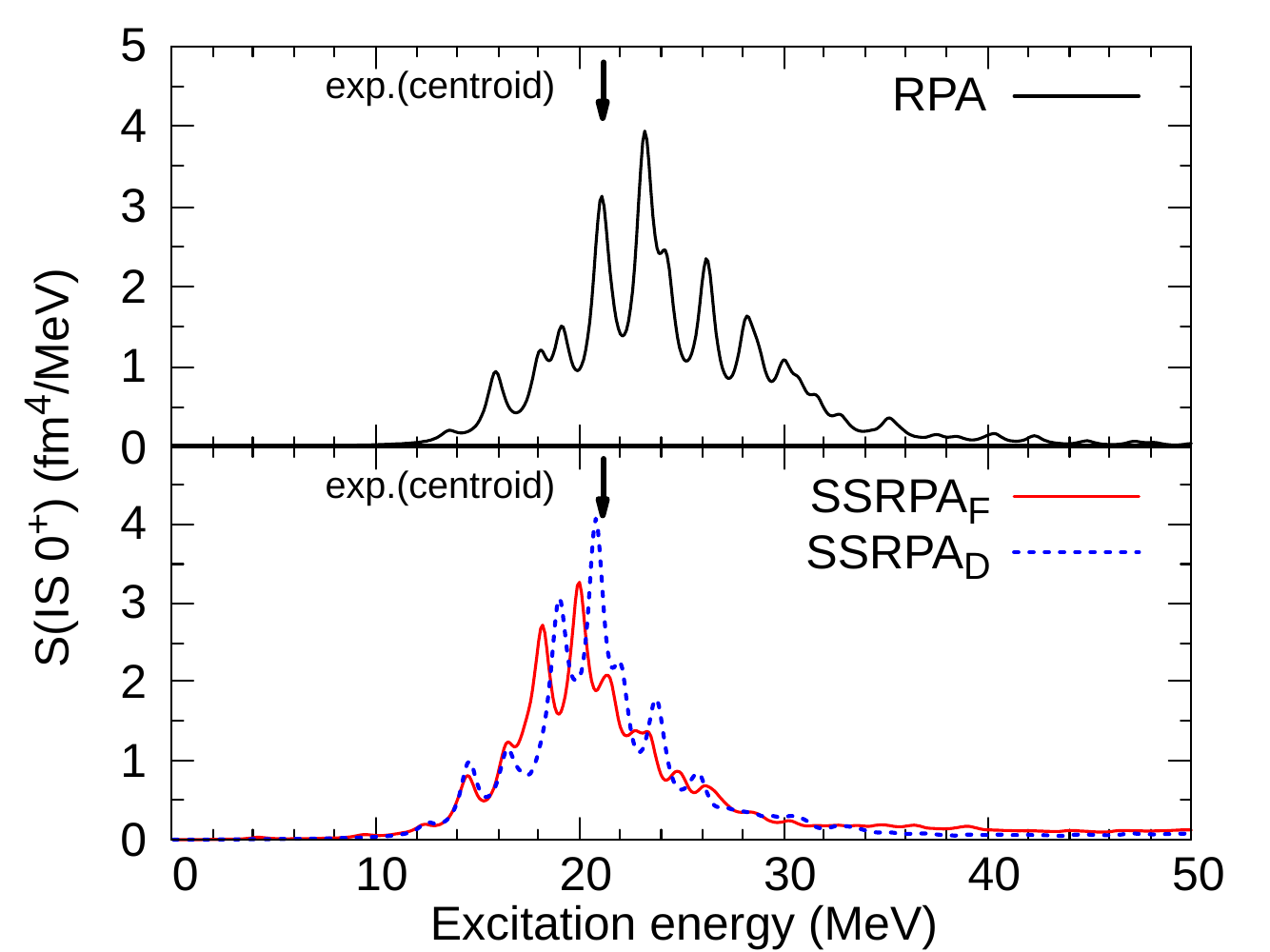}
\caption{Strength distribution of IS $0^{+}$ excitations in $^{16}$O calculated by RPA (top) and three types of SSRPA (bottom). The distributions are folded with a Lorentzian function of width $\Gamma_{0}=1$~MeV. The experimental centroid energy taken from Ref.~\cite{Lui2001} is shown by the arrow.}
\label{fig:O16_mono}
\end{figure}

Figure~\ref{fig:O16_mono} shows the IS monopole excitation strength distribution of $^{16}$O, calculated using RPA, SSRPA$_{\mathrm{F}}$, and SSRPA$_{\mathrm{D}}$ with the SGII interaction~\cite{Giai1981}.
The calculated strength distributions are folded with a Lorentzian function of width $\Gamma_{0}=1$~MeV.
Both RPA and SSRPA (SSRPA$_{\mathrm{F}}$ and SSRPA$_{\mathrm{D}}$) exhibit IS $0^{+}$ strength distributed over the excitation energy range of $\hbar\omega=10$-$40$~MeV.
The dominant 1p-1h configurations contributing the resonances in this energy range are $[np_{1/2}][0p_{1/2}]^{-1}$ and $[np_{3/2}][0p_{3/2}]^{-1}$ ($n=2, 3, 4$) for both neutron and proton.
\par
\begin{figure}
\centering
\includegraphics[width=0.99\linewidth]{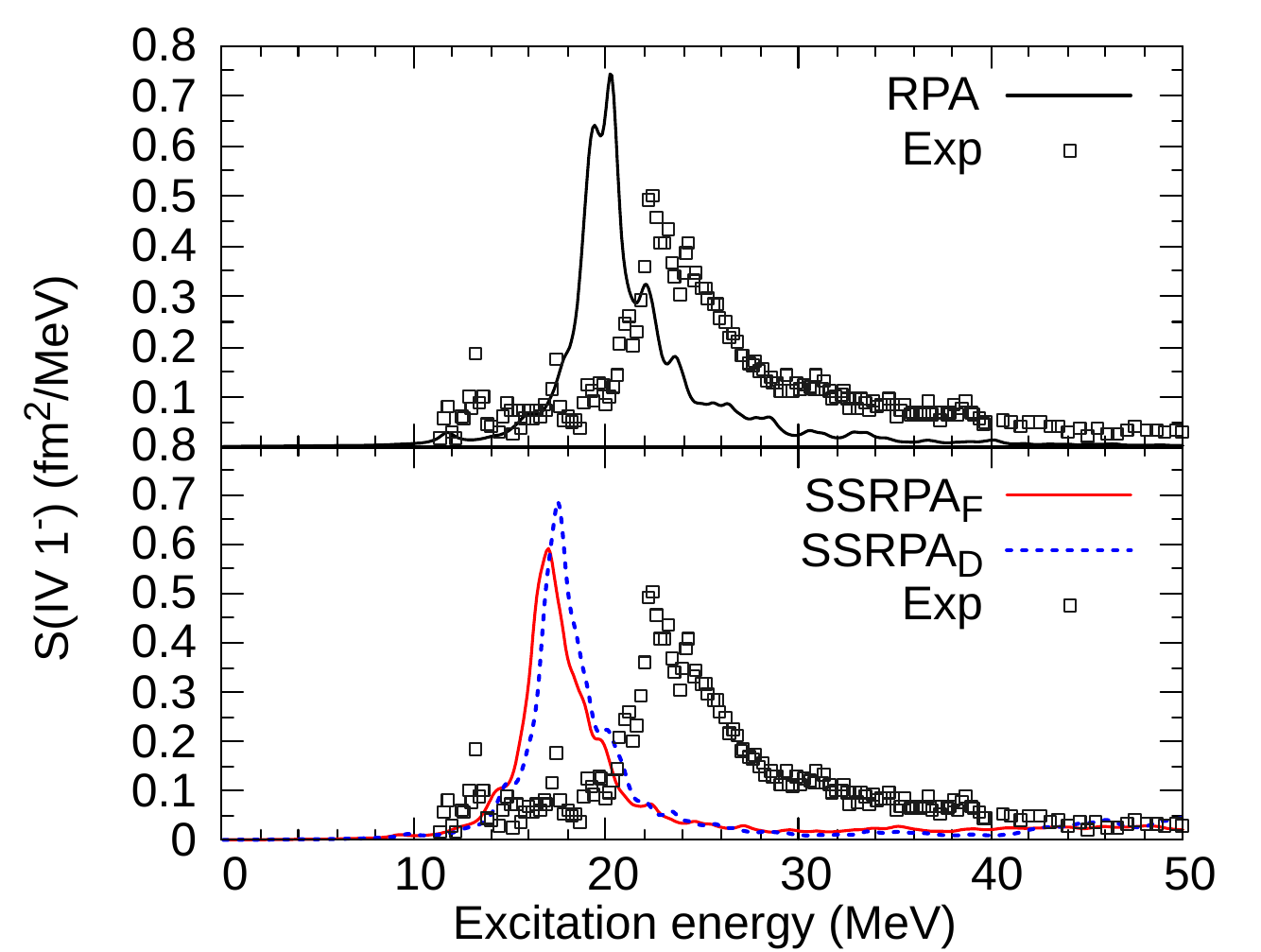}
\caption{Same as Fig.~\ref{fig:O16_mono}, but for IV $1^{-}$ excitations. The experimental data are taken from Ref.~\cite{Ahrens1975}.}
\label{fig:O16_dipo}
\end{figure}
\begin{figure}
\centering
\includegraphics[width=0.99\linewidth]{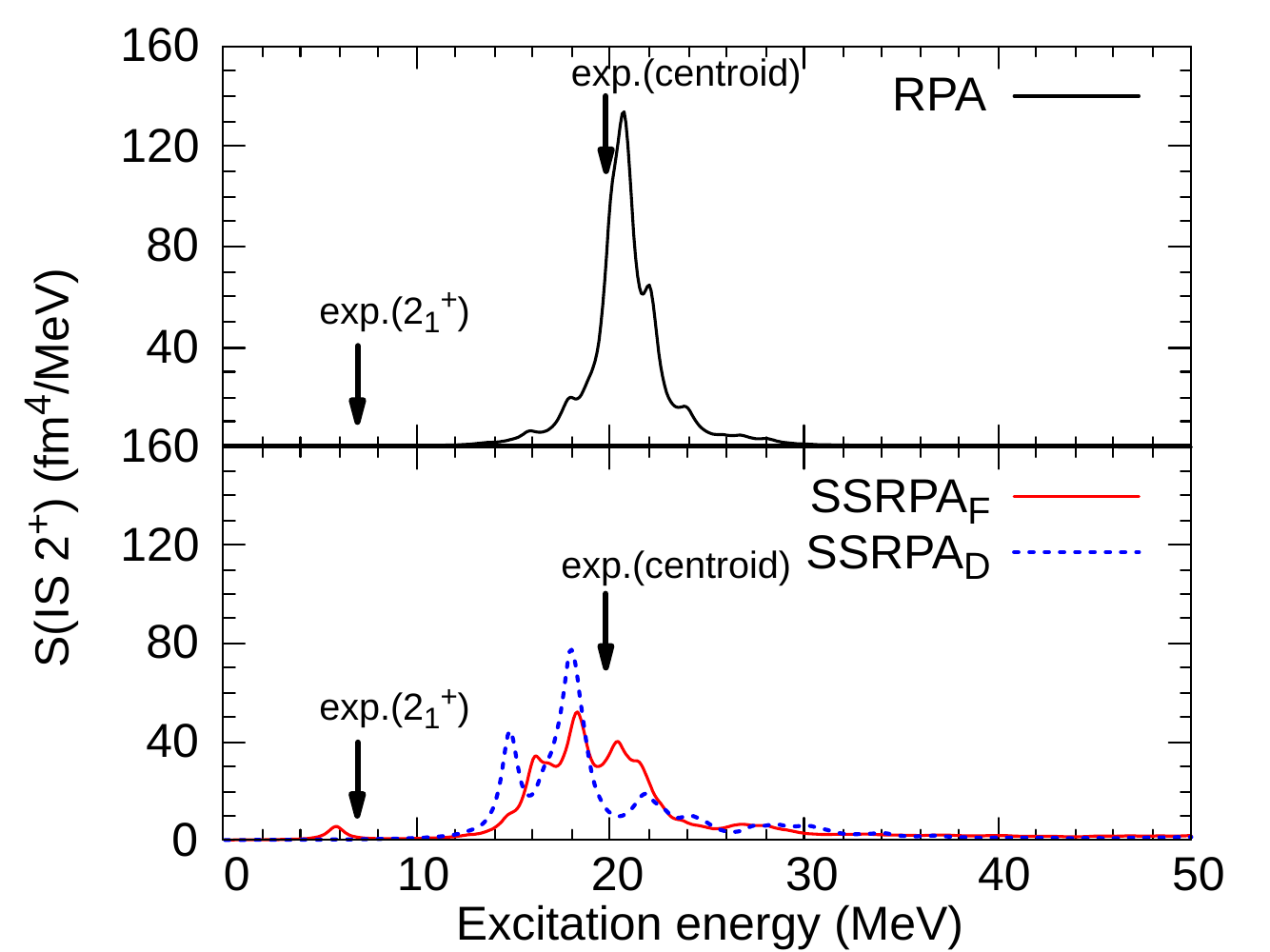}
\caption{Same as Fig.~\ref{fig:O16_mono}, but for IS $2^{+}$ excitations.}
\label{fig:O16_quad}
\end{figure}
\begin{figure}
\centering
\includegraphics[width=0.99\linewidth]{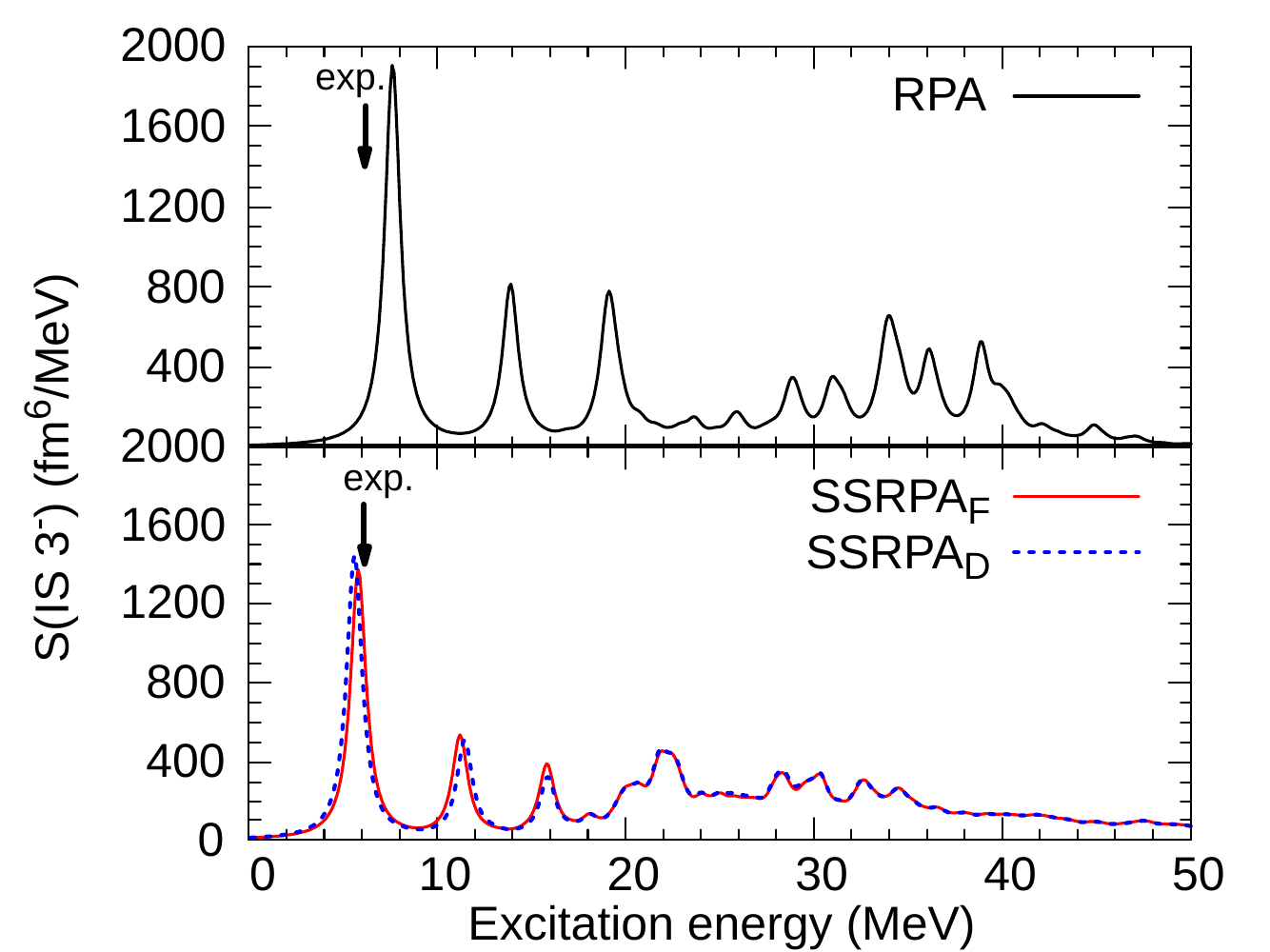}
\caption{Same as Fig.~\ref{fig:O16_mono}, but for IS $3^{-}$ excitations. 
The experimental datum of $3_{1}^{-}$ is taken from Ref.~\cite{Kibedi2002}. 
$\rm{B}(E3;0_{1}^{+} \rightarrow 3_{1}^{-}$ is 613, 577, 1480 $e^{2}$fm$^{6}$ for RPA, SSRPA$_{F}$, and experimental data, respectively.}
\label{fig:O16_octu}
\end{figure}

The damping effect, which arises from coupling to 2p-2h configurations, appears to be weak in SSRPA$_{\mathrm{F}}$ and SSRPA$_{\mathrm{D}}$.
This weak damping is understood as a cancellation between incoherent and coherent contributions to the self-energy for non-spin-flip isoscalar monopole excitations, as discussed in Ref.~\cite{Wambach1988}.
Although SSRPA$_{\mathrm{F}}$ and SSRPA$_{\mathrm{D}}$ exhibit noticeable differences, indicating that coupling among 2p-2h configurations has a non-negligible impact on the strength distributions, the overall structure is already captured reasonably well by SSRPA$_{\mathrm{D}}$.
Figure~\ref{fig:O16_dipo} shows the corresponding results for the IV $1^{-}$ excitation.
The dominant 1p-1h configurations contributing the resonances in the $\hbar\omega = 10$-$30$~MeV region are $[nd_{5/2}][0p_{3/2}]^{-1}$, $[nd_{3/2}][0p_{1/2}]^{-1}$ ($n=0, 1, 2$), and $[ns_{1/2}][0p_{3/2}]^{-1}$ and $[ns_{1/2}][0p_{1/2}]^{-1}$ ($n=1, 2$) for both neutron and proton.
Figure~\ref{fig:O16_quad} displays the results for the IS $2^{+}$ mode.
The qualitative behavior of RPA and SSRPA$_{\mathrm{F}}$ is similar to that seen in the IS $0^{+}$ case; however, in contrast to the monopole excitation, the IS $2^{+}$ mode exhibits a pronounced damping effect.
Figure~\ref{fig:O16_octu} displays the IS $3^{-}$ distribution.
The low-lying first $3^{-}$ state is found around $\hbar\omega=6$--$8$~MeV.
This is collective mode and the major 1p-1h configuration are proton and neutron $[0d_{5/2}][0p_{1/2}]^{-1}$.
\par
The center energies ($E_{\mathrm{L}}$) and widths ($\Gamma_{\mathrm{L}}$) of the strength distributions are also examined for four types of excitations.
These quantities were obtained by fitting $L_{J}$ in Eq.~\eqref{eq:foldL} with another Lorentzian function,
\begin{equation}
L_{\mathrm{fit}}(E)=N\times\frac{1}{\pi} \frac{2}{\Gamma}\frac{1}{(E-E_{\mathrm{L}})^2+(\Gamma/2)^{2}},
\label{eq:Lorentz}
\end{equation}
where the normalization factor $N$ is determined simultaneously with $E_{\mathrm{L}}$ and $\Gamma$ through a least-squares fitting procedure.
The physical width is then obtained following Ref.~\cite{Scamps2013} as
\begin{equation}
    \Gamma_{\mathrm{L}}\approx\Gamma-\Gamma_{0}.
\end{equation}
\par
A brief comparison with the available experimental data is presented.
Table~\ref{tab:centroid_width1} lists the center energies and widths extracted from the Lorentzian fits, as well as the centroid energies of the IS $0^{+}$, IV $1^{-}$, and IS $2^{+}$ resonances calculated using RPA and SSRPA$_{\mathrm{F}}$, together with the corresponding experimental data. 
Since the $3^{-}$ strength distribution is not well described by a Lorentzian shape, it is not included in the table.
To distinguish these quantities from those corresponding to the strength distributions induced by two-body external fields discussed in the next section, the superscript ``(1)'', namely $E_{\mathrm{L}}^{(1)}$ and $\Gamma_{\mathrm{L}}^{(1)}$, is used.
For the IS $0^{+}$ and IS $2^{+}$ modes, both RPA and SSRPA$_{\mathrm{F}}$ reasonably reproduce the experimental centroid energies. 
For the IV $1^{-}$ mode, however, both RPA and SSRPA$_{\mathrm{F}}$ underestimate the experimental centroid energies.
As shown in Fig.~\ref{fig:O16_dipo}, the calculated RPA and SSRPA$_{\mathrm{F}}$ strength distributions deviate from the experimental data, with their main peaks located about $4$--$5$~MeV lower in energy. 
This tendency is commonly observed for other Skyrme forces, whose parameters are primarily adjusted to nuclear ground-state properties. 
It should be noted, however, that better agreement can be achieved when the experimental distribution is included in the parameter adjustment of the Skyrme interaction~\cite{Lyutorovich2012}.
\par
RPA does not produce any significant strength corresponding to the experimental first $2^{+}$ state, whereas SSRPA$_{\mathrm{F}}$ yields a finite strength, as shown in Fig.~\ref{fig:O16_quad}.
For the first $3^{-}$ resonance, both RPA and SSRPA$_{\mathrm{F}}$ reasonably reproduce the experimental excitation energy of the $3^{-}_{1}$ state, as demonstrated in Fig.~\ref{fig:O16_octu}. 
However, the calculated $\mathrm{B}(E3;0_{1}^{+} \rightarrow 3_{1}^{-})$ values are $613$ and $577$ $e^{2}\mathrm{fm}^{6}$ for RPA and SSRPA$_{\mathrm{F}}$, respectively, underestimating the experimental value of $1480$ $e^{2}\mathrm{fm}^{6}$~\cite{Kibedi2002}. 
These results are, however, consistent with those reported in Refs.~\cite{Papa2009, Papa2010, Gambacurta2010, Gambacurta2015, Yang2021}.
\par
In Table~\ref{tab:centroid_width1}, center energies obtained with SSRPA$_{\mathrm{F}}$ are shifted to lower values by about $3.4$, $2.8$, and $1.6$~MeV for the IS $0^{+}$, IV $1^{-}$, and IS $2^{+}$ modes, respectively. 
This downward shift can be understood as a consequence of the coupling to 2p-2h configurations. 
Such coupling redistributes part of the strength to higher excitation energies, which in turn shifts the dominant resonance to lower energies so as to satisfy the relevant sum-rule constraint. 
Although SSRPA$_{\mathrm{F}}$ does not strictly fulfill the energy-weighted sum rule, the same argument applies to the inverse energy-weighted sum rule, which is fulfilled by SSRPA$_{\mathrm{F}}$~\cite{Gambacurta2015, Yang2021}.
The widths obtained with SSRPA$_{\mathrm{F}}$ are smaller than those from RPA for the IS $0^{+}$ mode, whereas the width for the IV $1^{-}$ mode remains essentially unchanged.
However, non-negligible strength exists beyond the main peak region and cannot be fully captured by the Lorentzian fit adopted in Eq.~\eqref{eq:Lorentz}. 
Therefore, the smaller Lorentzian widths do not necessarily indicate a weaker effect of 2p-2h configuration mixing. 
Indeed, the centroid energies $m_{1}/m_{0}$ obtained with SSRPA$_{\mathrm{F}}$ are systematically larger than those obtained with RPA, reflecting the redistribution of strength to higher excitation energies. 
In contrast, the width obtained with SSRPA$_{\mathrm{F}}$ is significantly larger than that from RPA for the IS $2^{+}$ mode, where the 2p-2h configuration mixing directly affects the main peak structure.
These results will be compared with the center energies, widths, and centroid energies of the strength distributions induced by two-body external fields in the next section.
\begin{table*}[t]
\caption{Center energies $E_{\mathrm{L}}^{(1)}$ and widths $\Gamma_{\mathrm{L}}^{(1)}$ extracted from the Lorentzian fits of the strength distributions, and centroid energies ($m_{1}/m_{0}$) of the IS $0^+$, IV $1^-$, and IS $2^+$ obtained from RPA and SSRPA$_{\rm{F}}$. Experimental data of centroid energies and widths are taken from Refs.~\cite{Lui2001}.}
\begin{ruledtabular}
\begin{tabular}{lccccccccc}
Mode 
& \multicolumn{3}{c}{$E_{\mathrm{L}}^{(1)}$ (MeV)} 
& \multicolumn{3}{c}{$\Gamma_{\mathrm{L}}^{(1)}$ (MeV)}
& \multicolumn{2}{c}{$m_{1}/m_{0}$ (MeV)}\\
 & RPA & SSRPA$_{\rm{F}}$  & Exp. & RPA & SSRPA$_{\rm{F}}$ & Exp. & RPA & SSRPA$_{\rm{F}}$\\
\colrule
$\mathrm{IS}\;0^{+}$ & 23.3 & 19.9 & 21.13$\pm$0.49 & 6.9 & 5.2 & 8.76$\pm$1.82 & 24.8 & 30.9 \\
$\mathrm{IV}\;1^{-}$ & 20.0 & 17.2 & 21.67$\pm$0.69 & 2.0 & 2.0 & 7.10$\pm$0.52 & 21.9 & 27.6 \\
$\mathrm{IS}\;2^{+}$ & 20.6 & 19.0 & 19.76$\pm$0.22 & 1.1 & 4.8 & 5.11$\pm$0.17 & 20.9 & 25.3 \\
%$\mathrm{IS}\;3^{-}$ & 16.7 & 21.0 & -- &41.3 & 40.8 & -\\
\end{tabular}
\end{ruledtabular}
\label{tab:centroid_width1}
\end{table*}
\subsection{Strength distributions of two-body external fields for $^{16}$O}
\label{sec:III-B}
The strength distributions of the two-body external fields constitute one of the main focuses of this work.
However, the basic properties of these strength distributions are not necessarily well understood.
Therefore, the results obtained with the unperturbed calculation are presented first, as they provide the essential physical picture of the nuclear responses to double-multipole excitations.
The characteristics of the strength distributions calculated with SSRPA$_{\mathrm{F}}$ are then discussed.
\begin{figure*}[htb]
\centering
\includegraphics[width=0.99\linewidth]{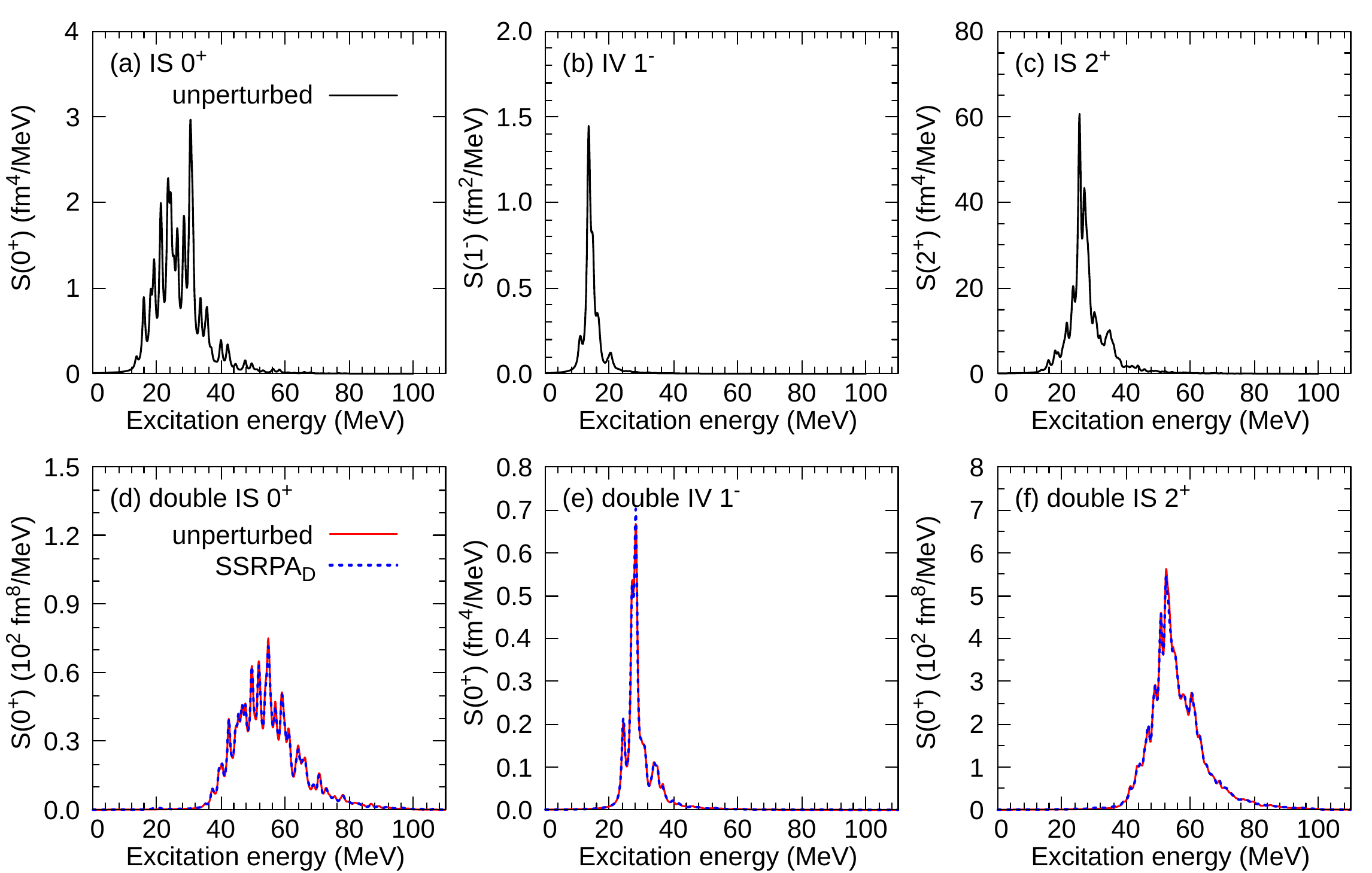}
\caption{
(a)–(c) Strength distributions of the one-body external fields for the IS $0^{+}$, IV $1^{-}$, and IS $2^{+}$ modes obtained with the unperturbed calculation.
(d)–(f) Strength distributions of the corresponding double multipole modes, defined in Eqs.~\eqref{eq:DISM}–\eqref{eq:DISQ}, obtained with unperturbed calculation and SSRPA$_{\mathrm{D}}$.}
\label{fig:O16all_HF}
\end{figure*}
\subsubsection{unperturbed calculation} 
Figure~\ref{fig:O16all_HF} shows the strength distributions for the IS $0^{+}$, IV $1^{-}$, and IS $2^{+}$ modes of obtained from the unperturbed calculation (the upper panels), and the $0^{+}$ components of the two-body external fields defined in Eqs.~\eqref{eq:DISM}-\eqref{eq:DISQ} for $^{16}$O obtained from the unperturbed and SSRPA$_{\mathrm{D}}$ calculations (the lower panels).
Since the the unperturbed results represent non-collective excitations, one may be concerned about their dependence on the model space, particularly on the box size (i.e., the treatment of continuum states).
The calculation confirms that, although the detailed fine structures of the strength distributions are sensitive to the model space, the overall locations of the main structures remain essentially unchanged.
\par
Intuitively, the strength distributions of the two-body external fields obtained from the the unperturbed calculation are expected to appear at approximately twice the excitation energies of the corresponding one-body external fields.
This tendency is indeed observed when comparing the strength distributions of the one-body and two-body external fields shown in Fig.~\ref{fig:O16all_HF}.
To quantify this observation, the center energies and widths estimated from Eq.~\eqref{eq:Lorentz} are summarized in Table~\ref{tab:centroid_width_HF}.
The center energies of the two-body external fields are found to be close to twice those of the corresponding one-body external fields, consistent with the intuitive expectation.
In contrast, the widths do not exhibit a universal behavior: while the width for the double IS $2^{+}$ mode increases by more than a factor of two, those for the IS $0^{+}$ and the IV $1^{-}$ modes show only moderate increases.
\par
One also finds that the results of SSRPA$_{\mathrm{D}}$ nearly overlap with those of the unperturbed calculation.
According to Eq.~\eqref{eq:SRPA}, the difference between SSRPA$_{\mathrm{D}}$ and the unperturbed calculation originates from the residual interaction in $A_{11'}$ and $A_{12'}$ ($A_{21'}$), both of which are absent in the unperturbed calculation.
This indicates that the contributions of these matrix elements are rather limited for the strength distributions induced by two-body external fields.
This behavior contrasts with the case of one-body external fields, where the coupling between 1p–1h and 2p–2h configurations plays a significant role, as seen in Figs.~\ref{fig:O16_mono}–\ref{fig:O16_quad}.
The present results suggests that transitions from 1p-1h to the 2p-2h configurations dominate over the reverse process, effectively leading to an effective decoupling between the two spaces.
In this sense, treating the 1p-1h and 2p-2h blocks as decoupled in the SSRPA equation serves as a good approximation for describing the responses to two-body external fields.
However, this approximation offers little computational benefit, since the dimension of the 2p-2h space is much larger than that of the 1p-1h space.
\begin{table}[t]
\caption{
Center energies and widths of the IS $0^{+}$, IV $1^{-}$, and IS $2^{+}$ modes ($E^{(1)}_{\mathrm{L}}$ and $\Gamma_{\mathrm{L}}^{(1)}$), and of the $0^{+}$ component of the corresponding double multipole modes obtained from unperturbed calculations ($E^{(2)}_{\mathrm{L}}$ and $\Gamma_{\mathrm{L}}^{(2)}$) for $^{16}$O. All values are given in MeV.
}
\label{tab:centroid_width_HF}
\begin{ruledtabular}
\begin{tabular}{lcclcc}
\multicolumn{6}{c}{\textbf{unperturbed}}\\ 
& \multicolumn{2}{c}{one-body} & & \multicolumn{2}{c}{two-body}\\
Mode & $E_{\mathrm{L}}^{(1)}$ & $\Gamma_{\mathrm{L}}^{(1)}$ & Mode & $E_{\mathrm{L}}^{(2)}$ & $\Gamma_{\mathrm{L}}^{(2)}$ \\
\hline
  IS $0^{+}$ & 26.0 & 12.1 & double IS $0^{+}$ & 52.8 & 15.4 \\
  IV $1^{-}$ & 13.8 &  1.2 & double IV $1^{-}$ & 27.8 &  1.8 \\
  IS $2^{+}$ & 26.3 &  3.9 & double IS $2^{+}$ & 53.5 &  9.9 \\
%  IS $3^{-}$ & 33.2 & 28.6 & double IS $3^{-}$ & 63.6 & 38.5 \\
\end{tabular}
\end{ruledtabular}
\end{table}
\subsubsection{SSRPA calculation}
\label{sec:SSRPA}
\begin{figure*}
\centering
\includegraphics[width=0.99\linewidth]{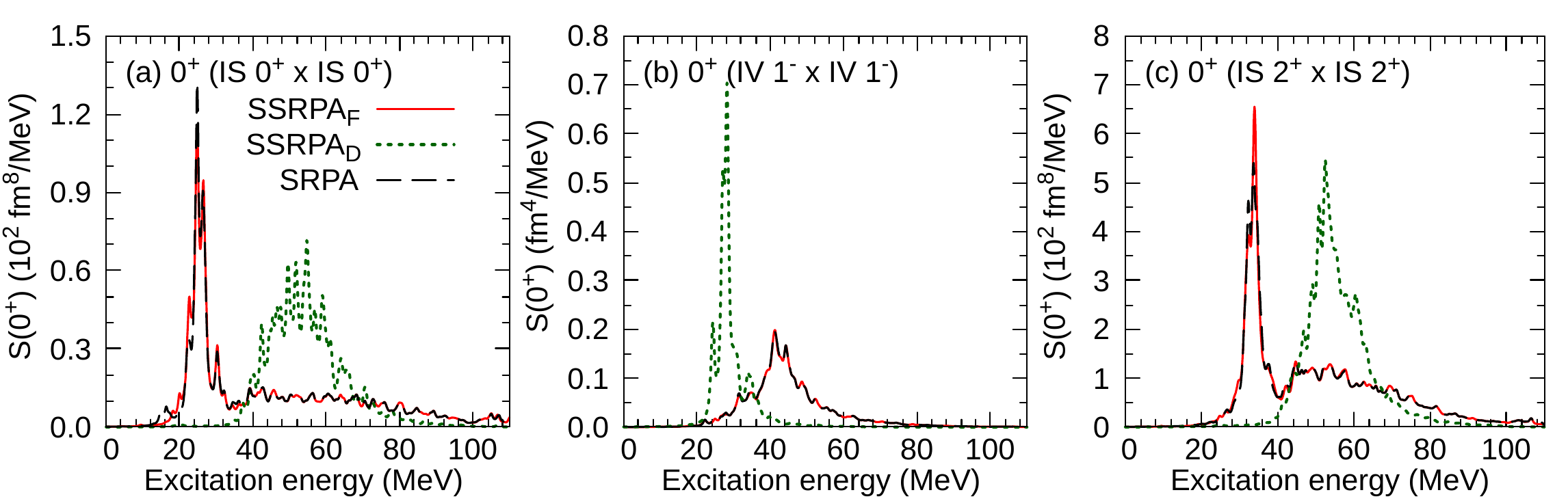}
\caption{
Strength distributions for the two-body external fields defined in Eqs.~\eqref{eq:DISM}--\eqref{eq:DISQ} in $^{16}$O. Results from SSRPA$_{\mathrm{F}}$, SSRPA$_{\mathrm{D}}$, and SRPA calculations are compared.
}
\label{fig:O16all}
\end{figure*}
Figure~\ref{fig:O16all} shows the strength distributions of the two-body external fields defined in Eqs.~\eqref{eq:DISM}--\eqref{eq:DISQ} in $^{16}$O, calculated by SSRPA.
Figure~\ref{fig:O16all}(a) presents the strength distribution of the $J^{\pi}=0^{+}$ component of the double IS $0^{+}$ excitations given in Eq.~\eqref{eq:DISM}.
The strength distribution obtained with SSRPA$_{\mathrm{F}}$ differs significantly from that of SSRPA$_{\mathrm{D}}$.
Since the primary distinction between these two calculations is the presence of the residual interaction among 2p-2h configurations, this comparison demonstrates that the couplings among different 2p-2h states plays a crucial role.
As a consequence, the main strength of the double IS $0^{+}$ excitation is shifted considerably toward lower energies relative to SSRPA$_{\mathrm{D}}$, in a manner analogous to the coherent 1p-1h excitations of IS $0^{+}$ mode.
Similarly, in Fig.~\ref{fig:O16all}(b), the SSRPA$_{\mathrm{F}}$ strength distribution of the double IV mode is shifted to higher excitation energies than the SSRPA$_{\mathrm{D}}$ result, analogous to the behavior observed for the IV dipole mode.
A qualitatively similar trends is also found for the double IS $2^{+}$ excitations shown in Fig.~\ref{fig:O16all}(c).
\par
\begin{figure}
\includegraphics[width=0.7\linewidth]{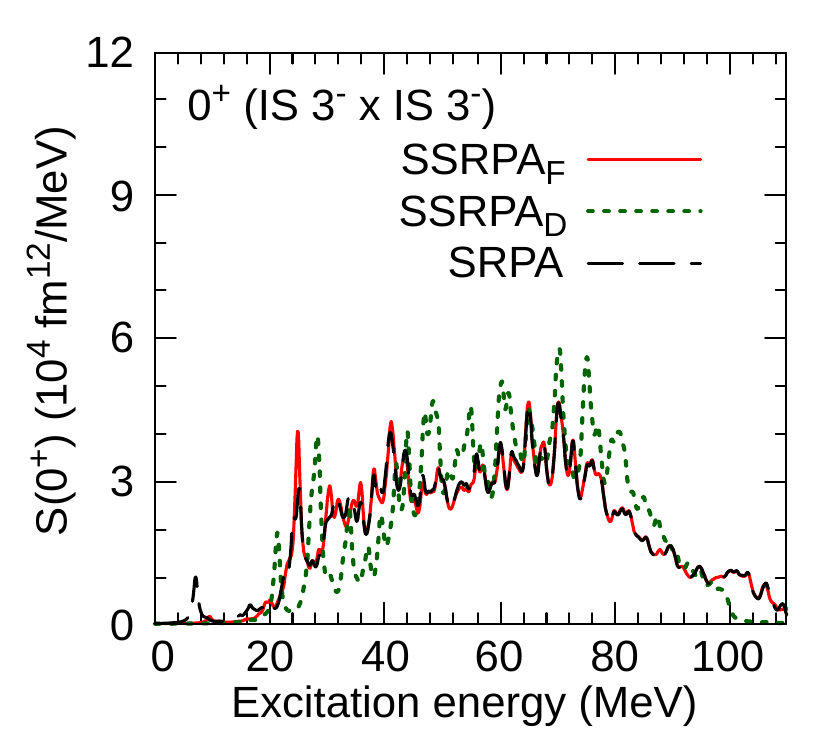}
\caption{
Strength distributions for the two-body external fields defined in Eq.~\eqref{eq:DISO} in $^{16}$O. Results from SSRPA$_{\mathrm{F}}$, SRPA$_{\mathrm{D}}$, and SRPA calculations are compared.}
\label{fig:O16all_octu}
\end{figure}

Figure~\ref{fig:O16all_octu} shows the strength distribution of the double $3^{-}$ external fields defined in Eq.~\eqref{eq:DISO}.
The results are quantitatively similar to those discussed in Fig.~\ref{fig:O16all}, particularly, regarding the differences between SRPA and SSRPA$_{\mathrm{F}}$. 
Compared with the responses induced by other double external fields, the modification of the strength distribution in SSRPA$_{\mathrm{F}}$ relative to SSRPA$_{\mathrm{D}}$ is not very pronounced.
However, this tendency is already found at the level of 1p-1h excitations (not discussed in this paper) and is therefore not surprising. 
These results suggest that the effect of the residual interaction depends on the specific combination of one-body external fields that forms the two-body external field.
\par
As shown in Figs.~\ref{fig:O16_mono}--\ref{fig:O16_quad}, that coupling among the 2p-2h configurations also affects the strength distributions induced by one-body external fields.
Its impact on responses to two-body external fields is, however, even more pronounced, and the underlying physics appears to be more intricate.
For instance, Fig.~\ref{fig:O16all}(a) shows main peaks around $25$~MeV together with substantial strength above $80$~MeV, well beyond the main peak region of SSRPA$_{\mathrm{D}}$, which appears around $\hbar\omega=50$~MeV. 
In contrast, in Fig.~\ref{fig:O16all}(b) the SSRPA$_{\mathrm{F}}$ strength distribution is shifted to higher excitation energies relative to SSRPA$_{\mathrm{D}}$, indicating that the dominant contribution of the residual interaction acts repulsively.
These results demonstrate that double-phonon excitations cannot always be described by a simple folding of one-body responses, as discussed in Ref.~\cite{Emling1994}.
\par
To further quantify this behavior, the matrix elements of the residual interaction are examined.
Although these distributions are not presented explicitly in figures or tables, they are found to be centered near zero, with a slight negative bias for interactions among $nn$ and $pp$ configurations and a slight positive bias for those among $np$ configurations.
Although a complete analysis is difficult due to the enormous configuration space and the complicated couplings involved, these trends are consistent with the behavior observed in Fig.~\ref{fig:O16all} and help us to gain a qualitative understanding of the observed resonance shifts.
\par
Table~\ref{tab:centroid_width} summarizes the center energies and widths extracted from the Lorentzian fits ($E^{(2)}_{\mathrm{L}}$ and $\Gamma_{\mathrm{L}}^{(2)}$), and centroid energies ($m_{1}/m_{0}$) of the $0^{+}$ component of the double-multipole modes obtained from SSRPA$_{\mathrm{F}}$. 
Because the double IS $0^{+}$ and $2^{+}$ distributions in Fig.~\ref{fig:O16all} can be separated into two groups (a low-energy group around $\hbar\omega\sim 30$~MeV and a broader high-energy component), the corresponding quantities for the first and second groups are listed separately.
One finds that the center energies of the double IS $0^{+}$ and $2^{+}$ modes are significantly smaller than twice those of the corresponding one-body excitations obtained with RPA and SSRPA$_{\mathrm{F}}$ (Table~\ref{tab:centroid_width1}).
These observations suggest that the residual interaction among 2p-2h configurations induces non-negligible correlations between the two underlying 1p-1h excitations, thereby going beyond a simple picture of independent double-phonon excitations, at least for the $^{16}$O channels considered here.
In contrast, the center energy of the double IV $1^{-}$ mode is only slightly larger than twice the corresponding one-body value.
This result is consistent with many previous studies~\cite{Chomaz1992, Ponomarev1994, Nishizaki1995, Lanza1997, Nishizaki1998, Dang1999}, which indicate that the anharmonic effect on the energy of double giant dipole resonances is small.
When focusing on the centroid energies of $m_{1}/m_{0}$, the results for the double IS $0^{+}$ and IV $1^{-}$ modes obtained with SSRPA$_{\mathrm{F}}$ are close to twice the corresponding RPA values listed in Table~\ref{tab:centroid_width1}. 
In contrast, the centroid energy of the double IS $2^{+}$ mode exceeds twice the RPA value, but remains close to twice the corresponding SSRPA$_{\mathrm{F}}$ result.
\begin{table}[t]
\caption{
Center energies $E^{(2)}_{\mathrm{L}}$, widths $\Gamma_{\mathrm{L}}^{(2)}$, and centroid energies $m_{1}/m_{0}$ for the $0^{+}$ components of the corresponding double multipole modes obtained with SSRPA$_{\mathrm{F}}$ for $^{16}$O. The center energies and widths are extracted from Lorentzian fits. For the double IS $0^{+}$ and $2^{+}$ modes, whose strength distributions in Fig.~\ref{fig:O16all} can be separated into two groups, the corresponding quantities are listed for the first and second groups separately. All values are given in MeV.
}
\label{tab:centroid_width}
\begin{ruledtabular}
\begin{tabular}{lccccc}
& \multicolumn{5}{c}{\textbf{SSRPA$_{\mathrm{F}}$}} \\ 
& \multicolumn{2}{c}{first group} & \multicolumn{2}{c}{second group} 
& centroid ene.\\
Mode
& $E_{\mathrm{L}}^{(2)}$ & $\Gamma_{\mathrm{L}}^{(2)}$ 
& $E_{\mathrm{L}}^{(2)}$ & $\Gamma_{\mathrm{L}}^{(2)}$ 
& $m_{1}/m_{0}$ \\
\hline
double IS $0^{+}$ & 25.3 & 2.7 & 57.8 & 43.1 & 46.6 \\
double IV $1^{-}$ & 42.5 & 11.7 & - & -      & 44.8 \\
double IS $2^{+}$ & 33.7 & 2.9 & 55.5 & 28.8 & 50.3
%double IS $3^{-}$ & 59.9 & 47.4 & - & - & 60.1 \\
\end{tabular}
\end{ruledtabular}
\end{table}
\subsubsection{Effectiveness of subtraction method}
As discussed in the previous section, the 1p-1h excitations can be effectively decoupled from the 2p-2h ones and the results of strength distributions are almost the same between the unperturbed calculation and SSRPA$_{\mathrm{D}}$.
This indicates that the subtraction method has essentially no impact on the strength distributions induced by two-body external fields, because the subtraction modifies only the residual interaction among the 1p-1h configurations, as shown in Eq.~\eqref{eq:subtract}.
To verify this explicitly, the same calculation without the subtraction method, denoted as SRPA, is carried out and the results are shown in Fig.~\ref{fig:O16all}.
The strength distributions obtained with SRPA and SSRPA$_{F}$ are almost indistinguishable, confirming that the subtraction method does not influence the response to two-body external fields.
Only a minor modification is found for the double IS $0^{+}$ excitation around $\hbar\omega=16$~MeV.
\subsubsection{2p-2h model space dependence}
\label{sec:III-B-4}
Thus far, I have discussed the $0^{+}$ component of double-phonon excitations using unperturbed 2p-2h excitation energies up to $E_{2p2h}=100$~MeV.
This cut off energy is larger than 
To obtain converged strength distributions induced by two-body excitations, a substantially larger model space is required than in the case of one-body external fields.
In terms of the cutoff energy for unperturbed 2p-2h configurations, one-body external fields typically require only $E_{2p2h}=50$--$60$~MeV~\cite{Gambacurta2016, Yang2021}, whereas the present two-body external fields require a cutoff of about $E_{2p2h}=100$~MeV.
\par
\begin{figure}
\includegraphics[width=0.99\linewidth]{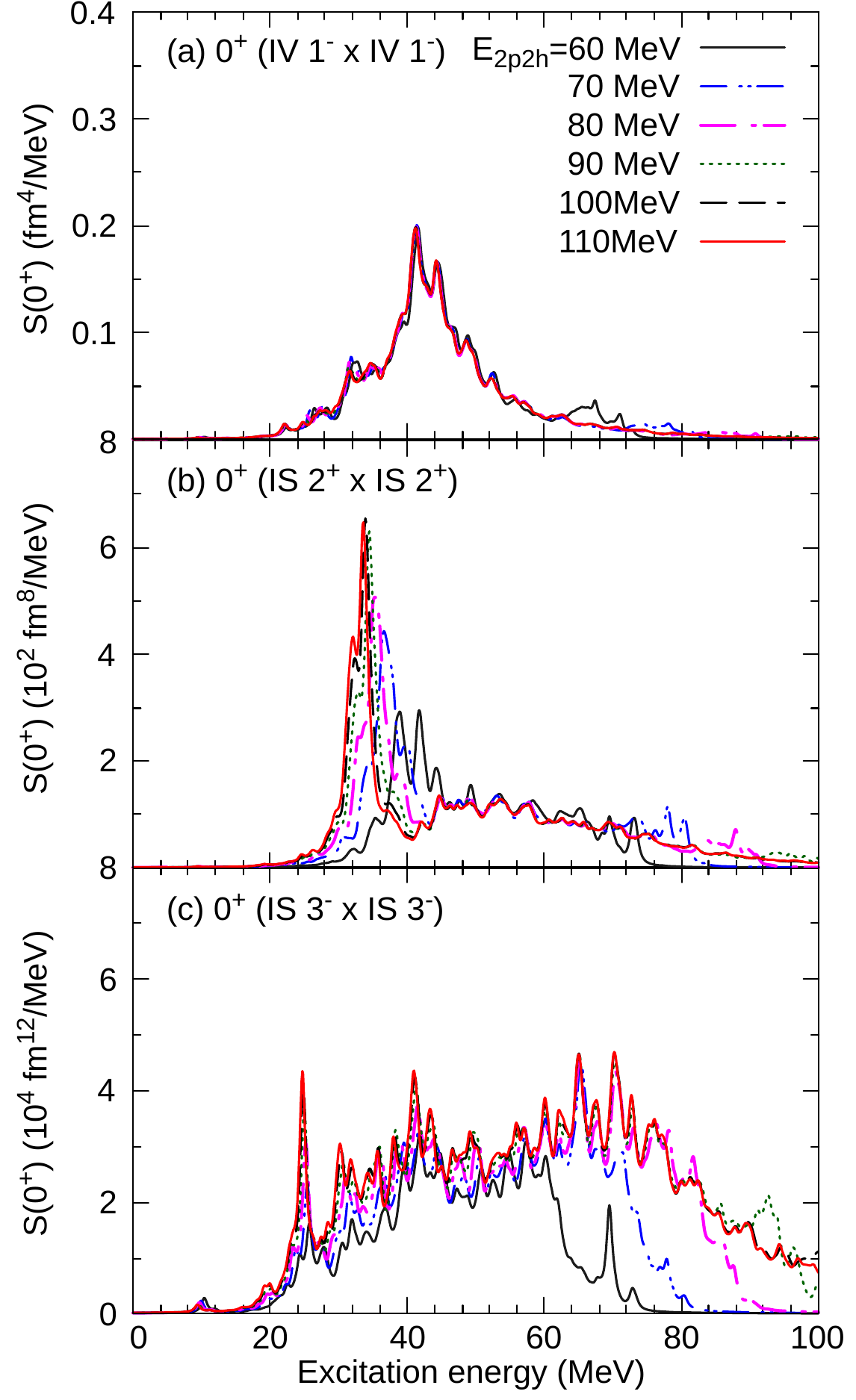}
\caption{Strength distributions of (a) double IV $1^{-}$, (b) double IS $2^{+}$, and (c) double IS $3^{-}$ excitations as a function of excitation energy ($\hbar\omega$) for different unperturbed 2p-2h unperturbed excitation energies ($E_{2p2h}$) in the SSRPA$_{\mathrm{F}}$ model space. Unperturbed 1p-1h excitation energies is set to be $100$~MeV.}
\label{fig:converge}
\end{figure}

Figure~\ref{fig:converge} illustrates the dependence of the strength distributions on $E_{2p2h}$ for the double IV $1^{-}$, double IS $2^{+}$, and double IS $3^{-}$ excitations.
The strength distribution of the double IV $1^{-}$ excitation is already sufficiently converged at $E_{2p2h}=60$~MeV in the main peak (around $\hbar\omega=45$~MeV), whereas a much larger model space is required for the double IS $2^{+}$ excitation.
This difference is attributed to the fact that the $1^{-}$ excitation involves mixing of double $1\hbar\omega$ configurations, while the double IS $2^{+}$ state involves mixing of double $2\hbar\omega$ configurations, which extend to significantly higher excitation energies than the double $1\hbar\omega$ ones.
A similar behavior was observed for the double IS $0^{+}$ excitation.
For the double IS $3^{-}$ excitation, a truncation of the model space at $E_{2p2h}\le80$~MeV leads to a significant underestimation of the strengths above $\hbar\omega=60$~MeV.
The overall strength distribution reasonably emerges when $E_{2p2h}\ge90$~MeV, and good convergence is achieved for $E_{2p2h}\ge100$~MeV.
Thus, $E_{2p2h}=100$~MeV is adopted in this work.

\subsection{Collectivity of resonances induced by two-body external fields}
\label{sect:collective}
The single transition amplitudes defined in Eqs.~\eqref{eq:1tran} and \eqref{eq:2tran} provide detailed information on the structure of each resonance, including its degree of collectivity and the dominant particle-hole configurations that contribute to the excited state. 
This section focuses on several representative resonances induced by the two-body external fields shown in Fig.~\ref{fig:O16all} and analyze their single transition amplitudes.
\par
\begin{figure}
\centering
\includegraphics[width=0.99\linewidth]{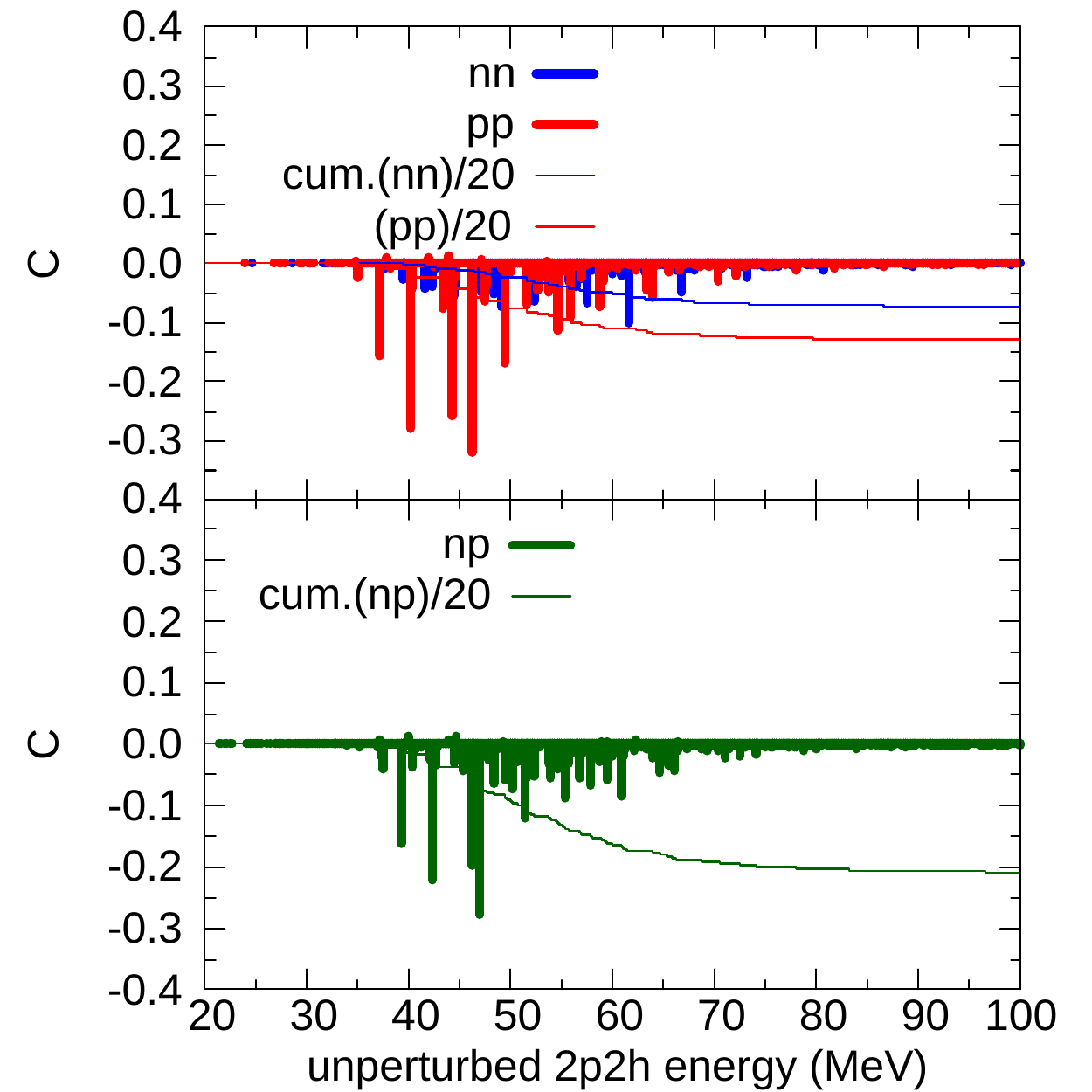}
\caption{
Single transition amplitude $C_{\nu;p_{1}p_{2}h_{1}h_{2}}$ contributing to the double IS $0^{+}$ resonance at $\hbar\omega=24.8$~MeV. 
The upper panel shows the proton-proton (pp) and neutron-neutron (nn) 2p-2h configurations, while the lower panel displays the neutron-proton (np) 2p-2h configurations.
The solid curve represents the cumulative sum of the amplitudes, divided by $20$, as a function of the unperturbed 2p-2h energy. 
}
\label{fig:O16_mono00ampL}
\end{figure}

I first discuss the low-lying resonance at $\hbar\omega=24.8$~MeV, which correspond to the main peak of the $0^{+}$ component of the double IS $0^{+}$ excitation.
Figure~\ref{fig:O16_mono00ampL} shows the single transition amplitudes $C$, defined in Eq.~\eqref{eq:t2p2hamp}, as a function of the unperturbed 2p-2h energy.
As explained in the previous section, two-body excitations can be classified into three types of 2p-2h configurations: proton-proton ($pp$), neutron-neutron ($nn$), and neutron-proton ($np$).
The upper panel of Fig.~\ref{fig:O16_mono00ampL} displays the $nn$ and $pp$ configurations, while the lower panel shows the $np$ configurations.
The solid line denotes the cumulative sum of the single transition amplitudes up to a given unperturbed 2p-2h energy.
One finds that the single transition amplitude receives sizable contributions from a wide range of unperturbed energies, and that most amplitudes have the same sign, indicating that this resonance is formed constructively from many 2p-2h configurations.
This behavior is consistent with the fact that the double IS $0^{+}$ mode involves $2\hbar\omega$ configurations and therefore can draw non-negligible strength even from relatively high-lying 2p-2h states.
Furthermore, this $24.8$~MeV resonance does not emerge when the calculation is performed within a reduced model space.
The cumulative contributions from the $nn$+$pp$ configurations are comparable in magnitude to those from the $np$ configurations: $\sum C=-4.024$ for $nn$+$pp$ and $\sum C=-4.155$ for $np$.
\par
As discussed in Sec.~\ref{sec:SSRPA}, the residual interaction among $nn$ and $pp$ configurations exhibits a slight attractive tendency (i.e., matrix elements biased toward negative values), which favors a downward shift of the resonance relative to SSRPA$_{\mathrm{D}}$.
In contrast, the $np$ configurations tend to act repulsively and push the strength to higher excitation energies.
More generally, it is found that the direction of the energy shift correlates with the balance of the single transition amplitudes: when the cumulative contribution from $nn$ + $pp$ configurations is comparable to that from $np$ configurations, the strength tends to be shifted to lower energies than in SSRPA$_{\mathrm{D}}$.
Conversely, when the $np$ contribution is dominant, the strength distribution is pushed toward higher excitation energies.
\par
Table~\ref{tab:amp00L} lists representative 2p-2h configurations contributing to the low-lying resonance of the double IS $0^{+}$ mode at $\hbar\omega=24.8$~MeV shown in Fig.~\ref{fig:O16_mono00ampL}.
In particular, the proton $[2p_{1/2}][0p_{1/2}]^{-1}$ configuration, which is one of the dominant components of the IS $0^{+}$ strength observed in Fig.~\ref{fig:O16_mono}, play an important role in this resonance through its coupling with other 1p-1h configurations.
Thus, this resonance can be interpreted, to a good approximation, as a double IS $0^{+}$ excitation built on top of proton $[2p_{1/2}][0p_{1/2}]^{-1}$ configuration.
\begin{table}[t]
\caption{
Single transition amplitude of double IS $0^{+}$ excitation at for $\hbar\omega=24.8$ MeV.
}
\begin{ruledtabular}
\begin{tabular}{lcc}
\multicolumn{1}{c}{2p-2h configuration} & $E_{2p2h}$ &
\multicolumn{1}{c}{$C_{\nu;p_{1}p_{2}h_{1}h_{2}}$} \\
\hline
$([2p_{1/2}][0p_{3/2}]^{-1})_{\pi}([3p_{3/2}][0p_{1/2}]^{-1})_{\pi}$ & $40.15$ & $-0.281$ \\
$([3p_{3/2}][0p_{3/2}]^{-1})_{\nu}([2p_{1/2}][0p_{1/2}]^{-1})_{\pi}$ & $42.22$ & $-0.222$\\
$([2p_{1/2}][0p_{3/2}]^{-1})_{\pi}([4p_{3/2}][0p_{1/2}]^{-1})_{\pi}$ & $44.22$ & $-0.258$\\
$([4p_{3/2}][0p_{3/2}]^{-1})_{\nu}([2p_{1/2}][0p_{1/2}]^{-1})_{\pi}$ & $46.24$ & $-0.196$ \\
$([1s_{1/2}][0s_{1/2}]^{-1})_{\pi}([2p_{1/2}][0p_{1/2}]^{-1})_{\pi}$ & $46.25$ & $-0.320$\\
$([1s_{1/2}][0s_{1/2}]^{-1})_{\nu}([2p_{1/2}][0p_{1/2}]^{-1})_{\pi}$ & $46.96$ & $-0.278$\\
\end{tabular}
\end{ruledtabular}
\label{tab:amp00L}
\end{table}
\par
\begin{figure}
\centering
\includegraphics[width=0.99\linewidth]{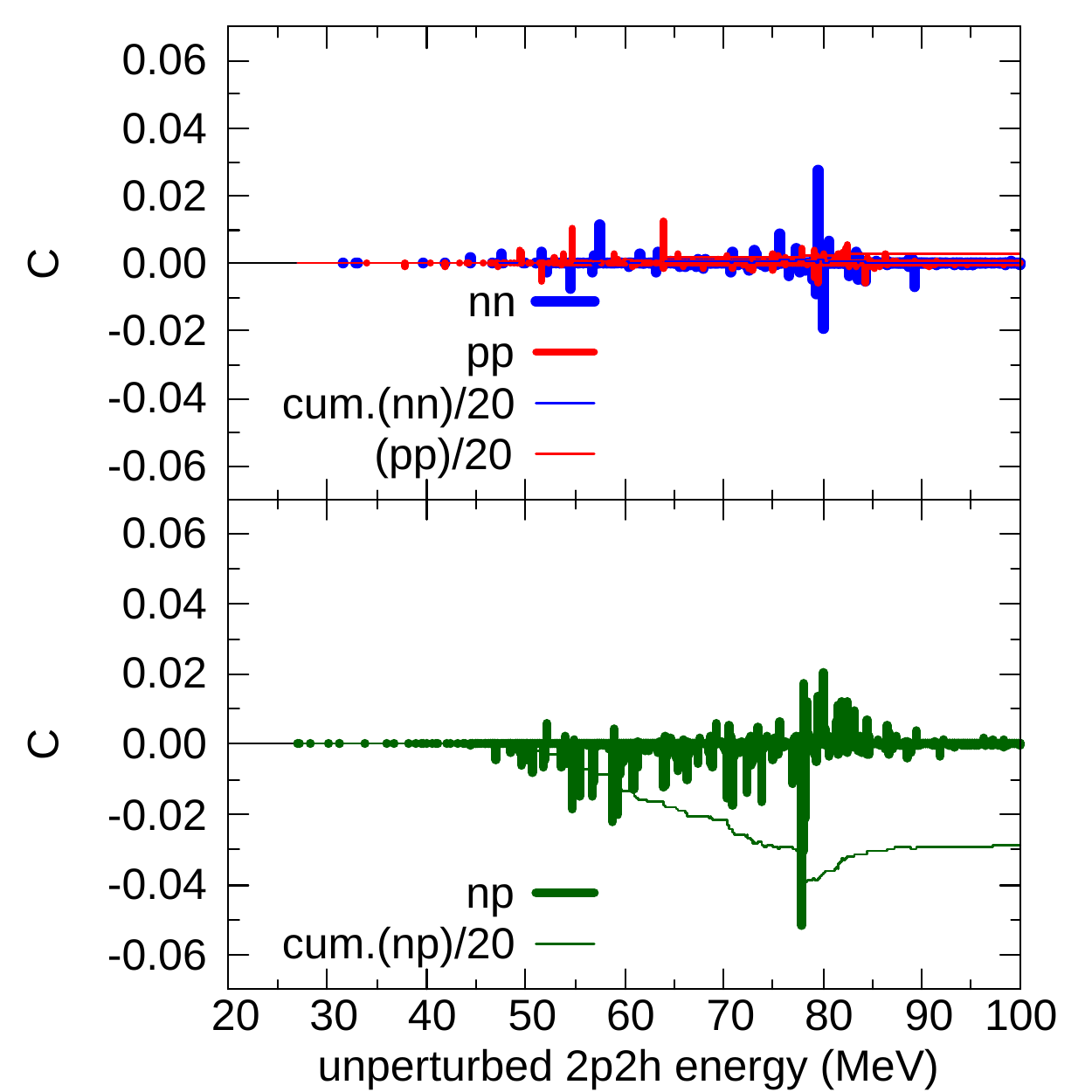}
\caption{Same as Fig.~\ref{fig:O16_mono00ampL}, but for the double IS $0^{+}$ excitation at $\hbar\omega=80.6$~MeV.}
\label{fig:O16_mono00ampH}
\end{figure}

Figure~\ref{fig:O16_mono00ampH} shows the single transition amplitudes and their cumulative sum for the higher-lying resonance at $\hbar\omega=80.6$~MeV in the double IS $0^{+}$ excitation.
In this case, the amplitudes associated with the $np$ configurations are dominant, while those from the $nn$ and $pp$ configurations are comparatively small.
As a consequence, the corresponding strength is shifted to higher excitation energies relative to the SSRPA$_{\mathrm{D}}$ result.
Similar to the low-lying resonance shown in Fig.~\ref{fig:O16_mono00ampL}, the amplitudes contributing to this high-lying resonance are distributed over a wide range of configurations, with comparable magnitudes.
However, unlike the low-lying case, a non-negligible fraction of the amplitudes has the opposite sign.
This indicates that the high-lying resonance is less collective.
A similar pattern is found for the double IS $2^{+}$ excitation, both for the low-lying main peak and for the higher-lying resonances.
Therefore, I do not repeat the discussion here.
\par
\begin{figure}
\centering
\includegraphics[width=0.99\linewidth]{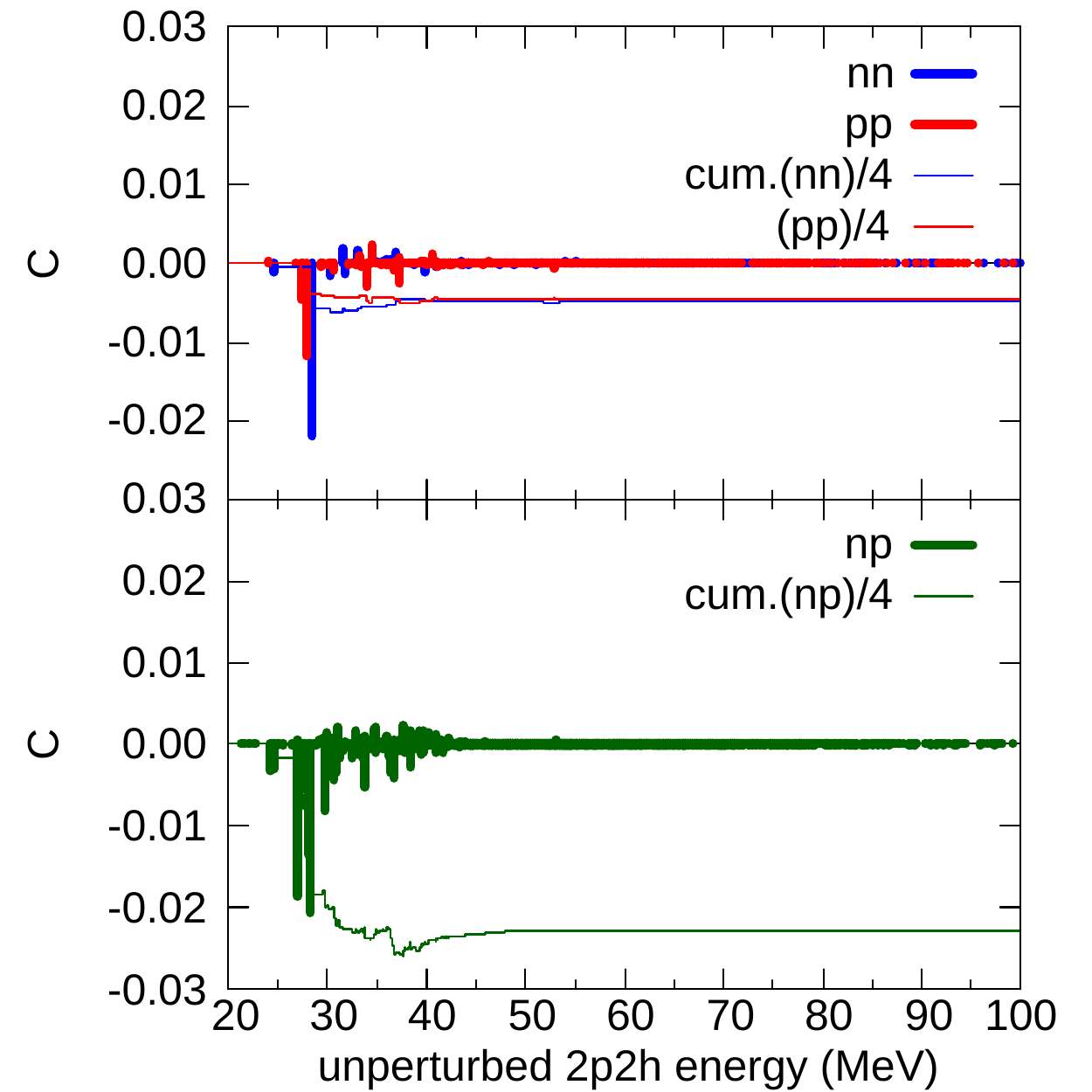}
\caption{2p-2h configurations contributing the resonance of double IV $1^{-}$ excitation at $\hbar\omega=40.4$~MeV. 
The solid line shows the cumulative sum divided by $4$ up to a given unperturbed 2p-2h energy.}
\label{fig:O16_mono11amp}
\end{figure}

Figure~\ref{fig:O16_mono11amp} shows the single transition amplitudes and their cumulative sums of the double IV $1^{-}$ excitation at $\hbar\omega=40.4$~MeV.
Most of the individual 2p-2h transition amplitudes for the $nn$, $pp$, and $np$ configurations are negative, and the resonance is formed constructively from these amplitudes.
The amplitudes associated with the $np$ configurations are clearly larger than those from the $nn$ and $pp$ configurations, indicating that the double IV $1^{-}$ mode is dominated by particle-hole excitations involving different nucleon species.
The total transition amplitude summed over 2p-2h configurations is $\sum C_{\nu, p_{1}p_{2}h_{1}h_{2}}=-0.0373$ for the $nn$+$pp$ configurations, whereas it is $\sum C_{\nu, p_{1}p_{2}h_{1}h_{2}}=-0.0909$ for the $np$ configurations.
As discussed above, the predominantly repulsive character of the $np$ channel shifts the strength to higher excitation energies compared with the SSRPA$_{\mathrm{D}}$ result.
\par
Table~\ref{tab:amp11} lists representative leading configurations for this resonance. 
The neutron and proton $[0d_{5/2}][0p_{3/2}]^{-1}$ configurations, which are one of the dominant components of the IV $1^{-}$ strength observed in Fig.~\ref{fig:O16_dipo}, play an important role in this resonance through their coherent coupling with other 1p-1h configurations.
Thus, similar to the mechanism found in the double IS $0^{+}$ excitation, this resonance may be interpreted as a double IV $1^{-}$ excitation built on top of these underlying single IV dipole configurations.
\begin{table}[t]
\caption{
Single transition amplitude of double IV $1^{-}$ excitation at for $\hbar\omega=40.4$ MeV.
}
\begin{ruledtabular}
\begin{tabular}{lcc}
\multicolumn{1}{c}{2p-2h configuration} & $E_{2p2h}$ &
\multicolumn{1}{c}{$C_{\nu;p_{1}p_{2}h_{1}h_{2}}$} \\
\hline
$([0d_{5/2}][0p_{3/2}]^{-1})_{\nu}([0d_{5/2}][0p_{3/2}]^{-1})_{\pi}$ & $26.96$ & $-0.0186$\\
$([0d_{5/2}][0p_{1/2}]^{-1})_{\nu}([1d_{3/2}][0p_{3/2}]^{-1})_{\pi}$ & $28.05$ & $-0.0135$ \\
$([0d_{5/2}][0p_{3/2}]^{-1})_{\pi}([1d_{3/2}][0p_{1/2}]^{-1})_{\pi}$ & $27.92$ & $-0.0117$ \\
$([0d_{5/2}][0p_{3/2}]^{-1})_{\nu}([0d_{3/2}][0p_{1/2}]^{-1})_{\nu}$ & $28.48$ & $-0.0220$\\
$([0d_{3/2}][0p_{3/2}]^{-1})_{\nu}([0d_{5/2}][0p_{1/2}]^{-1})_{\pi}$ & $28.35$ & $-0.0205$\\
\end{tabular}
\end{ruledtabular}
\label{tab:amp11}
\end{table}
\par
\begin{figure}
\centering
\includegraphics[width=0.99\linewidth]{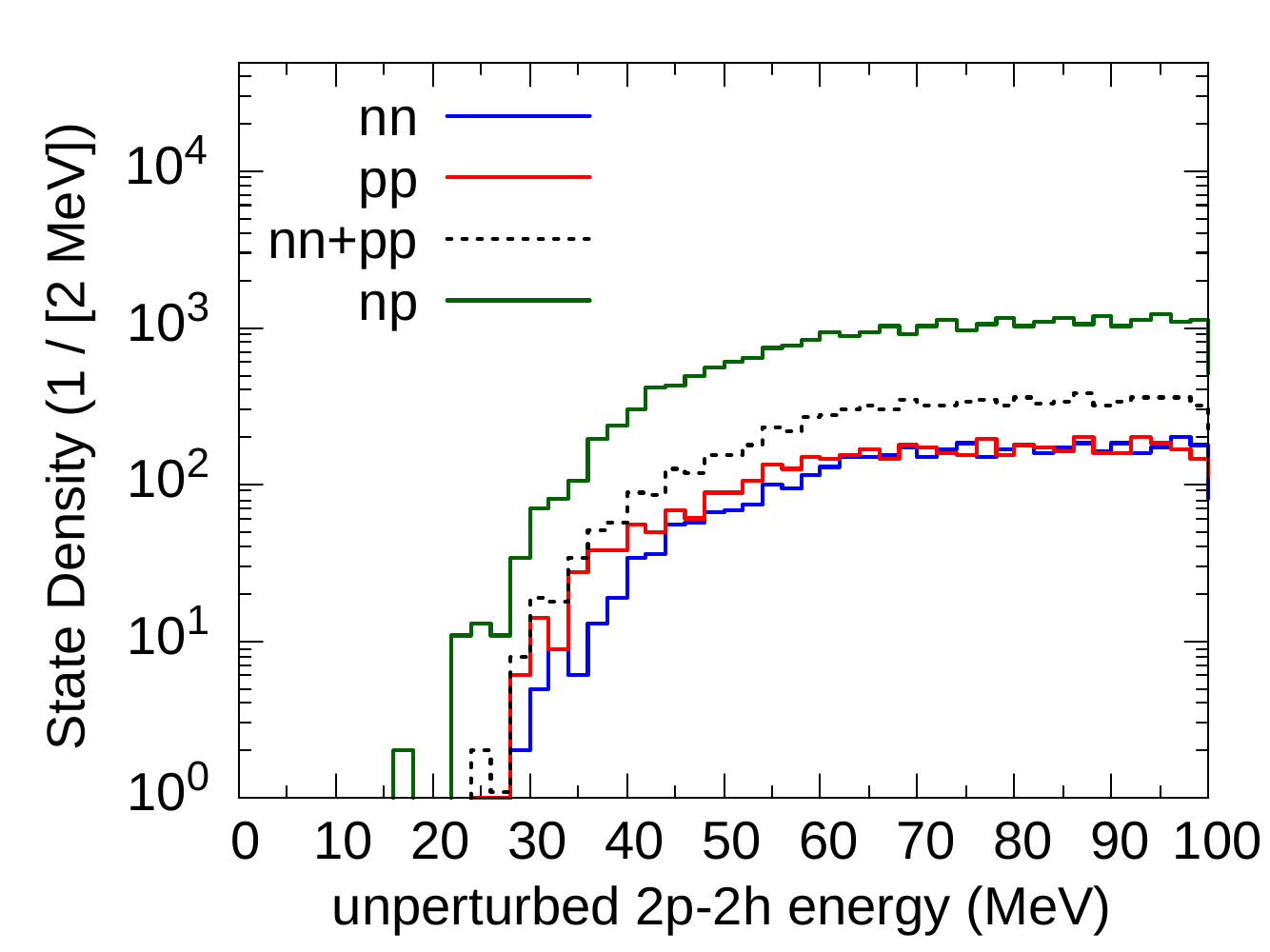}
\caption{$0^{+}$ state densities of neutron-neutron ($nn$), proton-proton ($pp$), their sum ($nn+pp$) and neutron-proton ($np$) 2p-2h configurations as functions of the unperturbed 2p-2h energy in $^{16}$O. The energy bin width is $2$~MeV.}
\label{fig:StateDensity}
\end{figure}

The enhanced contribution of the $np$ configurations to the high-lying resonances can be qualitatively understood from the behavior of the corresponding state densities.
Fig.~\ref{fig:StateDensity} shows the state densities of the $^{16}$O($0^{+}$) 2p-2h configurations for the $nn$, $pp$, and $np$ sectors as a function of unperturbed 2p-2h energy, using an energy bin of $2$~MeV.
Up to $30$--$35$~MeV, the combined state density of the $nn$ and $pp$ configurations ($nn$+$pp$) is comparable to that of the $np$ configurations.
Above $40$~MeV, however, the state density of the $np$ configurations clearly becomes larger by about a factor of $3$ compared with the sum of the $nn$+$pp$ contributions.
This behavior arises naturally from the Pauli exclusion principle.
Once a nucleon is excited to from 1p-1h configuration, the occupied and unoccupied states for a second identical nucleon ($nn$ or $pp$) are restricted, reducing the number of available 2p-2h configurations.
In contrast, such blocking does not occur between neutrons and protons, so the number of possible $np$ configurations remains much larger, especially at higher excitation energies.
As a consequence, high-lying resonances are predominantly built from neutron-proton 2p-2h configurations.
The same behavior is observed in $^{40}$Ca, as will be discussed in the next section.
It should be noted that in the present calculation the continuum single-particle spectrum is discretized by using a finite box, and the resulting state density is constructed from these discretized levels.
Nevertheless, the relative behavior, namely the significantly larger $np$ state density compared with $nn$ and $pp$ at high energies, is expected to remain qualitatively unchanged even under an exact treatment of the continuum.
\subsection{Examination of the qualitative features in $^{40}$Ca}
The discussion obtained so far is limited to $^{16}$O.
In order to see whether the same features is obtained in other nuclei, the same calculation is carried out for $^{40}$Ca.
\par
\begin{figure}
\includegraphics[width=0.98\linewidth]{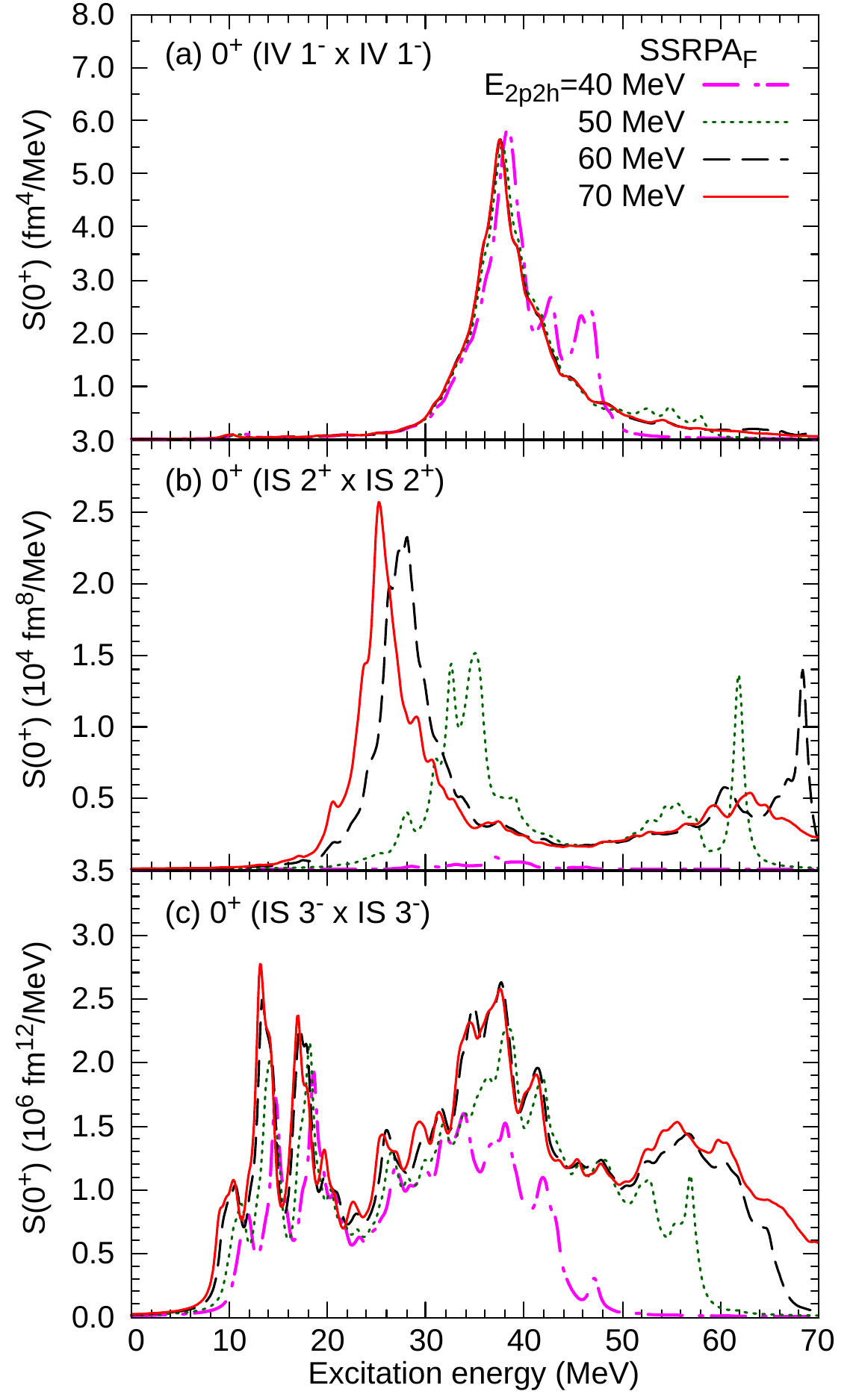}
\caption{
%Strength distributions for the one-body external fields calculated within RPA ((a)--(c)) and those for the corresponding two-body external fields calculated within SSRPA$_{\mathrm{F}}$ ((d)--(f)) defined in Eq.~\eqref{eq:DIVD} in $^{40}$Ca.  
Strength distributions for the two-body external fields calculated within SSRPA$_{\mathrm{F}}$ in $^{40}$Ca, where the external fields are defined in Eq.~\eqref{eq:DIVD}.
The results obtained with different cutoffs for the unperturbed 2p-2h excitation energies are compared.
%Experimental data of the centroid energy are taken from Refs.~\cite{Youngblood2001, Jingo2018, Ahrens1975}.
}
\label{fig:Ca40}
\end{figure}

Figure~\ref{fig:Ca40} shows the strength distributions for the two-body external fields (double IV $1^{-}$, IS $2^{+}$, and IS $3^{-}$ modes) calculated within SSRPA$_{\mathrm{F}}$ in $^{40}$Ca, where the external fields are defined in Eq.~\eqref{eq:DIVD}.
The results obtained with different cutoffs for the unperturbed 2p-2h excitation energies are compared.
For the double IV $1^{-}$ mode, the strength distribution is already converged even for a small value of $E_{2p2h}$, while for the double IS $2^{+}$ and IS $3^{-}$ modes, the strength distributions depend sensitively on $E_{2p2h}$.
Although the model space is truncated at $E_{2p2h}=70$~MeV, the result for the double IS $2^{+}$ mode is not expected to be fully converged.
Nevertheless, as in $^{16}$O, the strength distribution exhibits a main peak around $\hbar\omega=26$~MeV, together with non-negligible strength above $\hbar\omega=40$~MeV.
For the double IS $3^{-}$ mode, the low-lying distribution below $\hbar\omega=30$~MeV is not sensitive to the cutoff energy of the unperturbed 2p-2h configurations, whereas significant missing strength remains at higher excitation energies. 
Those results are consistent with those obtained in $^{16}$O.
\par
Table~\ref{tab:centroid_width_HF2} lists the center energies and widths of the strength distributions induced by one- and two-body external fields in the unperturbed calculations.
The results are qualitatively similar to those for $^{16}$O, as seen in Table~\ref{tab:centroid_width_HF2}.
The center energies for the two-body external fields are close to twice those of the corresponding one-body external fields.
The width for the double IS $2^{+}$ mode increases by more than a factor of two, while those for the IS $0^{+}$ and the IV $1^{-}$ modes show moderate increases by about a factor of $1.5$--$1.6$.
\begin{table}[t]
\caption{Center energies and widths of the IS $0^{+}$, IV $1^{-}$, and IS $2^{+}$ modes ($E^{(1)}_{\mathrm{L}}$ and $\Gamma_{\mathrm{L}}^{(1)}$), and of the $0^{+}$ component of the corresponding double multipole modes obtained from unperturbed calculations ($E^{(2)}_{\mathrm{L}}$ and $\Gamma_{\mathrm{L}}^{(2)}$) for $^{40}$Ca. All values are given in MeV.}
\label{tab:centroid_width_HF2}
\begin{ruledtabular}
\begin{tabular}{lcclcc}
\multicolumn{6}{c}{\textbf{unperturbed}}\\ 
& \multicolumn{2}{c}{one-body} & & \multicolumn{2}{c}{two-body}\\
Mode & $E_{\mathrm{L}}^{(1)}$ & $\Gamma_{\mathrm{L}}^{(1)}$ & Mode & $E_{\mathrm{L}}^{(2)}$ & $\Gamma_{\mathrm{L}}^{(2)}$ \\
\hline
  IS $0^{+}$ & 25.8 & 7.6 & double IS $0^{+}$ & 50.1 & 12.2 \\
  IV $1^{-}$ & 12.8 &  1.4 & double IV $1^{-}$ & 24.6 &  2.3 \\
  IS $2^{+}$ & 23.7 &  2.4 & double IS $2^{+}$ & 47.9 &  6.4 \\
\end{tabular}
\end{ruledtabular}
\end{table}
%exp $E_{L}=19.7, \Gamma_{L}=4.7$(Jingo)
%$E_{L}=20.0, \Gamma_{L}=5.5$(Ahrens)
%
\par
The center energies $E_{\mathrm{L}}$ and widths $\Gamma_{\mathrm{L}}$ of the strength distribution induced by one-body external fields and by the corresponding two-body external fields in $^{40}$Ca calculated within RPA and SSRPA$_{\mathrm{F}}$ are listed in Table~\ref{tab:centroid_width2}.
%The centroid energies $m_{1}/m_{0}$ calculated with SSRPA$_{\mathrm{F}}$ are also shown.
For the double IS $0^+$ and IS $2^+$ modes, $E_{\mathrm{L}}$ and $\Gamma_{\mathrm{L}}$ are extracted from the lower-energy peak, referred to as the first group.
Similar to the results obtained in $^{16}$O, the center energies of the double IS $0^{+}$ and IS $2^{+}$ modes are lower than twice those of the corresponding one-body excitations, indicating deviations from the simple harmonic double-phonon picture.
In contrast, the center energy of the double IV $1^{-}$ mode is approximately twice the corresponding one-body value, suggesting that the simple double-phonon picture remains valid for this mode.
Although the detailed shape of strength distribution differs somewhat because a different basis is used, the overall behavior of the present result is consistent with that reported by Nishizaki and Wambach~\cite{Nishizaki1995}.
Regarding the widths, the width of the double IS $0^{+}$ mode is smaller than that of the corresponding one-body fields, whereas that of the double IS $2^{+}$ mode is larger.
The width of the double IV $1^{-}$ mode is about three times larger than that of the corresponding one-body response.
Those results are again consistent with those obtained in $^{16}$O.
%When focusing on the centroid energies $m_{1}/m_{0}$, the values for the double IS $0^{+}$ and IS $2^{+}$ modes are close to twice those of the corresponding the one-body fields.
%
\begin{table}[t]
\caption{
Center energies $E_{\mathrm{L}}$ and widths $\Gamma_{\mathrm{L}}$ for the $0^{+}$ components of the corresponding double multipole modes in $^{40}$Ca obtained with SSRPA$_{\mathrm{F}}$. 
For the double IS $0^+$ and IS $2^+$ modes, $E_{\mathrm{L}}$ and $\Gamma_{\mathrm{L}}$ are extracted from the lower-energy peak, referred to as the first group in the text.
The centroid energies $m_{1}/m_{0}$ calculated with SSRPA$_{\mathrm{F}}$ are also shown.
All values are given in MeV.
}
\label{tab:centroid_width2}
\begin{ruledtabular}
\begin{tabular}{lcclcc}
  \multicolumn{3}{c}{\textbf{RPA}}
& \multicolumn{3}{c}{\textbf{SSRPA$_{\mathrm{F}}$}} \\ 
Mode
& $E_{\mathrm{L}}^{(1)}$ & $\Gamma_{\mathrm{L}}^{(1)}$ 
& Mode
& $E_{\mathrm{L}}^{(2)}$ & $\Gamma_{\mathrm{L}}^{(2)}$ \\
%& $m_{1}/m_{0}$ \\
\hline
IS $0^{+}$ & 21.4 & 4.6 & double IS $0^{+}$ & 18.1 & 1.4 \\%& 66.0 \\
IV $1^{-}$ & 18.3 & 1.8 & double IV $1^{-}$ & 37.8 & 5.2 \\%& 39.9 \\
IS $2^{+}$ & 17.3 & 0.4 & double IS $2^{+}$ & 25.8 & 4.7 \\%& 62.6 \\
%IS $3^{-}$ & & & double IS $3^{-}$ & 36.7 & 39.3 & \\
\end{tabular}
\end{ruledtabular}
\end{table}
\begin{figure}
\centering
\includegraphics[width=0.99\linewidth]{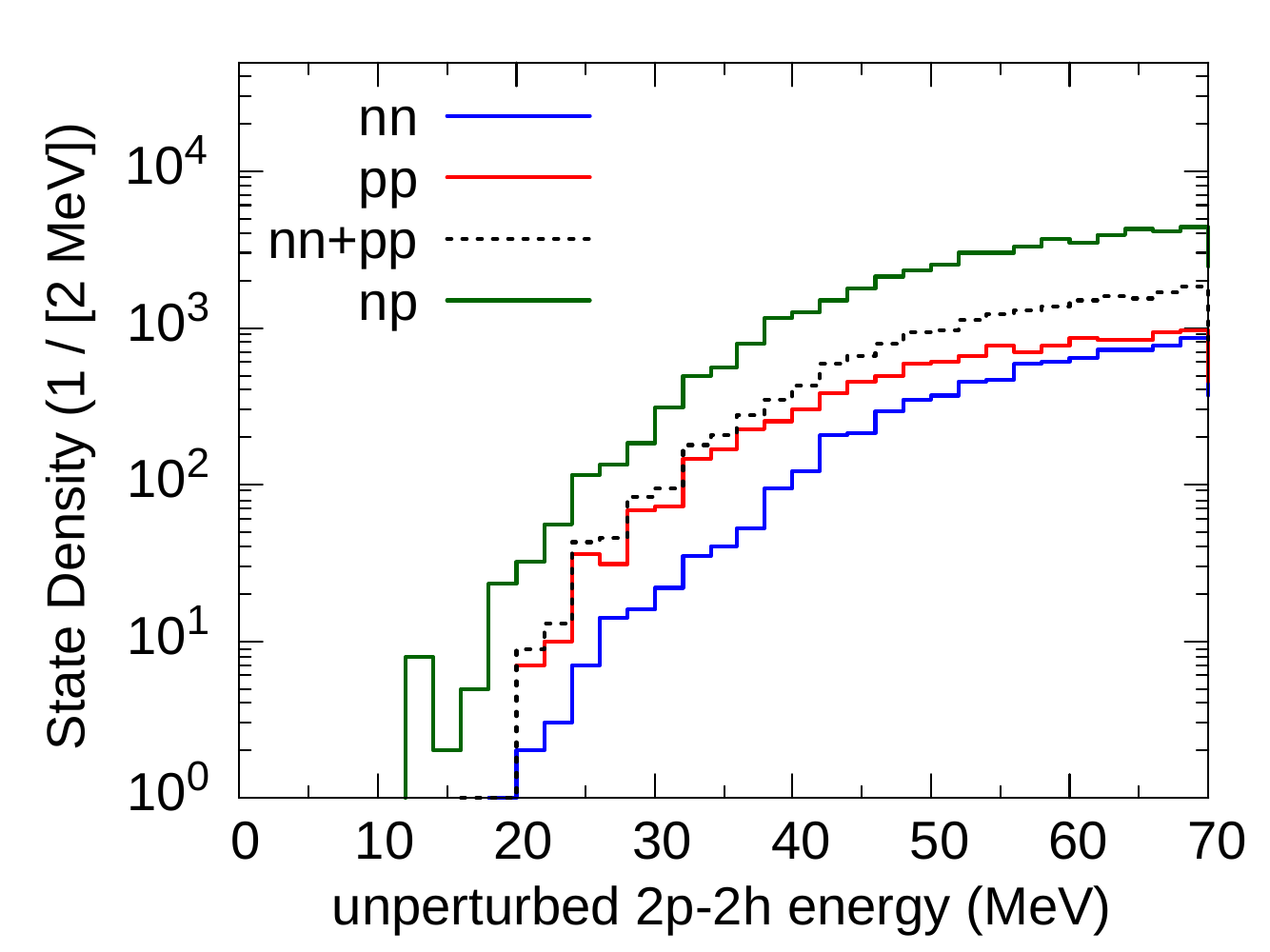}
\caption{Same as Fig.~\ref{fig:StateDensity}, but for $^{40}$Ca.}
\label{fig:StateDensity_Ca40}
\end{figure}
\par
Figure~\ref{fig:StateDensity_Ca40} shows the state densities of the $^{40}$Ca($0^{+}$) 2p-2h configurations for the $nn$, $pp$, and $np$ sectors as a function of unperturbed 2p-2h energy, using an energy bin of $2$~MeV.
The state density of the $nn+pp$ configurations is comparable to that of the $np$ configurations at lower energies, although the $np$ state density is larger over the entire energy region.
As already discussed for $^{16}$O, the $nn$ and $np$ configurations contribute attractively, while the $np$ configurations contribute repulsively.
The behavior may underlie the formation of the strength distributions observed in Fig.~\ref{fig:Ca40}.
\par
The results obtained in $^{40}$Ca are qualitatively similar to those found in $^{16}$O. 
This suggests that the qualitative features of double-phonon excitations discussed above are not specific to $^{16}$O, but may persist in other nuclei, at least up to this mass region.
\section{Summary}
\label{sect:summary}

Nuclear responses to two-body external fields, interpreted as double-phonon excitations, in $^{16}$O and $^{40}$Ca were investigated within the subtracted second random-phase approximation (SSRPA).
To clarify the baseline behavior, I first compared Hartree–Fock results with those obtained using SSRPA with the diagonal-approximation.
Their strength distributions nearly coincided for the two-body fields, indicating that residual interactions in the 1p-1h sector, as well as the subtraction procedure (which modifies only that sector), have only a limited influence in this channel. 
This behavior contrasted with one-body excitations, where the coupling between 1p-1h and 2p-2h configurations is essential for generating collectivity and damping patterns.
\par
When the full SSRPA calculation was performed, including couplings among different 2p-2h configurations through the residual interaction in $A_{22'}$, the two-body strengh distributions were significantly reshaped.
The double IS $0^+$ and $2^+$ modes of $^{16}$O showed strong redistribution: their main peaks are shifted to lower energies and substantial additional strength emerged at high excitation energies, suggesting the coexistence of attractive and repulsive components in the residual interaction.
In contrast, the double IV $1^-$ mode is pushed to higher energies, consistent with a predominantly repulsive contribution, particularly in channels dominated by neutron-proton configurations. 
\par
An analysis of single transition amplitudes revealed that low-lying resonances are built constructively from many 2p-2h configurations spread over a wide unperturbed energy range, with comparable cumulative contributions from $nn$/$pp$ and $np$ components, whereas high-lying resonances were dominated by $np$ configurations owing to their larger state density. 
Overall, these results demonstrate that double-phonon responses are not captured by a simple folding of one-body strength functions, and that explicit 2p-2h mixing treated in SSRPA is indispensable for a microscopic description.
\par
The same calculation for $^{40}$Ca as for $^{16}$O was performed to examine whether the qualitative feature found in $^{16}$O are specific to this nucleus.
Our analysis showed that the results for $^{40}$Ca are qualitatively similar to those for $^{16}$O.
This suggests that the qualitative features observed in both nuclei may also appear in other nuclei.
\par
This work focused on $^{16}$O and $^{40}$Ca, and restricted our analysis to the $0^{+}$ components of the excited states.
To obtain a more general perspective, it will be important to extend the study to heavier nuclei and other $J^{\pi}$ multipoles.
Nevertheless, it is expected that the present findings provide a useful basis for future investigations of additional nuclei and excited states.
As a next step, I plan to explore microscopic two-body current effects in muon-capture reactions within the same theoretical framework.
\section*{Acknowledgments}
The author thanks H. Sagawa and C. L. Bai for their technical support with the SSRPA formalism.
This work was supported by JSPS KAKENHI Grant Numbers JP23K03426, JP24K00647, and 25H00641, 26H00410, and by JST ERATO Grant No. JPMJER2304, Japan.
\bibliography{ref}

\begin{thebibliography}{79}%
\makeatletter
\providecommand \@ifxundefined [1]{%
 \@ifx{#1\undefined}
}%
\providecommand \@ifnum [1]{%
 \ifnum #1\expandafter \@firstoftwo
 \else \expandafter \@secondoftwo
 \fi
}%
\providecommand \@ifx [1]{%
 \ifx #1\expandafter \@firstoftwo
 \else \expandafter \@secondoftwo
 \fi
}%
\providecommand \natexlab [1]{#1}%
\providecommand \enquote  [1]{``#1''}%
\providecommand \bibnamefont  [1]{#1}%
\providecommand \bibfnamefont [1]{#1}%
\providecommand \citenamefont [1]{#1}%
\providecommand \href@noop [0]{\@secondoftwo}%
\providecommand \href [0]{\begingroup \@sanitize@url \@href}%
\providecommand \@href[1]{\@@startlink{#1}\@@href}%
\providecommand \@@href[1]{\endgroup#1\@@endlink}%
\providecommand \@sanitize@url [0]{\catcode `\\12\catcode `\$12\catcode
  `\&12\catcode `\#12\catcode `\^12\catcode `\_12\catcode `\%12\relax}%
\providecommand \@@startlink[1]{}%
\providecommand \@@endlink[0]{}%
\providecommand \url  [0]{\begingroup\@sanitize@url \@url }%
\providecommand \@url [1]{\endgroup\@href {#1}{\urlprefix }}%
\providecommand \urlprefix  [0]{URL }%
\providecommand \Eprint [0]{\href }%
\providecommand \doibase [0]{https://doi.org/}%
\providecommand \selectlanguage [0]{\@gobble}%
\providecommand \bibinfo  [0]{\@secondoftwo}%
\providecommand \bibfield  [0]{\@secondoftwo}%
\providecommand \translation [1]{[#1]}%
\providecommand \BibitemOpen [0]{}%
\providecommand \bibitemStop [0]{}%
\providecommand \bibitemNoStop [0]{.\EOS\space}%
\providecommand \EOS [0]{\spacefactor3000\relax}%
\providecommand \BibitemShut  [1]{\csname bibitem#1\endcsname}%
\let\auto@bib@innerbib\@empty
%</preamble>
\bibitem [{\citenamefont {Bohm}\ and\ \citenamefont {Pines}(1953)}]{BohmPines}%
  \BibitemOpen
  \bibfield  {author} {\bibinfo {author} {\bibfnamefont {D.}~\bibnamefont
  {Bohm}}\ and\ \bibinfo {author} {\bibfnamefont {D.}~\bibnamefont {Pines}},\
  }\bibfield  {title} {\bibinfo {title} {A collective description of electron
  interactions: Iii. coulomb interactions in a degenerate electron gas},\
  }\href {https://doi.org/10.1103/PhysRev.92.609} {\bibfield  {journal}
  {\bibinfo  {journal} {Phys. Rev.}\ }\textbf {\bibinfo {volume} {92}},\
  \bibinfo {pages} {609} (\bibinfo {year} {1953})}\BibitemShut {NoStop}%
\bibitem [{\citenamefont {Bohr}\ and\ \citenamefont
  {Mottelson}(1975)}]{BohrMottelson}%
  \BibitemOpen
  \bibfield  {author} {\bibinfo {author} {\bibfnamefont {A.}~\bibnamefont
  {Bohr}}\ and\ \bibinfo {author} {\bibfnamefont {B.}~\bibnamefont
  {Mottelson}},\ }\href@noop {} {\emph {\bibinfo {title} {Nuclear Structure
  Volume II: Nuclear Deformations}}}\ (\bibinfo  {publisher} {Benjamin, New
  York},\ \bibinfo {year} {1975})\BibitemShut {NoStop}%
\bibitem [{\citenamefont {Ring}\ and\ \citenamefont
  {Schuck}(1980)}]{RingandSchuck}%
  \BibitemOpen
  \bibfield  {author} {\bibinfo {author} {\bibfnamefont {P.}~\bibnamefont
  {Ring}}\ and\ \bibinfo {author} {\bibfnamefont {P.}~\bibnamefont {Schuck}},\
  }\href@noop {} {\emph {\bibinfo {title} {The Nuclear Many-Body Problem}}}\
  (\bibinfo  {publisher} {Springer-Verlag, Berlin},\ \bibinfo {year}
  {1980})\BibitemShut {NoStop}%
\bibitem [{\citenamefont {Bertsch}\ and\ \citenamefont
  {Tsai}(1975)}]{Bertsch1975}%
  \BibitemOpen
  \bibfield  {author} {\bibinfo {author} {\bibfnamefont {G.}~\bibnamefont
  {Bertsch}}\ and\ \bibinfo {author} {\bibfnamefont {S.}~\bibnamefont {Tsai}},\
  }\bibfield  {title} {\bibinfo {title} {A study of the nuclear response
  function},\ }\href
  {https://doi.org/https://doi.org/10.1016/0370-1573(75)90003-4} {\bibfield
  {journal} {\bibinfo  {journal} {Physics Reports}\ }\textbf {\bibinfo {volume}
  {18}},\ \bibinfo {pages} {125} (\bibinfo {year} {1975})}\BibitemShut
  {NoStop}%
\bibitem [{\citenamefont {Ichimura}\ \emph {et~al.}(2006)\citenamefont
  {Ichimura}, \citenamefont {Sakai},\ and\ \citenamefont
  {Wakasa}}]{Ichimura2013}%
  \BibitemOpen
  \bibfield  {author} {\bibinfo {author} {\bibfnamefont {M.}~\bibnamefont
  {Ichimura}}, \bibinfo {author} {\bibfnamefont {H.}~\bibnamefont {Sakai}},\
  and\ \bibinfo {author} {\bibfnamefont {T.}~\bibnamefont {Wakasa}},\
  }\bibfield  {title} {\bibinfo {title} {Spin-isospin responses via ($p,n$) and
  ($n,p$) reactions},\ }\href {https://doi.org/10.1016/j.ppnp.2005.09.001}
  {\bibfield  {journal} {\bibinfo  {journal} {Progress in Particle and Nuclear
  Physics}\ }\textbf {\bibinfo {volume} {56}},\ \bibinfo {pages} {446}
  (\bibinfo {year} {2006})}\BibitemShut {NoStop}%
\bibitem [{\citenamefont {Matsuo}(2015)}]{Matsuo2015}%
  \BibitemOpen
  \bibfield  {author} {\bibinfo {author} {\bibfnamefont {M.}~\bibnamefont
  {Matsuo}},\ }\bibfield  {title} {\bibinfo {title} {Continuum quasiparticle
  random-phase approximation for astrophysical direct neutron capture reactions
  on neutron-rich nuclei},\ }\href {https://doi.org/10.1103/PhysRevC.91.034604}
  {\bibfield  {journal} {\bibinfo  {journal} {Phys. Rev. C}\ }\textbf {\bibinfo
  {volume} {91}},\ \bibinfo {pages} {034604} (\bibinfo {year}
  {2015})}\BibitemShut {NoStop}%
\bibitem [{\citenamefont {Saito}\ and\ \citenamefont
  {Matsuo}(2023)}]{Saito2023}%
  \BibitemOpen
  \bibfield  {author} {\bibinfo {author} {\bibfnamefont {T.}~\bibnamefont
  {Saito}}\ and\ \bibinfo {author} {\bibfnamefont {M.}~\bibnamefont {Matsuo}},\
  }\bibfield  {title} {\bibinfo {title} {Continuum random-phase approximation
  for $(n,\ensuremath{\gamma})$ reactions on neutron-rich nuclei: Collective
  effects and resonances},\ }\href
  {https://doi.org/10.1103/PhysRevC.107.064607} {\bibfield  {journal} {\bibinfo
   {journal} {Phys. Rev. C}\ }\textbf {\bibinfo {volume} {107}},\ \bibinfo
  {pages} {064607} (\bibinfo {year} {2023})}\BibitemShut {NoStop}%
\bibitem [{\citenamefont {Sasaki}\ \emph {et~al.}(2025)\citenamefont {Sasaki},
  \citenamefont {Kawano},\ and\ \citenamefont {Dupuis}}]{Sasaki2025}%
  \BibitemOpen
  \bibfield  {author} {\bibinfo {author} {\bibfnamefont {H.}~\bibnamefont
  {Sasaki}}, \bibinfo {author} {\bibfnamefont {T.}~\bibnamefont {Kawano}},\
  and\ \bibinfo {author} {\bibfnamefont {M.}~\bibnamefont {Dupuis}},\
  }\bibfield  {title} {\bibinfo {title} {Modeling direct and pre-equilibrium
  processes of neutron-induced reactions with the noniterative finite amplitude
  method and with the distorted-wave born approximation},\ }\href
  {https://doi.org/10.1103/jjd9-9hrs} {\bibfield  {journal} {\bibinfo
  {journal} {Phys. Rev. C}\ }\textbf {\bibinfo {volume} {112}},\ \bibinfo
  {pages} {054607} (\bibinfo {year} {2025})}\BibitemShut {NoStop}%
\bibitem [{\citenamefont {Dupuis}(2017)}]{Dupuis2017}%
  \BibitemOpen
  \bibfield  {author} {\bibinfo {author} {\bibfnamefont {M.}~\bibnamefont
  {Dupuis}},\ }\bibfield  {title} {\bibinfo {title} {Microscopic description of
  elastic and direct inelastic nucleon scattering off spherical nuclei},\
  }\href {https://doi.org/10.1140/epja/i2017-12293-6} {\bibfield  {journal}
  {\bibinfo  {journal} {Eur. Phys. J. A}\ }\textbf {\bibinfo {volume} {53}},\
  \bibinfo {pages} {1} (\bibinfo {year} {2017})}\BibitemShut {NoStop}%
\bibitem [{\citenamefont {Dupuis}\ \emph {et~al.}(2024)\citenamefont {Dupuis},
  \citenamefont {Capote}, \citenamefont {Kawano}, \citenamefont {Kerveno},
  \citenamefont {Dessagne}, \citenamefont {Henning},\ and\ \citenamefont
  {Hilaire}}]{Dupuis2024}%
  \BibitemOpen
  \bibfield  {author} {\bibinfo {author} {\bibfnamefont {M.}~\bibnamefont
  {Dupuis}}, \bibinfo {author} {\bibfnamefont {R.}~\bibnamefont {Capote}},
  \bibinfo {author} {\bibfnamefont {T.}~\bibnamefont {Kawano}}, \bibinfo
  {author} {\bibfnamefont {M.}~\bibnamefont {Kerveno}}, \bibinfo {author}
  {\bibfnamefont {P.}~\bibnamefont {Dessagne}}, \bibinfo {author}
  {\bibfnamefont {G.}~\bibnamefont {Henning}},\ and\ \bibinfo {author}
  {\bibfnamefont {S.}~\bibnamefont {Hilaire}},\ }\bibfield  {title} {\bibinfo
  {title} {Microscopic modeling of direct pre-equilibrium emission: Impact on
  exclusive and inclusive ($n,xn$) and fission channels},\ }\href
  {https://doi.org/10.1051/epjconf/202429204003} {\bibfield  {journal}
  {\bibinfo  {journal} {EPJ Web Conf.}\ }\textbf {\bibinfo {volume} {292}},\
  \bibinfo {pages} {04003} (\bibinfo {year} {2024})}\BibitemShut {NoStop}%
\bibitem [{\citenamefont {{Da Providência}}(1965)}]{Providencia1965}%
  \BibitemOpen
  \bibfield  {author} {\bibinfo {author} {\bibfnamefont {J.}~\bibnamefont {{Da
  Providência}}},\ }\bibfield  {title} {\bibinfo {title} {{Variational
  approach to the many-body problem}},\ }\href
  {https://doi.org/https://doi.org/10.1016/0029-5582(65)90937-5} {\bibfield
  {journal} {\bibinfo  {journal} {Nuclear Physics}\ }\textbf {\bibinfo {volume}
  {61}},\ \bibinfo {pages} {87} (\bibinfo {year} {1965})}\BibitemShut {NoStop}%
\bibitem [{\citenamefont {Wambach}(1988)}]{Wambach1988}%
  \BibitemOpen
  \bibfield  {author} {\bibinfo {author} {\bibfnamefont {J.}~\bibnamefont
  {Wambach}},\ }\bibfield  {title} {\bibinfo {title} {Damping of
  small-amplitude nuclear collective motion},\ }\href
  {https://doi.org/10.1088/0034-4885/51/7/002} {\bibfield  {journal} {\bibinfo
  {journal} {Reports on Progress in Physics}\ }\textbf {\bibinfo {volume}
  {51}},\ \bibinfo {pages} {989} (\bibinfo {year} {1988})}\BibitemShut
  {NoStop}%
\bibitem [{\citenamefont {Dro\.{z}d\.{z}}\ \emph {et~al.}(1990)\citenamefont
  {Dro\.{z}d\.{z}}, \citenamefont {Nishizaki}, \citenamefont {Speth},\ and\
  \citenamefont {Wambach}}]{Drozdz1990}%
  \BibitemOpen
  \bibfield  {author} {\bibinfo {author} {\bibfnamefont {S.}~\bibnamefont
  {Dro\.{z}d\.{z}}}, \bibinfo {author} {\bibfnamefont {S.}~\bibnamefont
  {Nishizaki}}, \bibinfo {author} {\bibfnamefont {J.}~\bibnamefont {Speth}},\
  and\ \bibinfo {author} {\bibfnamefont {J.}~\bibnamefont {Wambach}},\
  }\bibfield  {title} {\bibinfo {title} {The nuclear response within extended
  rpa theories},\ }\href
  {https://doi.org/https://doi.org/10.1016/0370-1573(90)90084-F} {\bibfield
  {journal} {\bibinfo  {journal} {Physics Reports}\ }\textbf {\bibinfo {volume}
  {197}},\ \bibinfo {pages} {1} (\bibinfo {year} {1990})}\BibitemShut {NoStop}%
\bibitem [{\citenamefont {Ait-Tahar}\ and\ \citenamefont
  {Brink}(1993)}]{Ait-Tahar1993}%
  \BibitemOpen
  \bibfield  {author} {\bibinfo {author} {\bibfnamefont {S.}~\bibnamefont
  {Ait-Tahar}}\ and\ \bibinfo {author} {\bibfnamefont {D.}~\bibnamefont
  {Brink}},\ }\bibfield  {title} {\bibinfo {title} {Second rpa calculations of
  the charge-exchange quadrupole response functions in closed-shell nuclei},\
  }\href {https://doi.org/https://doi.org/10.1016/0375-9474(93)90170-3}
  {\bibfield  {journal} {\bibinfo  {journal} {Nuclear Physics A}\ }\textbf
  {\bibinfo {volume} {560}},\ \bibinfo {pages} {765} (\bibinfo {year}
  {1993})}\BibitemShut {NoStop}%
\bibitem [{\citenamefont {Gambacurta}\ \emph {et~al.}(2011)\citenamefont
  {Gambacurta}, \citenamefont {Grasso},\ and\ \citenamefont
  {Catara}}]{Gambacurta2011}%
  \BibitemOpen
  \bibfield  {author} {\bibinfo {author} {\bibfnamefont {D.}~\bibnamefont
  {Gambacurta}}, \bibinfo {author} {\bibfnamefont {M.}~\bibnamefont {Grasso}},\
  and\ \bibinfo {author} {\bibfnamefont {F.}~\bibnamefont {Catara}},\
  }\bibfield  {title} {\bibinfo {title} {Residual interaction in second
  random-phase approximation with density-dependent forces: rearrangement
  terms},\ }\href {https://doi.org/10.1088/0954-3899/38/3/035103} {\bibfield
  {journal} {\bibinfo  {journal} {Journal of Physics G: Nuclear and Particle
  Physics}\ }\textbf {\bibinfo {volume} {38}},\ \bibinfo {pages} {035103}
  (\bibinfo {year} {2011})}\BibitemShut {NoStop}%
\bibitem [{\citenamefont {Papakonstantinou}\ and\ \citenamefont
  {Roth}(2009)}]{Papa2009}%
  \BibitemOpen
  \bibfield  {author} {\bibinfo {author} {\bibfnamefont {P.}~\bibnamefont
  {Papakonstantinou}}\ and\ \bibinfo {author} {\bibfnamefont {R.}~\bibnamefont
  {Roth}},\ }\bibfield  {title} {\bibinfo {title} {Second random phase
  approximation and renormalized realistic interactions},\ }\href
  {https://doi.org/https://doi.org/10.1016/j.physletb.2008.12.037} {\bibfield
  {journal} {\bibinfo  {journal} {Physics Letters B}\ }\textbf {\bibinfo
  {volume} {671}},\ \bibinfo {pages} {356} (\bibinfo {year}
  {2009})}\BibitemShut {NoStop}%
\bibitem [{\citenamefont {Papakonstantinou}\ and\ \citenamefont
  {Roth}(2010)}]{Papa2010}%
  \BibitemOpen
  \bibfield  {author} {\bibinfo {author} {\bibfnamefont {P.}~\bibnamefont
  {Papakonstantinou}}\ and\ \bibinfo {author} {\bibfnamefont {R.}~\bibnamefont
  {Roth}},\ }\bibfield  {title} {\bibinfo {title} {Large-scale second
  random-phase approximation calculations with finite-range interactions},\
  }\href {https://doi.org/10.1103/PhysRevC.81.024317} {\bibfield  {journal}
  {\bibinfo  {journal} {Phys. Rev. C}\ }\textbf {\bibinfo {volume} {81}},\
  \bibinfo {pages} {024317} (\bibinfo {year} {2010})}\BibitemShut {NoStop}%
\bibitem [{\citenamefont {Tselyaev}(2007)}]{Tselyaev2007}%
  \BibitemOpen
  \bibfield  {author} {\bibinfo {author} {\bibfnamefont {V.~I.}\ \bibnamefont
  {Tselyaev}},\ }\bibfield  {title} {\bibinfo {title} {Quasiparticle time
  blocking approximation within the framework of generalized green function
  formalism},\ }\href {https://doi.org/10.1103/PhysRevC.75.024306} {\bibfield
  {journal} {\bibinfo  {journal} {Phys. Rev. C}\ }\textbf {\bibinfo {volume}
  {75}},\ \bibinfo {pages} {024306} (\bibinfo {year} {2007})}\BibitemShut
  {NoStop}%
\bibitem [{\citenamefont {Gambacurta}\ \emph {et~al.}(2015)\citenamefont
  {Gambacurta}, \citenamefont {Grasso},\ and\ \citenamefont
  {Engel}}]{Gambacurta2015}%
  \BibitemOpen
  \bibfield  {author} {\bibinfo {author} {\bibfnamefont {D.}~\bibnamefont
  {Gambacurta}}, \bibinfo {author} {\bibfnamefont {M.}~\bibnamefont {Grasso}},\
  and\ \bibinfo {author} {\bibfnamefont {J.}~\bibnamefont {Engel}},\ }\bibfield
   {title} {\bibinfo {title} {Subtraction method in the second random-phase
  approximation: First applications with a skyrme energy functional},\ }\href
  {https://doi.org/10.1103/PhysRevC.92.034303} {\bibfield  {journal} {\bibinfo
  {journal} {Phys. Rev. C}\ }\textbf {\bibinfo {volume} {92}},\ \bibinfo
  {pages} {034303} (\bibinfo {year} {2015})}\BibitemShut {NoStop}%
\bibitem [{\citenamefont {Yang}\ \emph {et~al.}(2021)\citenamefont {Yang},
  \citenamefont {Bai}, \citenamefont {Sagawa},\ and\ \citenamefont
  {Zhang}}]{Yang2021}%
  \BibitemOpen
  \bibfield  {author} {\bibinfo {author} {\bibfnamefont {M.~J.}\ \bibnamefont
  {Yang}}, \bibinfo {author} {\bibfnamefont {C.~L.}\ \bibnamefont {Bai}},
  \bibinfo {author} {\bibfnamefont {H.}~\bibnamefont {Sagawa}},\ and\ \bibinfo
  {author} {\bibfnamefont {H.~Q.}\ \bibnamefont {Zhang}},\ }\bibfield  {title}
  {\bibinfo {title} {Effects of the skyrme tensor force on ${0}^{+}$,
  ${2}^{+}$, and ${3}^{\ensuremath{-}}$ states in $^{16}\mathrm{O}$ and
  $^{40}\mathrm{Ca}$ nuclei within the second random-phase approximation},\
  }\href {https://doi.org/10.1103/PhysRevC.103.054308} {\bibfield  {journal}
  {\bibinfo  {journal} {Phys. Rev. C}\ }\textbf {\bibinfo {volume} {103}},\
  \bibinfo {pages} {054308} (\bibinfo {year} {2021})}\BibitemShut {NoStop}%
\bibitem [{\citenamefont {Gambacurta}\ \emph {et~al.}(2018)\citenamefont
  {Gambacurta}, \citenamefont {Grasso},\ and\ \citenamefont
  {Vasseur}}]{Gambacurta2018}%
  \BibitemOpen
  \bibfield  {author} {\bibinfo {author} {\bibfnamefont {D.}~\bibnamefont
  {Gambacurta}}, \bibinfo {author} {\bibfnamefont {M.}~\bibnamefont {Grasso}},\
  and\ \bibinfo {author} {\bibfnamefont {O.}~\bibnamefont {Vasseur}},\
  }\bibfield  {title} {\bibinfo {title} {{Electric dipole strength and dipole
  polarizability in $^{48}$Ca within a fully self-consistent second random
  phase approximation}},\ }\href
  {https://doi.org/https://doi.org/10.1016/j.physletb.2017.12.026} {\bibfield
  {journal} {\bibinfo  {journal} {Physics Letters B}\ }\textbf {\bibinfo
  {volume} {777}},\ \bibinfo {pages} {163} (\bibinfo {year}
  {2018})}\BibitemShut {NoStop}%
\bibitem [{\citenamefont {Gambacurta}\ \emph {et~al.}(2020)\citenamefont
  {Gambacurta}, \citenamefont {Grasso},\ and\ \citenamefont
  {Engel}}]{Gambacurta2020}%
  \BibitemOpen
  \bibfield  {author} {\bibinfo {author} {\bibfnamefont {D.}~\bibnamefont
  {Gambacurta}}, \bibinfo {author} {\bibfnamefont {M.}~\bibnamefont {Grasso}},\
  and\ \bibinfo {author} {\bibfnamefont {J.}~\bibnamefont {Engel}},\ }\bibfield
   {title} {\bibinfo {title} {{Gamow-Teller Strength in $^{48}\mathrm{Ca}$ and
  $^{78}\mathrm{Ni}$ with the Charge-Exchange Subtracted Second Random-Phase
  Approximation}},\ }\href {https://doi.org/10.1103/PhysRevLett.125.212501}
  {\bibfield  {journal} {\bibinfo  {journal} {Phys. Rev. Lett.}\ }\textbf
  {\bibinfo {volume} {125}},\ \bibinfo {pages} {212501} (\bibinfo {year}
  {2020})}\BibitemShut {NoStop}%
\bibitem [{\citenamefont {Yang}\ \emph {et~al.}(2022)\citenamefont {Yang},
  \citenamefont {Bai}, \citenamefont {Sagawa},\ and\ \citenamefont
  {Zhang}}]{Yang2022}%
  \BibitemOpen
  \bibfield  {author} {\bibinfo {author} {\bibfnamefont {M.~J.}\ \bibnamefont
  {Yang}}, \bibinfo {author} {\bibfnamefont {C.~L.}\ \bibnamefont {Bai}},
  \bibinfo {author} {\bibfnamefont {H.}~\bibnamefont {Sagawa}},\ and\ \bibinfo
  {author} {\bibfnamefont {H.~Q.}\ \bibnamefont {Zhang}},\ }\bibfield  {title}
  {\bibinfo {title} {Gamow-teller transitions in magic nuclei calculated by the
  charge-exchange subtracted second random-phase approximation},\ }\href
  {https://doi.org/10.1103/PhysRevC.106.014319} {\bibfield  {journal} {\bibinfo
   {journal} {Phys. Rev. C}\ }\textbf {\bibinfo {volume} {106}},\ \bibinfo
  {pages} {014319} (\bibinfo {year} {2022})}\BibitemShut {NoStop}%
\bibitem [{\citenamefont {Gambacurta}\ and\ \citenamefont
  {Grasso}(2022)}]{Gambacurta2022}%
  \BibitemOpen
  \bibfield  {author} {\bibinfo {author} {\bibfnamefont {D.}~\bibnamefont
  {Gambacurta}}\ and\ \bibinfo {author} {\bibfnamefont {M.}~\bibnamefont
  {Grasso}},\ }\bibfield  {title} {\bibinfo {title} {{Quenching of Gamow-Teller
  strengths and two-particle--two-hole configurations}},\ }\href
  {https://doi.org/10.1103/PhysRevC.105.014321} {\bibfield  {journal} {\bibinfo
   {journal} {Phys. Rev. C}\ }\textbf {\bibinfo {volume} {105}},\ \bibinfo
  {pages} {014321} (\bibinfo {year} {2022})}\BibitemShut {NoStop}%
\bibitem [{\citenamefont {Yang}\ \emph {et~al.}(2023)\citenamefont {Yang},
  \citenamefont {Sagawa}, \citenamefont {Bai},\ and\ \citenamefont
  {Zhang}}]{Yang2023}%
  \BibitemOpen
  \bibfield  {author} {\bibinfo {author} {\bibfnamefont {M.~J.}\ \bibnamefont
  {Yang}}, \bibinfo {author} {\bibfnamefont {H.}~\bibnamefont {Sagawa}},
  \bibinfo {author} {\bibfnamefont {C.~L.}\ \bibnamefont {Bai}},\ and\ \bibinfo
  {author} {\bibfnamefont {H.~Q.}\ \bibnamefont {Zhang}},\ }\bibfield  {title}
  {\bibinfo {title} {Effects of two-particle--two-hole configurations and
  tensor force on $\ensuremath{\beta}$ decay of magic nuclei},\ }\href
  {https://doi.org/10.1103/PhysRevC.107.014325} {\bibfield  {journal} {\bibinfo
   {journal} {Phys. Rev. C}\ }\textbf {\bibinfo {volume} {107}},\ \bibinfo
  {pages} {014325} (\bibinfo {year} {2023})}\BibitemShut {NoStop}%
\bibitem [{\citenamefont {Col\`o}\ \emph {et~al.}(2010)\citenamefont {Col\`o},
  \citenamefont {Sagawa},\ and\ \citenamefont {Bortignon}}]{Colo2010}%
  \BibitemOpen
  \bibfield  {author} {\bibinfo {author} {\bibfnamefont {G.}~\bibnamefont
  {Col\`o}}, \bibinfo {author} {\bibfnamefont {H.}~\bibnamefont {Sagawa}},\
  and\ \bibinfo {author} {\bibfnamefont {P.~F.}\ \bibnamefont {Bortignon}},\
  }\bibfield  {title} {\bibinfo {title} {Effect of particle-vibration coupling
  on single-particle states: A consistent study within the skyrme framework},\
  }\href {https://doi.org/10.1103/PhysRevC.82.064307} {\bibfield  {journal}
  {\bibinfo  {journal} {Phys. Rev. C}\ }\textbf {\bibinfo {volume} {82}},\
  \bibinfo {pages} {064307} (\bibinfo {year} {2010})}\BibitemShut {NoStop}%
\bibitem [{\citenamefont {Niu}\ \emph {et~al.}(2012)\citenamefont {Niu},
  \citenamefont {Col\`o}, \citenamefont {Brenna}, \citenamefont {Bortignon},\
  and\ \citenamefont {Meng}}]{Niu2012}%
  \BibitemOpen
  \bibfield  {author} {\bibinfo {author} {\bibfnamefont {Y.~F.}\ \bibnamefont
  {Niu}}, \bibinfo {author} {\bibfnamefont {G.}~\bibnamefont {Col\`o}},
  \bibinfo {author} {\bibfnamefont {M.}~\bibnamefont {Brenna}}, \bibinfo
  {author} {\bibfnamefont {P.~F.}\ \bibnamefont {Bortignon}},\ and\ \bibinfo
  {author} {\bibfnamefont {J.}~\bibnamefont {Meng}},\ }\bibfield  {title}
  {\bibinfo {title} {Gamow-teller response within skyrme random-phase
  approximation plus particle-vibration coupling},\ }\href
  {https://doi.org/10.1103/PhysRevC.85.034314} {\bibfield  {journal} {\bibinfo
  {journal} {Phys. Rev. C}\ }\textbf {\bibinfo {volume} {85}},\ \bibinfo
  {pages} {034314} (\bibinfo {year} {2012})}\BibitemShut {NoStop}%
\bibitem [{\citenamefont {Litvinova}\ \emph {et~al.}(2008)\citenamefont
  {Litvinova}, \citenamefont {Ring},\ and\ \citenamefont
  {Tselyaev}}]{Litvinova2008}%
  \BibitemOpen
  \bibfield  {author} {\bibinfo {author} {\bibfnamefont {E.}~\bibnamefont
  {Litvinova}}, \bibinfo {author} {\bibfnamefont {P.}~\bibnamefont {Ring}},\
  and\ \bibinfo {author} {\bibfnamefont {V.}~\bibnamefont {Tselyaev}},\
  }\bibfield  {title} {\bibinfo {title} {Relativistic quasiparticle time
  blocking approximation: Dipole response of open-shell nuclei},\ }\href
  {https://doi.org/10.1103/PhysRevC.78.014312} {\bibfield  {journal} {\bibinfo
  {journal} {Phys. Rev. C}\ }\textbf {\bibinfo {volume} {78}},\ \bibinfo
  {pages} {014312} (\bibinfo {year} {2008})}\BibitemShut {NoStop}%
\bibitem [{\citenamefont {Catara}\ \emph {et~al.}(1992)\citenamefont {Catara},
  \citenamefont {Chomaz},\ and\ \citenamefont {{Van Giai}}}]{Catara1992}%
  \BibitemOpen
  \bibfield  {author} {\bibinfo {author} {\bibfnamefont {F.}~\bibnamefont
  {Catara}}, \bibinfo {author} {\bibfnamefont {P.}~\bibnamefont {Chomaz}},\
  and\ \bibinfo {author} {\bibfnamefont {N.}~\bibnamefont {{Van Giai}}},\
  }\bibfield  {title} {\bibinfo {title} {Electromagnetic decay of two-phonon
  states},\ }\href
  {https://doi.org/https://doi.org/10.1016/0370-2693(92)90946-2} {\bibfield
  {journal} {\bibinfo  {journal} {Physics Letters B}\ }\textbf {\bibinfo
  {volume} {277}},\ \bibinfo {pages} {1} (\bibinfo {year} {1992})}\BibitemShut
  {NoStop}%
\bibitem [{\citenamefont {Severyukhin}\ and\ \citenamefont
  {Van~Giai}(2004)}]{Severyukhin2004}%
  \BibitemOpen
  \bibfield  {author} {\bibinfo {author} {\bibfnamefont {V.~V.}\ \bibnamefont
  {Severyukhin}, \bibfnamefont {A.~P. ad~Voronov}}\ and\ \bibinfo {author}
  {\bibfnamefont {N.}~\bibnamefont {Van~Giai}},\ }\bibfield  {title} {\bibinfo
  {title} {Effects of phonon-phonon coupling on low-lying states in
  neutron-rich sn isotopes},\ }\href
  {https://doi.org/10.1140/epja/i2004-10048-2} {\bibfield  {journal} {\bibinfo
  {journal} {The European Physical Journal A - Hadrons and Nuclei}\ }\textbf
  {\bibinfo {volume} {22}},\ \bibinfo {pages} {397} (\bibinfo {year}
  {2004})}\BibitemShut {NoStop}%
\bibitem [{\citenamefont {Litvinova}\ \emph {et~al.}(2013)\citenamefont
  {Litvinova}, \citenamefont {Ring},\ and\ \citenamefont
  {Tselyaev}}]{Litvinova2013}%
  \BibitemOpen
  \bibfield  {author} {\bibinfo {author} {\bibfnamefont {E.}~\bibnamefont
  {Litvinova}}, \bibinfo {author} {\bibfnamefont {P.}~\bibnamefont {Ring}},\
  and\ \bibinfo {author} {\bibfnamefont {V.}~\bibnamefont {Tselyaev}},\
  }\bibfield  {title} {\bibinfo {title} {Relativistic two-phonon model for the
  low-energy nuclear response},\ }\href
  {https://doi.org/10.1103/PhysRevC.88.044320} {\bibfield  {journal} {\bibinfo
  {journal} {Phys. Rev. C}\ }\textbf {\bibinfo {volume} {88}},\ \bibinfo
  {pages} {044320} (\bibinfo {year} {2013})}\BibitemShut {NoStop}%
\bibitem [{\citenamefont {Gambacurta}\ \emph {et~al.}(2016)\citenamefont
  {Gambacurta}, \citenamefont {Catara}, \citenamefont {Grasso}, \citenamefont
  {Sambataro}, \citenamefont {Andr\'es},\ and\ \citenamefont
  {Lanza}}]{Gambacurta2016}%
  \BibitemOpen
  \bibfield  {author} {\bibinfo {author} {\bibfnamefont {D.}~\bibnamefont
  {Gambacurta}}, \bibinfo {author} {\bibfnamefont {F.}~\bibnamefont {Catara}},
  \bibinfo {author} {\bibfnamefont {M.}~\bibnamefont {Grasso}}, \bibinfo
  {author} {\bibfnamefont {M.}~\bibnamefont {Sambataro}}, \bibinfo {author}
  {\bibfnamefont {M.~V.}\ \bibnamefont {Andr\'es}},\ and\ \bibinfo {author}
  {\bibfnamefont {E.~G.}\ \bibnamefont {Lanza}},\ }\bibfield  {title} {\bibinfo
  {title} {Nuclear excitations as coupled one and two
  random-phase-approximation modes},\ }\href
  {https://doi.org/10.1103/PhysRevC.93.024309} {\bibfield  {journal} {\bibinfo
  {journal} {Phys. Rev. C}\ }\textbf {\bibinfo {volume} {93}},\ \bibinfo
  {pages} {024309} (\bibinfo {year} {2016})}\BibitemShut {NoStop}%
\bibitem [{\citenamefont {Arsenyev}\ and\ \citenamefont
  {Severyukhin}(2023)}]{Arsenyev2023}%
  \BibitemOpen
  \bibfield  {author} {\bibinfo {author} {\bibfnamefont {N.~N.}\ \bibnamefont
  {Arsenyev}}\ and\ \bibinfo {author} {\bibfnamefont {A.~P.}\ \bibnamefont
  {Severyukhin}},\ }\bibfield  {title} {\bibinfo {title} {{Isoscalar giant
  monopole resonance in $^{40,48}$Ca}},\ }\href
  {https://doi.org/10.1088/1742-6596/2586/1/012047} {\bibfield  {journal}
  {\bibinfo  {journal} {Journal of Physics: Conference Series}\ }\textbf
  {\bibinfo {volume} {2586}},\ \bibinfo {pages} {012047} (\bibinfo {year}
  {2023})}\BibitemShut {NoStop}%
\bibitem [{\citenamefont {Tohyama}(2001)}]{Tohyama2001}%
  \BibitemOpen
  \bibfield  {author} {\bibinfo {author} {\bibfnamefont {M.}~\bibnamefont
  {Tohyama}},\ }\bibfield  {title} {\bibinfo {title} {Collectivity of double
  giant resonances in extended rpa theories},\ }\href
  {https://doi.org/10.1103/PhysRevC.64.067304} {\bibfield  {journal} {\bibinfo
  {journal} {Phys. Rev. C}\ }\textbf {\bibinfo {volume} {64}},\ \bibinfo
  {pages} {067304} (\bibinfo {year} {2001})}\BibitemShut {NoStop}%
\bibitem [{\citenamefont {Emling}(1994)}]{Emling1994}%
  \BibitemOpen
  \bibfield  {author} {\bibinfo {author} {\bibfnamefont {H.}~\bibnamefont
  {Emling}},\ }\bibfield  {title} {\bibinfo {title} {Electromagnetic excitation
  of the two-phonon giant dipole resonance},\ }\href
  {https://doi.org/https://doi.org/10.1016/0146-6410(94)90052-3} {\bibfield
  {journal} {\bibinfo  {journal} {Progress in Particle and Nuclear Physics}\
  }\textbf {\bibinfo {volume} {33}},\ \bibinfo {pages} {729} (\bibinfo {year}
  {1994})}\BibitemShut {NoStop}%
\bibitem [{\citenamefont {Chomaz}\ and\ \citenamefont
  {Frascaria}(1995)}]{Chomaz1995}%
  \BibitemOpen
  \bibfield  {author} {\bibinfo {author} {\bibfnamefont {P.}~\bibnamefont
  {Chomaz}}\ and\ \bibinfo {author} {\bibfnamefont {N.}~\bibnamefont
  {Frascaria}},\ }\bibfield  {title} {\bibinfo {title} {Multiple phonon
  excitation in nuclei: experimental results and theoretical descriptions},\
  }\href {https://doi.org/https://doi.org/10.1016/0370-1573(94)00079-I}
  {\bibfield  {journal} {\bibinfo  {journal} {Physics Reports}\ }\textbf
  {\bibinfo {volume} {252}},\ \bibinfo {pages} {275} (\bibinfo {year}
  {1995})}\BibitemShut {NoStop}%
\bibitem [{\citenamefont {Dowell}\ \emph {et~al.}(1983)\citenamefont {Dowell},
  \citenamefont {Feldman}, \citenamefont {Snover}, \citenamefont {Sandorfi},\
  and\ \citenamefont {Collins}}]{Dowell1983}%
  \BibitemOpen
  \bibfield  {author} {\bibinfo {author} {\bibfnamefont {D.~H.}\ \bibnamefont
  {Dowell}}, \bibinfo {author} {\bibfnamefont {G.}~\bibnamefont {Feldman}},
  \bibinfo {author} {\bibfnamefont {K.~A.}\ \bibnamefont {Snover}}, \bibinfo
  {author} {\bibfnamefont {A.~M.}\ \bibnamefont {Sandorfi}},\ and\ \bibinfo
  {author} {\bibfnamefont {M.~T.}\ \bibnamefont {Collins}},\ }\bibfield
  {title} {\bibinfo {title} {Excited-state giant dipole resonances in ($p,
  \ensuremath{\gamma}$): A new probe of single-particle strengths},\ }\href
  {https://doi.org/10.1103/PhysRevLett.50.1191} {\bibfield  {journal} {\bibinfo
   {journal} {Phys. Rev. Lett.}\ }\textbf {\bibinfo {volume} {50}},\ \bibinfo
  {pages} {1191} (\bibinfo {year} {1983})}\BibitemShut {NoStop}%
\bibitem [{\citenamefont {Bertulani}\ and\ \citenamefont
  {Baur}(1988)}]{Bertulani1988}%
  \BibitemOpen
  \bibfield  {author} {\bibinfo {author} {\bibfnamefont {C.~A.}\ \bibnamefont
  {Bertulani}}\ and\ \bibinfo {author} {\bibfnamefont {G.}~\bibnamefont
  {Baur}},\ }\bibfield  {title} {\bibinfo {title} {Electromagnetic processes in
  relativistic heavy ion collisions},\ }\href
  {https://doi.org/https://doi.org/10.1016/0370-1573(88)90142-1} {\bibfield
  {journal} {\bibinfo  {journal} {Physics Reports}\ }\textbf {\bibinfo {volume}
  {163}},\ \bibinfo {pages} {299} (\bibinfo {year} {1988})}\BibitemShut
  {NoStop}%
\bibitem [{\citenamefont {Bertulani}\ and\ \citenamefont
  {Ponomarev}(1999)}]{Bertulani1999}%
  \BibitemOpen
  \bibfield  {author} {\bibinfo {author} {\bibfnamefont {C.}~\bibnamefont
  {Bertulani}}\ and\ \bibinfo {author} {\bibfnamefont {V.}~\bibnamefont
  {Ponomarev}},\ }\bibfield  {title} {\bibinfo {title} {Microscopic studies on
  two-phonon giant resonances},\ }\href
  {https://doi.org/https://doi.org/10.1016/S0370-1573(99)00038-1} {\bibfield
  {journal} {\bibinfo  {journal} {Physics Reports}\ }\textbf {\bibinfo {volume}
  {321}},\ \bibinfo {pages} {139} (\bibinfo {year} {1999})}\BibitemShut
  {NoStop}%
\bibitem [{\citenamefont {Esbensen}(2005)}]{Esbensen2005}%
  \BibitemOpen
  \bibfield  {author} {\bibinfo {author} {\bibfnamefont {H.}~\bibnamefont
  {Esbensen}},\ }\bibfield  {title} {\bibinfo {title} {Sensitivity to
  multi-phonon excitations in heavy-ion fusion reactions},\ }\href
  {https://doi.org/10.1103/PhysRevC.72.054607} {\bibfield  {journal} {\bibinfo
  {journal} {Phys. Rev. C}\ }\textbf {\bibinfo {volume} {72}},\ \bibinfo
  {pages} {054607} (\bibinfo {year} {2005})}\BibitemShut {NoStop}%
\bibitem [{\citenamefont {Auerbach}(1990)}]{Auerbach1990}%
  \BibitemOpen
  \bibfield  {author} {\bibinfo {author} {\bibfnamefont {N.}~\bibnamefont
  {Auerbach}},\ }\bibfield  {title} {\bibinfo {title} {Isotensor double giant
  resonances in double charge-exchange reactions},\ }\href
  {https://doi.org/https://doi.org/10.1016/0003-4916(90)90216-B} {\bibfield
  {journal} {\bibinfo  {journal} {Annals of Physics}\ }\textbf {\bibinfo
  {volume} {197}},\ \bibinfo {pages} {376} (\bibinfo {year}
  {1990})}\BibitemShut {NoStop}%
\bibitem [{\citenamefont {Muto}(1992)}]{Muto1992}%
  \BibitemOpen
  \bibfield  {author} {\bibinfo {author} {\bibfnamefont {K.}~\bibnamefont
  {Muto}},\ }\bibfield  {title} {\bibinfo {title} {Sum rules for double gamow -
  teller excitation},\ }\href
  {https://doi.org/https://doi.org/10.1016/0370-2693(92)90948-4} {\bibfield
  {journal} {\bibinfo  {journal} {Physics Letters B}\ }\textbf {\bibinfo
  {volume} {277}},\ \bibinfo {pages} {13} (\bibinfo {year} {1992})}\BibitemShut
  {NoStop}%
\bibitem [{\citenamefont {Sagawa}\ and\ \citenamefont
  {Uesaka}(2016)}]{Sagawa2016}%
  \BibitemOpen
  \bibfield  {author} {\bibinfo {author} {\bibfnamefont {H.}~\bibnamefont
  {Sagawa}}\ and\ \bibinfo {author} {\bibfnamefont {T.}~\bibnamefont
  {Uesaka}},\ }\bibfield  {title} {\bibinfo {title} {Sum rule study for double
  gamow-teller states},\ }\href {https://doi.org/10.1103/PhysRevC.94.064325}
  {\bibfield  {journal} {\bibinfo  {journal} {Phys. Rev. C}\ }\textbf {\bibinfo
  {volume} {94}},\ \bibinfo {pages} {064325} (\bibinfo {year}
  {2016})}\BibitemShut {NoStop}%
\bibitem [{\citenamefont {Sakaue}\ \emph {et~al.}(2024)\citenamefont {Sakaue},
  \citenamefont {Yako}, \citenamefont {Ota}, \citenamefont {Baba},
  \citenamefont {Chillery}, \citenamefont {Doornenbal}, \citenamefont {Dozono},
  \citenamefont {Ebina}, \citenamefont {Fukuda}, \citenamefont {Fukunishi},
  \citenamefont {Furuno}, \citenamefont {Hanai}, \citenamefont {Harada},
  \citenamefont {Hayakawa}, \citenamefont {Hijikata}, \citenamefont {Horikawa},
  \citenamefont {Huang}, \citenamefont {Imai}, \citenamefont {Itahashi},
  \citenamefont {Kobayashi}, \citenamefont {Kondo}, \citenamefont {Li},
  \citenamefont {Maeda}, \citenamefont {Matsui}, \citenamefont {Matsumoto},
  \citenamefont {Matsumura}, \citenamefont {Michimasa}, \citenamefont
  {Nakatsuka}, \citenamefont {Nishi}, \citenamefont {Sakanashi}, \citenamefont
  {Sasano}, \citenamefont {Sekiya}, \citenamefont {Shimizu}, \citenamefont
  {Shimizu}, \citenamefont {Shimoura}, \citenamefont {Sumikama}, \citenamefont
  {Suzuki}, \citenamefont {Suzuki}, \citenamefont {Takaki}, \citenamefont
  {Takeshige}, \citenamefont {Takeda}, \citenamefont {Tanaka}, \citenamefont
  {Tanaka}, \citenamefont {Togano}, \citenamefont {Tsuji}, \citenamefont
  {Yang}, \citenamefont {Yoshida}, \citenamefont {Yoshimoto}, \citenamefont
  {Zenihiro},\ and\ \citenamefont {Uesaka}}]{Sakaue2024}%
  \BibitemOpen
  \bibfield  {author} {\bibinfo {author} {\bibfnamefont {A.}~\bibnamefont
  {Sakaue}}, \bibinfo {author} {\bibfnamefont {K.}~\bibnamefont {Yako}},
  \bibinfo {author} {\bibfnamefont {S.}~\bibnamefont {Ota}}, \bibinfo {author}
  {\bibfnamefont {H.}~\bibnamefont {Baba}}, \bibinfo {author} {\bibfnamefont
  {T.}~\bibnamefont {Chillery}}, \bibinfo {author} {\bibfnamefont
  {P.}~\bibnamefont {Doornenbal}}, \bibinfo {author} {\bibfnamefont
  {M.}~\bibnamefont {Dozono}}, \bibinfo {author} {\bibfnamefont
  {N.}~\bibnamefont {Ebina}}, \bibinfo {author} {\bibfnamefont
  {N.}~\bibnamefont {Fukuda}}, \bibinfo {author} {\bibfnamefont
  {N.}~\bibnamefont {Fukunishi}}, \bibinfo {author} {\bibfnamefont
  {T.}~\bibnamefont {Furuno}}, \bibinfo {author} {\bibfnamefont
  {S.}~\bibnamefont {Hanai}}, \bibinfo {author} {\bibfnamefont
  {T.}~\bibnamefont {Harada}}, \bibinfo {author} {\bibfnamefont
  {S.}~\bibnamefont {Hayakawa}}, \bibinfo {author} {\bibfnamefont
  {Y.}~\bibnamefont {Hijikata}}, \bibinfo {author} {\bibfnamefont
  {K.}~\bibnamefont {Horikawa}}, \bibinfo {author} {\bibfnamefont {S.~W.}\
  \bibnamefont {Huang}}, \bibinfo {author} {\bibfnamefont {N.}~\bibnamefont
  {Imai}}, \bibinfo {author} {\bibfnamefont {K.}~\bibnamefont {Itahashi}},
  \bibinfo {author} {\bibfnamefont {N.}~\bibnamefont {Kobayashi}}, \bibinfo
  {author} {\bibfnamefont {Y.}~\bibnamefont {Kondo}}, \bibinfo {author}
  {\bibfnamefont {J.}~\bibnamefont {Li}}, \bibinfo {author} {\bibfnamefont
  {Y.}~\bibnamefont {Maeda}}, \bibinfo {author} {\bibfnamefont
  {T.}~\bibnamefont {Matsui}}, \bibinfo {author} {\bibfnamefont {S.~Y.}\
  \bibnamefont {Matsumoto}}, \bibinfo {author} {\bibfnamefont {R.}~\bibnamefont
  {Matsumura}}, \bibinfo {author} {\bibfnamefont {S.}~\bibnamefont
  {Michimasa}}, \bibinfo {author} {\bibfnamefont {N.}~\bibnamefont
  {Nakatsuka}}, \bibinfo {author} {\bibfnamefont {T.}~\bibnamefont {Nishi}},
  \bibinfo {author} {\bibfnamefont {K.}~\bibnamefont {Sakanashi}}, \bibinfo
  {author} {\bibfnamefont {M.}~\bibnamefont {Sasano}}, \bibinfo {author}
  {\bibfnamefont {R.}~\bibnamefont {Sekiya}}, \bibinfo {author} {\bibfnamefont
  {N.}~\bibnamefont {Shimizu}}, \bibinfo {author} {\bibfnamefont
  {Y.}~\bibnamefont {Shimizu}}, \bibinfo {author} {\bibfnamefont
  {S.}~\bibnamefont {Shimoura}}, \bibinfo {author} {\bibfnamefont
  {T.}~\bibnamefont {Sumikama}}, \bibinfo {author} {\bibfnamefont
  {D.}~\bibnamefont {Suzuki}}, \bibinfo {author} {\bibfnamefont
  {H.}~\bibnamefont {Suzuki}}, \bibinfo {author} {\bibfnamefont
  {M.}~\bibnamefont {Takaki}}, \bibinfo {author} {\bibfnamefont
  {S.}~\bibnamefont {Takeshige}}, \bibinfo {author} {\bibfnamefont
  {H.}~\bibnamefont {Takeda}}, \bibinfo {author} {\bibfnamefont
  {J.}~\bibnamefont {Tanaka}}, \bibinfo {author} {\bibfnamefont {Y.~K.}\
  \bibnamefont {Tanaka}}, \bibinfo {author} {\bibfnamefont {Y.}~\bibnamefont
  {Togano}}, \bibinfo {author} {\bibfnamefont {R.}~\bibnamefont {Tsuji}},
  \bibinfo {author} {\bibfnamefont {Z.~H.}\ \bibnamefont {Yang}}, \bibinfo
  {author} {\bibfnamefont {K.}~\bibnamefont {Yoshida}}, \bibinfo {author}
  {\bibfnamefont {M.}~\bibnamefont {Yoshimoto}}, \bibinfo {author}
  {\bibfnamefont {J.}~\bibnamefont {Zenihiro}},\ and\ \bibinfo {author}
  {\bibfnamefont {T.}~\bibnamefont {Uesaka}},\ }\bibfield  {title} {\bibinfo
  {title} {Candidate for the double gamow–teller giant resonance in $^{48}$ca
  studied by the ($^{12}$c, $^{12}$be(0$^{+}_{2}$)) reaction at 250
  mev/nucleon},\ }\href {https://doi.org/10.1093/ptep/ptae174} {\bibfield
  {journal} {\bibinfo  {journal} {Progress of Theoretical and Experimental
  Physics}\ }\textbf {\bibinfo {volume} {2024}},\ \bibinfo {pages} {123D03}
  (\bibinfo {year} {2024})},\ \Eprint
  {https://arxiv.org/abs/https://academic.oup.com/ptep/article-pdf/2024/12/123D03/61234893/ptae174.pdf}
  {https://academic.oup.com/ptep/article-pdf/2024/12/123D03/61234893/ptae174.pdf}
  \BibitemShut {NoStop}%
\bibitem [{\citenamefont {Lanza}\ \emph {et~al.}(1997)\citenamefont {Lanza},
  \citenamefont {Andrés}, \citenamefont {Catara}, \citenamefont {Chomaz},\
  and\ \citenamefont {Volpe}}]{Lanza1997}%
  \BibitemOpen
  \bibfield  {author} {\bibinfo {author} {\bibfnamefont {E.}~\bibnamefont
  {Lanza}}, \bibinfo {author} {\bibfnamefont {M.}~\bibnamefont {Andrés}},
  \bibinfo {author} {\bibfnamefont {F.}~\bibnamefont {Catara}}, \bibinfo
  {author} {\bibfnamefont {P.}~\bibnamefont {Chomaz}},\ and\ \bibinfo {author}
  {\bibfnamefont {C.}~\bibnamefont {Volpe}},\ }\bibfield  {title} {\bibinfo
  {title} {Role of anharmonicities and nonlinearities in heavy ion collisions a
  microscopic approach},\ }\href
  {https://doi.org/https://doi.org/10.1016/S0375-9474(96)00444-7} {\bibfield
  {journal} {\bibinfo  {journal} {Nuclear Physics A}\ }\textbf {\bibinfo
  {volume} {613}},\ \bibinfo {pages} {445} (\bibinfo {year}
  {1997})}\BibitemShut {NoStop}%
\bibitem [{\citenamefont {Andreozzi}\ \emph {et~al.}(2007)\citenamefont
  {Andreozzi}, \citenamefont {Knapp}, \citenamefont {Iudice}, \citenamefont
  {Porrino},\ and\ \citenamefont {Kvasil}}]{Andreozzi2007}%
  \BibitemOpen
  \bibfield  {author} {\bibinfo {author} {\bibfnamefont {F.}~\bibnamefont
  {Andreozzi}}, \bibinfo {author} {\bibfnamefont {F.}~\bibnamefont {Knapp}},
  \bibinfo {author} {\bibfnamefont {N.~L.}\ \bibnamefont {Iudice}}, \bibinfo
  {author} {\bibfnamefont {A.}~\bibnamefont {Porrino}},\ and\ \bibinfo {author}
  {\bibfnamefont {J.}~\bibnamefont {Kvasil}},\ }\bibfield  {title} {\bibinfo
  {title} {Exact formulation and solution of the nuclear eigenvalue problem in
  a microscopic multiphonon space},\ }\href
  {https://doi.org/10.1103/PhysRevC.75.044312} {\bibfield  {journal} {\bibinfo
  {journal} {Phys. Rev. C}\ }\textbf {\bibinfo {volume} {75}},\ \bibinfo
  {pages} {044312} (\bibinfo {year} {2007})}\BibitemShut {NoStop}%
\bibitem [{\citenamefont {Ponomarev}\ and\ \citenamefont
  {Voronov}(1992)}]{Ponomarev1992}%
  \BibitemOpen
  \bibfield  {author} {\bibinfo {author} {\bibfnamefont {V.}~\bibnamefont
  {Ponomarev}}\ and\ \bibinfo {author} {\bibfnamefont {V.}~\bibnamefont
  {Voronov}},\ }\bibfield  {title} {\bibinfo {title} {On double resonances in
  spherical nuclei},\ }\href
  {https://doi.org/https://doi.org/10.1016/0370-2693(92)91829-X} {\bibfield
  {journal} {\bibinfo  {journal} {Physics Letters B}\ }\textbf {\bibinfo
  {volume} {279}},\ \bibinfo {pages} {1} (\bibinfo {year} {1992})}\BibitemShut
  {NoStop}%
\bibitem [{\citenamefont {Ponomarev}\ \emph {et~al.}(1994)\citenamefont
  {Ponomarev}, \citenamefont {Vigezzi}, \citenamefont {Bortignon},
  \citenamefont {Broglia}, \citenamefont {Col\`o}, \citenamefont {Lazzari},
  \citenamefont {Voronov},\ and\ \citenamefont {Baur}}]{Ponomarev1994}%
  \BibitemOpen
  \bibfield  {author} {\bibinfo {author} {\bibfnamefont {V.~Y.}\ \bibnamefont
  {Ponomarev}}, \bibinfo {author} {\bibfnamefont {E.}~\bibnamefont {Vigezzi}},
  \bibinfo {author} {\bibfnamefont {P.~F.}\ \bibnamefont {Bortignon}}, \bibinfo
  {author} {\bibfnamefont {R.~A.}\ \bibnamefont {Broglia}}, \bibinfo {author}
  {\bibfnamefont {G.}~\bibnamefont {Col\`o}}, \bibinfo {author} {\bibfnamefont
  {G.}~\bibnamefont {Lazzari}}, \bibinfo {author} {\bibfnamefont {V.~V.}\
  \bibnamefont {Voronov}},\ and\ \bibinfo {author} {\bibfnamefont
  {G.}~\bibnamefont {Baur}},\ }\bibfield  {title} {\bibinfo {title} {Multiple
  excitation of giant dipole resonances in relativistic heavy ion collisions},\
  }\href {https://doi.org/10.1103/PhysRevLett.72.1168} {\bibfield  {journal}
  {\bibinfo  {journal} {Phys. Rev. Lett.}\ }\textbf {\bibinfo {volume} {72}},\
  \bibinfo {pages} {1168} (\bibinfo {year} {1994})}\BibitemShut {NoStop}%
\bibitem [{\citenamefont {Bertulani}\ and\ \citenamefont
  {Zelevinsky}(1994)}]{Bertulani1994}%
  \BibitemOpen
  \bibfield  {author} {\bibinfo {author} {\bibfnamefont {C.}~\bibnamefont
  {Bertulani}}\ and\ \bibinfo {author} {\bibfnamefont {V.}~\bibnamefont
  {Zelevinsky}},\ }\bibfield  {title} {\bibinfo {title} {Excitation of
  multiphonon giant resonance states in relativistic heavy-ion collisions},\
  }\href {https://doi.org/https://doi.org/10.1016/0375-9474(94)90368-9}
  {\bibfield  {journal} {\bibinfo  {journal} {Nuclear Physics A}\ }\textbf
  {\bibinfo {volume} {568}},\ \bibinfo {pages} {931} (\bibinfo {year}
  {1994})}\BibitemShut {NoStop}%
\bibitem [{\citenamefont {Dinh~Dang}\ \emph {et~al.}(1999)\citenamefont
  {Dinh~Dang}, \citenamefont {Tanabe},\ and\ \citenamefont {Arima}}]{Dang1999}%
  \BibitemOpen
  \bibfield  {author} {\bibinfo {author} {\bibfnamefont {N.}~\bibnamefont
  {Dinh~Dang}}, \bibinfo {author} {\bibfnamefont {K.}~\bibnamefont {Tanabe}},\
  and\ \bibinfo {author} {\bibfnamefont {A.}~\bibnamefont {Arima}},\ }\bibfield
   {title} {\bibinfo {title} {Damping of the double giant dipole resonance},\
  }\href {https://doi.org/10.1103/PhysRevC.59.3128} {\bibfield  {journal}
  {\bibinfo  {journal} {Phys. Rev. C}\ }\textbf {\bibinfo {volume} {59}},\
  \bibinfo {pages} {3128} (\bibinfo {year} {1999})}\BibitemShut {NoStop}%
\bibitem [{\citenamefont {Nishizaki}\ and\ \citenamefont
  {Wambach}(1995)}]{Nishizaki1995}%
  \BibitemOpen
  \bibfield  {author} {\bibinfo {author} {\bibfnamefont {S.}~\bibnamefont
  {Nishizaki}}\ and\ \bibinfo {author} {\bibfnamefont {J.}~\bibnamefont
  {Wambach}},\ }\bibfield  {title} {\bibinfo {title} {Double-dipole excitations
  in $^{40}$ca},\ }\href
  {https://doi.org/https://doi.org/10.1016/0370-2693(95)00229-E} {\bibfield
  {journal} {\bibinfo  {journal} {Physics Letters B}\ }\textbf {\bibinfo
  {volume} {349}},\ \bibinfo {pages} {7} (\bibinfo {year} {1995})}\BibitemShut
  {NoStop}%
\bibitem [{\citenamefont {Nishizaki}\ and\ \citenamefont
  {Wambach}(1998)}]{Nishizaki1998}%
  \BibitemOpen
  \bibfield  {author} {\bibinfo {author} {\bibfnamefont {S.}~\bibnamefont
  {Nishizaki}}\ and\ \bibinfo {author} {\bibfnamefont {J.}~\bibnamefont
  {Wambach}},\ }\bibfield  {title} {\bibinfo {title} {Double-giant-dipole
  resonance in ${}^{208}\mathrm{Pb}$},\ }\href
  {https://doi.org/10.1103/PhysRevC.57.1515} {\bibfield  {journal} {\bibinfo
  {journal} {Phys. Rev. C}\ }\textbf {\bibinfo {volume} {57}},\ \bibinfo
  {pages} {1515} (\bibinfo {year} {1998})}\BibitemShut {NoStop}%
\bibitem [{\citenamefont {Yeh}\ \emph {et~al.}(1996)\citenamefont {Yeh},
  \citenamefont {Garrett}, \citenamefont {McGrath}, \citenamefont {Yates},\
  and\ \citenamefont {Belgya}}]{Yeh1996}%
  \BibitemOpen
  \bibfield  {author} {\bibinfo {author} {\bibfnamefont {M.}~\bibnamefont
  {Yeh}}, \bibinfo {author} {\bibfnamefont {P.~E.}\ \bibnamefont {Garrett}},
  \bibinfo {author} {\bibfnamefont {C.~A.}\ \bibnamefont {McGrath}}, \bibinfo
  {author} {\bibfnamefont {S.~W.}\ \bibnamefont {Yates}},\ and\ \bibinfo
  {author} {\bibfnamefont {T.}~\bibnamefont {Belgya}},\ }\bibfield  {title}
  {\bibinfo {title} {Two-phonon octupole excitation in ${}^{208}$pb},\ }\href
  {https://doi.org/10.1103/PhysRevLett.76.1208} {\bibfield  {journal} {\bibinfo
   {journal} {Phys. Rev. Lett.}\ }\textbf {\bibinfo {volume} {76}},\ \bibinfo
  {pages} {1208} (\bibinfo {year} {1996})}\BibitemShut {NoStop}%
\bibitem [{\citenamefont {Bargioni}\ \emph {et~al.}(1995)\citenamefont
  {Bargioni}, \citenamefont {Bizzeti}, \citenamefont {Bizzeti-Sona},
  \citenamefont {Bazzacco}, \citenamefont {Lunardi}, \citenamefont {Pavan},
  \citenamefont {Rossi-Alvarez}, \citenamefont {de~Angelis}, \citenamefont
  {Maron},\ and\ \citenamefont {Rico}}]{Bargioni1995}%
  \BibitemOpen
  \bibfield  {author} {\bibinfo {author} {\bibfnamefont {L.}~\bibnamefont
  {Bargioni}}, \bibinfo {author} {\bibfnamefont {P.}~\bibnamefont {Bizzeti}},
  \bibinfo {author} {\bibfnamefont {A.}~\bibnamefont {Bizzeti-Sona}}, \bibinfo
  {author} {\bibfnamefont {D.}~\bibnamefont {Bazzacco}}, \bibinfo {author}
  {\bibfnamefont {S.}~\bibnamefont {Lunardi}}, \bibinfo {author} {\bibfnamefont
  {P.}~\bibnamefont {Pavan}}, \bibinfo {author} {\bibfnamefont
  {C.}~\bibnamefont {Rossi-Alvarez}}, \bibinfo {author} {\bibfnamefont
  {G.}~\bibnamefont {de~Angelis}}, \bibinfo {author} {\bibfnamefont
  {G.}~\bibnamefont {Maron}},\ and\ \bibinfo {author} {\bibfnamefont
  {J.}~\bibnamefont {Rico}},\ }\bibfield  {title} {\bibinfo {title}
  {Double-octupole excitations in the n=84 nuclei $^{144}\mathrm{Nd}$ and
  $^{146}\mathrm{Sm}$},\ }\href {https://doi.org/10.1103/PhysRevC.51.R1057}
  {\bibfield  {journal} {\bibinfo  {journal} {Phys. Rev. C}\ }\textbf {\bibinfo
  {volume} {51}},\ \bibinfo {pages} {R1057} (\bibinfo {year}
  {1995})}\BibitemShut {NoStop}%
\bibitem [{\citenamefont {Caballero}\ \emph {et~al.}(2005)\citenamefont
  {Caballero}, \citenamefont {Rubio}, \citenamefont {Kleinheinz}, \citenamefont
  {Algora}, \citenamefont {Dewald}, \citenamefont {Fitzler}, \citenamefont
  {Gadea}, \citenamefont {Jolie}, \citenamefont {Julin}, \citenamefont
  {Linnemann}, \citenamefont {Lunardi}, \citenamefont {Menegazzo},
  \citenamefont {M\"oller}, \citenamefont {N\`acher}, \citenamefont
  {Piiparinen},\ and\ \citenamefont {Yates}}]{Caballero2005}%
  \BibitemOpen
  \bibfield  {author} {\bibinfo {author} {\bibfnamefont {L.}~\bibnamefont
  {Caballero}}, \bibinfo {author} {\bibfnamefont {B.}~\bibnamefont {Rubio}},
  \bibinfo {author} {\bibfnamefont {P.}~\bibnamefont {Kleinheinz}}, \bibinfo
  {author} {\bibfnamefont {A.}~\bibnamefont {Algora}}, \bibinfo {author}
  {\bibfnamefont {A.}~\bibnamefont {Dewald}}, \bibinfo {author} {\bibfnamefont
  {A.}~\bibnamefont {Fitzler}}, \bibinfo {author} {\bibfnamefont
  {A.}~\bibnamefont {Gadea}}, \bibinfo {author} {\bibfnamefont
  {J.}~\bibnamefont {Jolie}}, \bibinfo {author} {\bibfnamefont
  {R.}~\bibnamefont {Julin}}, \bibinfo {author} {\bibfnamefont
  {A.}~\bibnamefont {Linnemann}}, \bibinfo {author} {\bibfnamefont
  {S.}~\bibnamefont {Lunardi}}, \bibinfo {author} {\bibfnamefont
  {R.}~\bibnamefont {Menegazzo}}, \bibinfo {author} {\bibfnamefont
  {O.}~\bibnamefont {M\"oller}}, \bibinfo {author} {\bibfnamefont
  {E.}~\bibnamefont {N\`acher}}, \bibinfo {author} {\bibfnamefont
  {M.}~\bibnamefont {Piiparinen}},\ and\ \bibinfo {author} {\bibfnamefont
  {S.}~\bibnamefont {Yates}},\ }\bibfield  {title} {\bibinfo {title}
  {Two‐phonon octupole excitation in ${}^{146}$gd},\ }\href
  {https://doi.org/10.1063/1.2140669} {\bibfield  {journal} {\bibinfo
  {journal} {AIP Conference Proceedings}\ }\textbf {\bibinfo {volume} {802}},\
  \bibinfo {pages} {287} (\bibinfo {year} {2005})}\BibitemShut {NoStop}%
\bibitem [{\citenamefont {Pysmenetska}\ \emph {et~al.}(2006)\citenamefont
  {Pysmenetska}, \citenamefont {Walter}, \citenamefont {Enders}, \citenamefont
  {Garrel}, \citenamefont {Karg}, \citenamefont {Kneissl}, \citenamefont
  {Kohstall}, \citenamefont {Neumann-Cosel}, \citenamefont {Pitz},
  \citenamefont {Ponomarev}, \citenamefont {Scheck}, \citenamefont {Stedile},\
  and\ \citenamefont {Volz}}]{Pysmenetska2006}%
  \BibitemOpen
  \bibfield  {author} {\bibinfo {author} {\bibfnamefont {I.}~\bibnamefont
  {Pysmenetska}}, \bibinfo {author} {\bibfnamefont {S.}~\bibnamefont {Walter}},
  \bibinfo {author} {\bibfnamefont {J.}~\bibnamefont {Enders}}, \bibinfo
  {author} {\bibfnamefont {H.~v.}\ \bibnamefont {Garrel}}, \bibinfo {author}
  {\bibfnamefont {O.}~\bibnamefont {Karg}}, \bibinfo {author} {\bibfnamefont
  {U.}~\bibnamefont {Kneissl}}, \bibinfo {author} {\bibfnamefont
  {C.}~\bibnamefont {Kohstall}}, \bibinfo {author} {\bibfnamefont {P.~v.}\
  \bibnamefont {Neumann-Cosel}}, \bibinfo {author} {\bibfnamefont {H.~H.}\
  \bibnamefont {Pitz}}, \bibinfo {author} {\bibfnamefont {V.~Y.}\ \bibnamefont
  {Ponomarev}}, \bibinfo {author} {\bibfnamefont {M.}~\bibnamefont {Scheck}},
  \bibinfo {author} {\bibfnamefont {F.}~\bibnamefont {Stedile}},\ and\ \bibinfo
  {author} {\bibfnamefont {S.}~\bibnamefont {Volz}},\ }\bibfield  {title}
  {\bibinfo {title} {Two-phonon ${1}^{\ensuremath{-}}$ state in
  $^{112}\mathrm{Sn}$ observed in resonant photon scattering},\ }\href
  {https://doi.org/10.1103/PhysRevC.73.017302} {\bibfield  {journal} {\bibinfo
  {journal} {Phys. Rev. C}\ }\textbf {\bibinfo {volume} {73}},\ \bibinfo
  {pages} {017302} (\bibinfo {year} {2006})}\BibitemShut {NoStop}%
\bibitem [{\citenamefont {Isaak}\ \emph {et~al.}(2023)\citenamefont {Isaak},
  \citenamefont {Savran}, \citenamefont {Pietralla}, \citenamefont {Tsoneva},
  \citenamefont {Zilges}, \citenamefont {Eberhardt}, \citenamefont {Geppert},
  \citenamefont {Gorges}, \citenamefont {Lenske},\ and\ \citenamefont
  {Renisch}}]{Isaak2023}%
  \BibitemOpen
  \bibfield  {author} {\bibinfo {author} {\bibfnamefont {J.}~\bibnamefont
  {Isaak}}, \bibinfo {author} {\bibfnamefont {D.}~\bibnamefont {Savran}},
  \bibinfo {author} {\bibfnamefont {N.}~\bibnamefont {Pietralla}}, \bibinfo
  {author} {\bibfnamefont {N.}~\bibnamefont {Tsoneva}}, \bibinfo {author}
  {\bibfnamefont {A.}~\bibnamefont {Zilges}}, \bibinfo {author} {\bibfnamefont
  {K.}~\bibnamefont {Eberhardt}}, \bibinfo {author} {\bibfnamefont
  {C.}~\bibnamefont {Geppert}}, \bibinfo {author} {\bibfnamefont
  {C.}~\bibnamefont {Gorges}}, \bibinfo {author} {\bibfnamefont
  {H.}~\bibnamefont {Lenske}},\ and\ \bibinfo {author} {\bibfnamefont
  {D.}~\bibnamefont {Renisch}},\ }\bibfield  {title} {\bibinfo {title} {Direct
  demonstration of the two-phonon structure of the
  ${J}^{\ensuremath{\pi}}={1}_{4742\phantom{\rule{0.16em}{0ex}}\mathrm{keV}}^{\ensuremath{-}}$
  state of $^{88}\mathrm{Sr}$},\ }\href
  {https://doi.org/10.1103/PhysRevC.108.L051301} {\bibfield  {journal}
  {\bibinfo  {journal} {Phys. Rev. C}\ }\textbf {\bibinfo {volume} {108}},\
  \bibinfo {pages} {L051301} (\bibinfo {year} {2023})}\BibitemShut {NoStop}%
\bibitem [{\citenamefont {Lifshitz}\ and\ \citenamefont
  {Singer}(1980)}]{Lifshitz1980}%
  \BibitemOpen
  \bibfield  {author} {\bibinfo {author} {\bibfnamefont {M.}~\bibnamefont
  {Lifshitz}}\ and\ \bibinfo {author} {\bibfnamefont {P.}~\bibnamefont
  {Singer}},\ }\bibfield  {title} {\bibinfo {title} {Nuclear excitation
  function and particle emission from complex nuclei following muon capture},\
  }\href {https://doi.org/10.1103/PhysRevC.22.2135} {\bibfield  {journal}
  {\bibinfo  {journal} {Phys. Rev. C}\ }\textbf {\bibinfo {volume} {22}},\
  \bibinfo {pages} {2135} (\bibinfo {year} {1980})}\BibitemShut {NoStop}%
\bibitem [{\citenamefont {Minato}\ \emph {et~al.}(2023)\citenamefont {Minato},
  \citenamefont {Naito},\ and\ \citenamefont {Iwamoto}}]{Minato2023}%
  \BibitemOpen
  \bibfield  {author} {\bibinfo {author} {\bibfnamefont {F.}~\bibnamefont
  {Minato}}, \bibinfo {author} {\bibfnamefont {T.}~\bibnamefont {Naito}},\ and\
  \bibinfo {author} {\bibfnamefont {O.}~\bibnamefont {Iwamoto}},\ }\bibfield
  {title} {\bibinfo {title} {Nuclear many-body effects on particle emission
  following muon capture on $^{28}\mathrm{Si}$ and $^{40}\mathrm{Ca}$},\ }\href
  {https://doi.org/10.1103/PhysRevC.107.054314} {\bibfield  {journal} {\bibinfo
   {journal} {Phys. Rev. C}\ }\textbf {\bibinfo {volume} {107}},\ \bibinfo
  {pages} {054314} (\bibinfo {year} {2023})}\BibitemShut {NoStop}%
\bibitem [{\citenamefont {Men\'endez}\ \emph {et~al.}(2011)\citenamefont
  {Men\'endez}, \citenamefont {Gazit},\ and\ \citenamefont
  {Schwenk}}]{Menendez2011}%
  \BibitemOpen
  \bibfield  {author} {\bibinfo {author} {\bibfnamefont {J.}~\bibnamefont
  {Men\'endez}}, \bibinfo {author} {\bibfnamefont {D.}~\bibnamefont {Gazit}},\
  and\ \bibinfo {author} {\bibfnamefont {A.}~\bibnamefont {Schwenk}},\
  }\bibfield  {title} {\bibinfo {title} {Chiral two-body currents in nuclei:
  Gamow-teller transitions and neutrinoless double-beta decay},\ }\href
  {https://doi.org/10.1103/PhysRevLett.107.062501} {\bibfield  {journal}
  {\bibinfo  {journal} {Phys. Rev. Lett.}\ }\textbf {\bibinfo {volume} {107}},\
  \bibinfo {pages} {062501} (\bibinfo {year} {2011})}\BibitemShut {NoStop}%
\bibitem [{\citenamefont {Bacca}\ and\ \citenamefont
  {Pastore}(2014)}]{Bacca2014}%
  \BibitemOpen
  \bibfield  {author} {\bibinfo {author} {\bibfnamefont {S.}~\bibnamefont
  {Bacca}}\ and\ \bibinfo {author} {\bibfnamefont {S.}~\bibnamefont
  {Pastore}},\ }\bibfield  {title} {\bibinfo {title} {Electromagnetic reactions
  on light nuclei},\ }\href {https://doi.org/10.1088/0954-3899/41/12/123002}
  {\bibfield  {journal} {\bibinfo  {journal} {Journal of Physics G: Nuclear and
  Particle Physics}\ }\textbf {\bibinfo {volume} {41}},\ \bibinfo {pages}
  {123002} (\bibinfo {year} {2014})}\BibitemShut {NoStop}%
\bibitem [{\citenamefont {Brase}\ \emph {et~al.}(2026)\citenamefont {Brase},
  \citenamefont {Miyagi}, \citenamefont {Men\'endez},\ and\ \citenamefont
  {Schwenk}}]{Brase2026}%
  \BibitemOpen
  \bibfield  {author} {\bibinfo {author} {\bibfnamefont {C.}~\bibnamefont
  {Brase}}, \bibinfo {author} {\bibfnamefont {T.}~\bibnamefont {Miyagi}},
  \bibinfo {author} {\bibfnamefont {J.}~\bibnamefont {Men\'endez}},\ and\
  \bibinfo {author} {\bibfnamefont {A.}~\bibnamefont {Schwenk}},\ }\bibfield
  {title} {\bibinfo {title} {Two-body currents at finite momentum transfer and
  applications to $m1$ transitions},\ }\href
  {https://doi.org/10.1103/4tky-t2h1} {\bibfield  {journal} {\bibinfo
  {journal} {Phys. Rev. C}\ }\textbf {\bibinfo {volume} {113}},\ \bibinfo
  {pages} {014317} (\bibinfo {year} {2026})}\BibitemShut {NoStop}%
\bibitem [{\citenamefont {Yannouleas}(1987)}]{Yannouleas1987}%
  \BibitemOpen
  \bibfield  {author} {\bibinfo {author} {\bibfnamefont {C.}~\bibnamefont
  {Yannouleas}},\ }\bibfield  {title} {\bibinfo {title} {Zero-temperature
  second random phase approximation and its formal properties},\ }\href
  {https://doi.org/10.1103/PhysRevC.35.1159} {\bibfield  {journal} {\bibinfo
  {journal} {Phys. Rev. C}\ }\textbf {\bibinfo {volume} {35}},\ \bibinfo
  {pages} {1159} (\bibinfo {year} {1987})}\BibitemShut {NoStop}%
\bibitem [{\citenamefont {{Van Giai}}\ and\ \citenamefont
  {Sagawa}(1981)}]{Giai1981}%
  \BibitemOpen
  \bibfield  {author} {\bibinfo {author} {\bibfnamefont {N.}~\bibnamefont {{Van
  Giai}}}\ and\ \bibinfo {author} {\bibfnamefont {H.}~\bibnamefont {Sagawa}},\
  }\bibfield  {title} {\bibinfo {title} {Spin-isospin and pairing properties of
  modified skyrme interactions},\ }\href
  {https://doi.org/https://doi.org/10.1016/0370-2693(81)90646-8} {\bibfield
  {journal} {\bibinfo  {journal} {Physics Letters B}\ }\textbf {\bibinfo
  {volume} {106}},\ \bibinfo {pages} {379} (\bibinfo {year}
  {1981})}\BibitemShut {NoStop}%
\bibitem [{\citenamefont {Vautherin}\ and\ \citenamefont
  {Brink}(1972)}]{VautherinBrink}%
  \BibitemOpen
  \bibfield  {author} {\bibinfo {author} {\bibfnamefont {D.}~\bibnamefont
  {Vautherin}}\ and\ \bibinfo {author} {\bibfnamefont {D.~M.}\ \bibnamefont
  {Brink}},\ }\bibfield  {title} {\bibinfo {title} {{Hartree-Fock Calculations
  with Skyrme's Interaction. I. Spherical Nuclei}},\ }\href
  {https://doi.org/10.1103/PhysRevC.5.626} {\bibfield  {journal} {\bibinfo
  {journal} {Phys. Rev. C}\ }\textbf {\bibinfo {volume} {5}},\ \bibinfo {pages}
  {626} (\bibinfo {year} {1972})}\BibitemShut {NoStop}%
\bibitem [{\citenamefont {Slater}(1951)}]{Slater1951}%
  \BibitemOpen
  \bibfield  {author} {\bibinfo {author} {\bibfnamefont {J.~C.}\ \bibnamefont
  {Slater}},\ }\bibfield  {title} {\bibinfo {title} {A simplification of the
  hartree-fock method},\ }\href {https://doi.org/10.1103/PhysRev.81.385}
  {\bibfield  {journal} {\bibinfo  {journal} {Phys. Rev.}\ }\textbf {\bibinfo
  {volume} {81}},\ \bibinfo {pages} {385} (\bibinfo {year} {1951})}\BibitemShut
  {NoStop}%
\bibitem [{\citenamefont {ROWE}(1968)}]{Rowe1968}%
  \BibitemOpen
  \bibfield  {author} {\bibinfo {author} {\bibfnamefont {D.~J.}\ \bibnamefont
  {ROWE}},\ }\bibfield  {title} {\bibinfo {title} {Equations-of-motion method
  and the extended shell model},\ }\href
  {https://doi.org/10.1103/RevModPhys.40.153} {\bibfield  {journal} {\bibinfo
  {journal} {Rev. Mod. Phys.}\ }\textbf {\bibinfo {volume} {40}},\ \bibinfo
  {pages} {153} (\bibinfo {year} {1968})}\BibitemShut {NoStop}%
\bibitem [{\citenamefont {Nakada}\ \emph {et~al.}(2009)\citenamefont {Nakada},
  \citenamefont {Mizuyama}, \citenamefont {Yamagami},\ and\ \citenamefont
  {Matsuo}}]{Nakada2009}%
  \BibitemOpen
  \bibfield  {author} {\bibinfo {author} {\bibfnamefont {H.}~\bibnamefont
  {Nakada}}, \bibinfo {author} {\bibfnamefont {K.}~\bibnamefont {Mizuyama}},
  \bibinfo {author} {\bibfnamefont {M.}~\bibnamefont {Yamagami}},\ and\
  \bibinfo {author} {\bibfnamefont {M.}~\bibnamefont {Matsuo}},\ }\bibfield
  {title} {\bibinfo {title} {Rpa calculations with gaussian expansion method},\
  }\href {https://doi.org/https://doi.org/10.1016/j.nuclphysa.2009.07.010}
  {\bibfield  {journal} {\bibinfo  {journal} {Nuclear Physics A}\ }\textbf
  {\bibinfo {volume} {828}},\ \bibinfo {pages} {283} (\bibinfo {year}
  {2009})}\BibitemShut {NoStop}%
\bibitem [{\citenamefont {Papakonstantinou}(2014)}]{Papa2014}%
  \BibitemOpen
  \bibfield  {author} {\bibinfo {author} {\bibfnamefont {P.}~\bibnamefont
  {Papakonstantinou}},\ }\bibfield  {title} {\bibinfo {title} {Second
  random-phase approximation, thouless' theorem, and the stability condition
  reexamined and clarified},\ }\href
  {https://doi.org/10.1103/PhysRevC.90.024305} {\bibfield  {journal} {\bibinfo
  {journal} {Phys. Rev. C}\ }\textbf {\bibinfo {volume} {90}},\ \bibinfo
  {pages} {024305} (\bibinfo {year} {2014})}\BibitemShut {NoStop}%
\bibitem [{\citenamefont {Mizuyama}\ and\ \citenamefont
  {Col\`o}(2012)}]{Mizuyama2012}%
  \BibitemOpen
  \bibfield  {author} {\bibinfo {author} {\bibfnamefont {K.}~\bibnamefont
  {Mizuyama}}\ and\ \bibinfo {author} {\bibfnamefont {G.}~\bibnamefont
  {Col\`o}},\ }\bibfield  {title} {\bibinfo {title} {Subtraction of the
  spurious translational mode from the random-phase approximation response
  function},\ }\href {https://doi.org/10.1103/PhysRevC.85.024307} {\bibfield
  {journal} {\bibinfo  {journal} {Phys. Rev. C}\ }\textbf {\bibinfo {volume}
  {85}},\ \bibinfo {pages} {024307} (\bibinfo {year} {2012})}\BibitemShut
  {NoStop}%
\bibitem [{\citenamefont {Inakura}\ \emph {et~al.}(2009)\citenamefont
  {Inakura}, \citenamefont {Nakatsukasa},\ and\ \citenamefont
  {Yabana}}]{Inakura2009}%
  \BibitemOpen
  \bibfield  {author} {\bibinfo {author} {\bibfnamefont {T.}~\bibnamefont
  {Inakura}}, \bibinfo {author} {\bibfnamefont {T.}~\bibnamefont
  {Nakatsukasa}},\ and\ \bibinfo {author} {\bibfnamefont {K.}~\bibnamefont
  {Yabana}},\ }\bibfield  {title} {\bibinfo {title} {Self-consistent
  calculation of nuclear photoabsorption cross sections: Finite amplitude
  method with skyrme functionals in the three-dimensional real space},\ }\href
  {https://doi.org/10.1103/PhysRevC.80.044301} {\bibfield  {journal} {\bibinfo
  {journal} {Phys. Rev. C}\ }\textbf {\bibinfo {volume} {80}},\ \bibinfo
  {pages} {044301} (\bibinfo {year} {2009})}\BibitemShut {NoStop}%
\bibitem [{\citenamefont {Knapp}\ \emph {et~al.}(2023)\citenamefont {Knapp},
  \citenamefont {Papakonstantinou}, \citenamefont {Vesel\'y}, \citenamefont
  {De~Gregorio}, \citenamefont {Herko},\ and\ \citenamefont
  {Lo~Iudice}}]{Knapp2023}%
  \BibitemOpen
  \bibfield  {author} {\bibinfo {author} {\bibfnamefont {F.}~\bibnamefont
  {Knapp}}, \bibinfo {author} {\bibfnamefont {P.}~\bibnamefont
  {Papakonstantinou}}, \bibinfo {author} {\bibfnamefont {P.}~\bibnamefont
  {Vesel\'y}}, \bibinfo {author} {\bibfnamefont {G.}~\bibnamefont
  {De~Gregorio}}, \bibinfo {author} {\bibfnamefont {J.}~\bibnamefont {Herko}},\
  and\ \bibinfo {author} {\bibfnamefont {N.}~\bibnamefont {Lo~Iudice}},\
  }\bibfield  {title} {\bibinfo {title} {Comparative analysis of formalisms and
  performances of three different beyond-mean-field approaches},\ }\href
  {https://doi.org/10.1103/PhysRevC.107.014305} {\bibfield  {journal} {\bibinfo
   {journal} {Phys. Rev. C}\ }\textbf {\bibinfo {volume} {107}},\ \bibinfo
  {pages} {014305} (\bibinfo {year} {2023})}\BibitemShut {NoStop}%
\bibitem [{\citenamefont {Gambacurta}\ \emph {et~al.}(2010)\citenamefont
  {Gambacurta}, \citenamefont {Grasso},\ and\ \citenamefont
  {Catara}}]{Gambacurta2010}%
  \BibitemOpen
  \bibfield  {author} {\bibinfo {author} {\bibfnamefont {D.}~\bibnamefont
  {Gambacurta}}, \bibinfo {author} {\bibfnamefont {M.}~\bibnamefont {Grasso}},\
  and\ \bibinfo {author} {\bibfnamefont {F.}~\bibnamefont {Catara}},\
  }\bibfield  {title} {\bibinfo {title} {Collective nuclear excitations with
  skyrme-second random-phase approximation},\ }\href
  {https://doi.org/10.1103/PhysRevC.81.054312} {\bibfield  {journal} {\bibinfo
  {journal} {Phys. Rev. C}\ }\textbf {\bibinfo {volume} {81}},\ \bibinfo
  {pages} {054312} (\bibinfo {year} {2010})}\BibitemShut {NoStop}%
\bibitem [{\citenamefont {Lui}\ \emph {et~al.}(2001)\citenamefont {Lui},
  \citenamefont {Clark},\ and\ \citenamefont {Youngblood}}]{Lui2001}%
  \BibitemOpen
  \bibfield  {author} {\bibinfo {author} {\bibfnamefont {Y.-W.}\ \bibnamefont
  {Lui}}, \bibinfo {author} {\bibfnamefont {H.~L.}\ \bibnamefont {Clark}},\
  and\ \bibinfo {author} {\bibfnamefont {D.~H.}\ \bibnamefont {Youngblood}},\
  }\bibfield  {title} {\bibinfo {title} {Giant resonances in
  ${}^{16}\mathrm{O}$},\ }\href {https://doi.org/10.1103/PhysRevC.64.064308}
  {\bibfield  {journal} {\bibinfo  {journal} {Phys. Rev. C}\ }\textbf {\bibinfo
  {volume} {64}},\ \bibinfo {pages} {064308} (\bibinfo {year}
  {2001})}\BibitemShut {NoStop}%
\bibitem [{\citenamefont {Ahrens}\ \emph {et~al.}(1975)\citenamefont {Ahrens},
  \citenamefont {Borchert}, \citenamefont {Czock}, \citenamefont {Eppler},
  \citenamefont {Gimm}, \citenamefont {Gundrum}, \citenamefont {Kr\"oning},
  \citenamefont {Riehn}, \citenamefont {{Sita Ram}}, \citenamefont {Zieger},\
  and\ \citenamefont {Ziegler}}]{Ahrens1975}%
  \BibitemOpen
  \bibfield  {author} {\bibinfo {author} {\bibfnamefont {J.}~\bibnamefont
  {Ahrens}}, \bibinfo {author} {\bibfnamefont {H.}~\bibnamefont {Borchert}},
  \bibinfo {author} {\bibfnamefont {K.}~\bibnamefont {Czock}}, \bibinfo
  {author} {\bibfnamefont {H.}~\bibnamefont {Eppler}}, \bibinfo {author}
  {\bibfnamefont {H.}~\bibnamefont {Gimm}}, \bibinfo {author} {\bibfnamefont
  {H.}~\bibnamefont {Gundrum}}, \bibinfo {author} {\bibfnamefont
  {M.}~\bibnamefont {Kr\"oning}}, \bibinfo {author} {\bibfnamefont
  {P.}~\bibnamefont {Riehn}}, \bibinfo {author} {\bibfnamefont
  {G.}~\bibnamefont {{Sita Ram}}}, \bibinfo {author} {\bibfnamefont
  {A.}~\bibnamefont {Zieger}},\ and\ \bibinfo {author} {\bibfnamefont
  {B.}~\bibnamefont {Ziegler}},\ }\bibfield  {title} {\bibinfo {title} {Total
  nuclear photon absorption cross sections for some light elements},\ }\href
  {https://doi.org/https://doi.org/10.1016/0375-9474(75)90543-6} {\bibfield
  {journal} {\bibinfo  {journal} {Nuclear Physics A}\ }\textbf {\bibinfo
  {volume} {251}},\ \bibinfo {pages} {479} (\bibinfo {year}
  {1975})}\BibitemShut {NoStop}%
\bibitem [{\citenamefont {Kib\'{e}di}\ and\ \citenamefont
  {Spear}(2002)}]{Kibedi2002}%
  \BibitemOpen
  \bibfield  {author} {\bibinfo {author} {\bibfnamefont {T.}~\bibnamefont
  {Kib\'{e}di}}\ and\ \bibinfo {author} {\bibfnamefont {R.}~\bibnamefont
  {Spear}},\ }\bibfield  {title} {\bibinfo {title} {Reduced electric-octupole
  transition probabilities, $b(e3;0_{1}^{+}\rightarrow 3_{1}^{-})$-an update},\
  }\href {https://doi.org/https://doi.org/10.1006/adnd.2001.0871} {\bibfield
  {journal} {\bibinfo  {journal} {Atomic Data and Nuclear Data Tables}\
  }\textbf {\bibinfo {volume} {80}},\ \bibinfo {pages} {35} (\bibinfo {year}
  {2002})}\BibitemShut {NoStop}%
\bibitem [{\citenamefont {Scamps}\ and\ \citenamefont
  {Lacroix}(2013)}]{Scamps2013}%
  \BibitemOpen
  \bibfield  {author} {\bibinfo {author} {\bibfnamefont {G.}~\bibnamefont
  {Scamps}}\ and\ \bibinfo {author} {\bibfnamefont {D.}~\bibnamefont
  {Lacroix}},\ }\bibfield  {title} {\bibinfo {title} {Systematics of isovector
  and isoscalar giant quadrupole resonances in normal and superfluid spherical
  nuclei},\ }\href {https://doi.org/10.1103/PhysRevC.88.044310} {\bibfield
  {journal} {\bibinfo  {journal} {Phys. Rev. C}\ }\textbf {\bibinfo {volume}
  {88}},\ \bibinfo {pages} {044310} (\bibinfo {year} {2013})}\BibitemShut
  {NoStop}%
\bibitem [{\citenamefont {Lyutorovich}\ \emph {et~al.}(2012)\citenamefont
  {Lyutorovich}, \citenamefont {Tselyaev}, \citenamefont {Speth}, \citenamefont
  {Krewald}, \citenamefont {Gr\"ummer},\ and\ \citenamefont
  {Reinhard}}]{Lyutorovich2012}%
  \BibitemOpen
  \bibfield  {author} {\bibinfo {author} {\bibfnamefont {N.}~\bibnamefont
  {Lyutorovich}}, \bibinfo {author} {\bibfnamefont {V.~I.}\ \bibnamefont
  {Tselyaev}}, \bibinfo {author} {\bibfnamefont {J.}~\bibnamefont {Speth}},
  \bibinfo {author} {\bibfnamefont {S.}~\bibnamefont {Krewald}}, \bibinfo
  {author} {\bibfnamefont {F.}~\bibnamefont {Gr\"ummer}},\ and\ \bibinfo
  {author} {\bibfnamefont {P.-G.}\ \bibnamefont {Reinhard}},\ }\bibfield
  {title} {\bibinfo {title} {Self-consistent calculations of the electric giant
  dipole resonances in light and heavy nuclei},\ }\href
  {https://doi.org/10.1103/PhysRevLett.109.092502} {\bibfield  {journal}
  {\bibinfo  {journal} {Phys. Rev. Lett.}\ }\textbf {\bibinfo {volume} {109}},\
  \bibinfo {pages} {092502} (\bibinfo {year} {2012})}\BibitemShut {NoStop}%
\bibitem [{\citenamefont {Chomaz}\ and\ \citenamefont {{Van
  Giai}}(1992)}]{Chomaz1992}%
  \BibitemOpen
  \bibfield  {author} {\bibinfo {author} {\bibfnamefont {P.}~\bibnamefont
  {Chomaz}}\ and\ \bibinfo {author} {\bibfnamefont {N.}~\bibnamefont {{Van
  Giai}}},\ }\bibfield  {title} {\bibinfo {title} {Comment on the width and the
  lifetime of multiphonon states},\ }\href
  {https://doi.org/https://doi.org/10.1016/0370-2693(92)90471-F} {\bibfield
  {journal} {\bibinfo  {journal} {Physics Letters B}\ }\textbf {\bibinfo
  {volume} {282}},\ \bibinfo {pages} {13} (\bibinfo {year} {1992})}\BibitemShut
  {NoStop}%
\end{thebibliography}%

\end{document}